\documentclass[aps,rmp,reprint,amsmath,amssymb,graphicx,longbibliography]{revtex4-1} 

\usepackage{hyperref}
\usepackage{bm}
\usepackage{array}
\usepackage{amsmath}
\usepackage{slashed}
\usepackage{xcolor}
\usepackage{graphicx}
\usepackage{caption}
\usepackage{subcaption}
\usepackage[normalem]{ulem}
\usepackage[utf8]{inputenc}
\usepackage[T1]{fontenc}
\newcolumntype{L}{>$l<$}

\newcommand{\MS}{{\overline{\rm MS}}}

\newcommand{\beq}{\begin{eqnarray}}
\newcommand{\eeq}{\end{eqnarray}}

\newcommand{\nn}{\nonumber}

\DeclareRobustCommand{\eq}[1]{Eq.~(\ref{eq:#1})}
\DeclareRobustCommand{\eqs}[2]{Eqs.~(\ref{eq:#1}) and (\ref{eq:#2})}
\DeclareRobustCommand{\sec}[1]{Sec.~\ref{sec:#1}}
\DeclareRobustCommand{\secs}[2]{Secs.~(\ref{sec:#1}) and (\ref{sec:#2})}
\DeclareRobustCommand{\fig}[1]{Fig.~\ref{fig:#1}}

\begin{document}

\title{Large-Momentum Effective Theory}

\author{Xiangdong Ji}
\email{xji@umd.edu}
\affiliation{Maryland Center for Fundamental Physics,
Department of Physics, University of Maryland,
College Park, Maryland 20742, USA}
\affiliation{Tsung-Dao Lee Institute, Shanghai Jiao Tong University, Shanghai, 200240, China}

\author{Yizhuang Liu}
\email{yizhuang.liu@sjtu.edu.cn}
\affiliation{Tsung-Dao Lee Institute, Shanghai Jiao Tong University, Shanghai, 200240, China}
\affiliation{Institut fur Theoretische Physik, Universitat Regensburg, D-93040 Regensburg, Germany}
\affiliation{Institute of Theoretical Physics,Jagiellonian University, 30-348 Kraków, Poland}

\author{Yu-Sheng Liu}
\email{mestelqure@gmail.com}
\affiliation{Tsung-Dao Lee Institute, Shanghai Jiao Tong University, Shanghai, 200240, China}

\author{Jian-Hui Zhang}
\email{zhangjianhui@bnu.edu.cn}
\affiliation{Center of Advanced Quantum Studies, Department of Physics, Beijing Normal University, Beijing 100875, China}

\author{Yong Zhao}
\email{yong.zhao@anl.gov}
\affiliation{Physics Department, Brookhaven National Laboratory Bldg. 510A, Upton, NY 11973, USA}
\affiliation{Physics Division, Argonne National Laboratory, Lemont, IL 60439, USA}

\date{\today{}}

\begin{abstract}
Since the parton model was introduced by Feynman more than fifty years ago,
we have learned much about the partonic structure of the proton through a
large body of high-energy experimental data and dedicated global fits. However,
calculating the partonic observables such as parton distribution function (PDFs)
from the fundamental theory of strong interactions, QCD, has made limited
progress. Recently, the authors have advocated a formalism, large-momentum effective theory (LaMET),
through which one can extract parton physics from the properties of the proton
travelling at a moderate boost-factor, e.g., $\gamma\sim (2-5)$.
The key observation behind this approach is that Lorentz symmetry allows the standard
formalism of partons in terms of light-front operators to be replaced by
an equivalent one with large-momentum states and time-independent operators of
a universality class. With LaMET, the PDFs, generalized PDFs or GPDs,
transverse-momentum-dependent PDFs, and light-front wave functions
can all be extracted in principle from lattice simulations of
QCD (or other non-perturbative methods) through standard effective field theory matching
and running. Future lattice QCD calculations with exa-scale computational
facilities can help to understand the experimental data related to the
hadronic structure, including those from the upcoming Electron-Ion Colliders
dedicated to exploring the partonic landscape of the proton. Here we review the progress
made in the past few years in development of the LaMET formalism and its applications,
particularly on the demonstration of its effectiveness from initial lattice QCD simulations.
\end{abstract}
\maketitle

\tableofcontents{}

\section{Introduction}
\label{sec:intro}
The proton and neutron, collectively called the nucleon, are the basic
building blocks of visible matter in the universe today. Ever since they were
discovered in laboratories nearly a century ago~\cite{Rutherford:1919fnt,Chadwick:1932ma}, their fundamental
properties have been vigorously explored:
from the determination of the spin through the specific heat of liquid hydrogen~\cite{Dennison},
to the measurement of the magnetic moments~\cite{Stern}, and the extraction of their electromagnetic sizes through elastic electron scattering
~\cite{Hofstadter:1956qs}. The most revealing discovery, however, came from the
electron deep-inelastic scattering (DIS) on the proton and nuclei at Stanford Linear Accelerator
Center (SLAC) in the late 1960s, in which the constituents of the
proton and neutron, quarks (and later gluons), were discovered~\cite{Bloom:1969kc}.
Soon after, quantum chromodynamics (QCD), a
quantum field theory (QFT) based on ``color'' SU(3) gauge symmetry,
was established as the fundamental theory of strong interactions~\cite{Fritzsch:1973pi,Gross:1973id,Politzer:1973fx},
and of the internal structure of the nucleon as well~\cite{Thomas:2001kw}.

\begin{figure}[htb]	
	\includegraphics[width=0.9\linewidth]{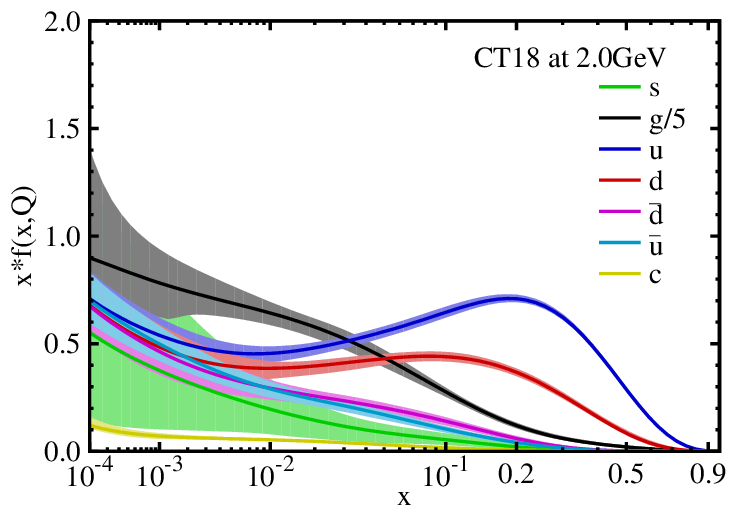}
	\caption{Phenomenological parton distributions obtained by the CTEQ-TEA collaboration (CT18) from
		fits to global high-energy scattering data~\cite{Hou:2019efy}, where $0\le x\le 1$ is the fraction of the proton's
		infinite momentum carried in a parton.}
	\label{fig:PDFS}
\end{figure}

During the last fifty years, significant progress has been made
in understanding the nucleon's internal structure in both experiment and theory.
Multiple experimental facilities have been built to study
high-energy collisions involving protons and nuclei, from which
a large amount of experimental data has been accumulated.
Based on the QCD factorization theorems~\cite{Collins:2011zzd}, derived from perturbative QCD
analyses beyond Feynman's parton model~\cite{Feynman:1973xc},
the parton distribution functions (PDFs), which characterize the
longitudinal momentum distributions of quarks and gluons in hadrons moving at
infinite momentum, have been obtained from global fits
to these data~\cite{Gao:2017yyd, Hou:2019efy, Harland-Lang:2014zoa, Ball:2017nwa}. A recent
result of the phenomenological proton PDFs is shown in Fig.~\ref{fig:PDFS} where $x$
is the momentum fraction of the proton carried by partons.
The PDFs provide a comprehensive
description of the quark and gluon content of the nucleon.
On the theoretical frontier, the Euclidean path-integral formalism of QCD, combined with the lattice
regularization and Monte Carlo simulations~\cite{Wilson:1974sk}, has
offered a systematic way of performing {\it ab initio} calculations of non-perturbative strong interactions.
The rapid rise in computational power and development of intelligent numerical
algorithms have made such a lattice QCD approach extremely successful in computing hadron
spectroscopy, the strong coupling, hadronic form factors,
etc., and even scattering phase shifts~\cite{Tanabashi:2018oca,Briceno:2017max,Aoki:2019cca}.

Despite these impressive achievements, we have not been able to systematically
explain the partonic structure of the proton
from first principles, or more explicitly,
we have not made fundamental progress in computing
the quark and gluon distributions starting from the QCD Lagrangian (see \sec{intro_other} for a brief
summary). There is actually a good reason behind it: The
standard formulation of parton physics in the textbooks~\cite{Sterman:1994ce,Collins:2011zzd}
is accomplished through the {\it dynamical correlators} of
quark and gluon fields on the light-front (LF) defined by $t-z={\rm const.}$,
which has the important feature of being independent of the proton's momentum.
On the other hand, lattice QCD is formulated in the Euclidean space with imaginary time,
and cannot be used to directly calculate the dynamical correlations that depend on real time. The standard lattice
approach to parton physics has been to calculate the lower moments of parton distributions,
which are matrix elements of local operators~\cite{Lin:2017snn}.
However, the limitations to the first few moments prohibit practitioners
from reproducing reliably the $x$-dependent structure as shown in Fig.~\ref{fig:PDFS},
other than fitting model functional forms.
Over the years, Hamiltonian diagonalization in LF quantization (LFQ)~\cite{Brodsky:1997de} and
Schwinger-Dyson equations~\cite{Maris:2003vk} have been proposed to solve the nucleon
structure as Minkowskian approaches. Although significant advances have been
made phenomenologically, a systematic approximation to calculate the nucleon PDFs is still missing.

A few years ago, some of the present authors proposed
a general approach to calculate $x$-dependent parton distributions based on
Feynman's original idea about partons: They
are the infinite-momentum limit of {\it static properties of the proton}
at large momentum, and therefore are intrinsically
Euclidean quantities accessible through lattice
QCD~\cite{Ji:2013dva,Ji:2013fga,Ji:2014gla}.
According to this, parton physics in an intermediate
range of $x_{\rm min}\sim 0.1 <x<x_{\rm max}\sim 0.9$ can be calculated
from the physical properties of the proton at a moderately-large
momentum, e.g., with a Lorentz boost factor $\gamma=2-5$. The theory
has been named as {\it large-momentum effective theory} (LaMET)
because a rigorous connection between the infinite-momentum frame (IMF) partons
and quarks and gluons at a finite momentum requires
proper account of the ultraviolet (UV) modes with large momentum
in effective field theory (EFT) and systematic power counting.

The basic principle for LaMET comes from an implicit
observation in the naive parton model:
The structure of the proton is approximately
independent of its momentum so long as it is much larger than
a typical strong-interaction scale $\Lambda_{\rm QCD}$, or its mass.
For example, the quark momentum distribution at moderate $x$ in the proton
at $P=|\vec{P}|=5$ GeV is not very different from that
at $P=50$ GeV or $P=5$ TeV. One might call this phenomenon
{\it large-momentum symmetry}, the nature of which is similar to that of the electronic
structure of the hydrogen atom is not sensitive to the proton mass, so long
as it is much larger than that of the electron.
The asymptotic behavior of the proton structure might be
controlled by an expansion in $\Lambda_{\rm QCD}/P$, but a justification would
require a better understanding of the underlying
dynamics. Assuming this, Feynman replaced the protons probed
at large but finite momenta in high-energy scattering with
the one at infinite momentum $P=\infty$,
corresponding to the leading term in the $\Lambda_{\rm QCD}/P$ expansion, and therefore
the idealized concepts of {\rm the proton in the IMF}
and its constituents---{\it partons}---were born.

In QFTs, however, the existence of the $P=\infty$ limit
depends on their UV behavior. In general, the infinite-momentum limit
does not commute with the UV cut-off limit $\Lambda_{\rm UV}\to \infty$.
While the physical limit is $(\Lambda_{\rm UV}\gg P) \to\infty$,
the parton model and subsequent QCD factorization
theorems use $(P\gg\Lambda_{\rm UV})\to\infty$, keeping
all PDFs with the finite support $|x|\le 1$ where negative $x$ is for antiquarks. Thus partons are an idealized concept
which does not exist in the real world.
Fortunately, because of asymptotic freedom, the above differences can be calculated in perturbative QCD.
Therefore, LaMET is an effective theory of partons, which uses the ordinary
field theoretical calculations $(\Lambda_{\rm UV}\gg P)\to\infty$ and
systematically takes into account non-commuting $P\to\infty$ limits through
EFT matching and running and finite $P$ effects by power corrections. Thus, the PDFs defined in the IMF
or on the LF can be accessed at moderate $x$ from the structure
calculations at $P\sim $ a few GeVs.

The first application of LaMET was to the total gluon helicity $\Delta G$ in the
polarized proton, a quantity of significant experimental interest
at the polarized RHIC~\cite{Bunce:2000uv}, but not within
theoretical reach for many years. In~\cite{Ji:2013fga}, we have shown that from a large-momentum
matrix element of the gluon spin operator in a physical gauge,
$\Delta G$ can be obtained through an EFT matching.
Following this success, LaMET
was applied to the collinear quark PDFs~\cite{Ji:2013dva}. This latter application has
generated considerable theoretical as well as numerical activities,
particularly for the flavor non-singlet $u-d$ distributions in the proton
and other hadrons. A general LaMET framework was subsequently introduced in~\cite{Ji:2014gla}.
More recently, the approach has been extended to the gluons as well~\cite{Zhang:2018diq,Li:2018tpe}.
Therefore, the PDFs can now be computed directly in lattice QCD at specific Feynman variable $x$,
without using LFQ. Besides, the partonic landscape of the proton is extremely
rich, and LaMET holds the promise of computing parton physics beyond the collinear PDFs.

In recent years, tremendous progress has been made in formulating new parton observables
for the proton. In particular, two parallel concepts have been developed
in characterizing the transverse structure of the proton. The first is the generalized
parton distributions (GPDs)~\cite{Mueller:1998fv,Ji:1996ek,Radyushkin:1998es}. The GPDs combine the features of the proton's
elastic form factors, which provide the transverse-space density of partons~\cite{Miller:2007uy},
and Feynman PDFs, and interpolate them. Given the joint
longitudinal-momentum and transverse-space distributions,
one can construct the orbital angular momentum (OAM) of partons, among others~\cite{Ji:1996ek}.
In general, the GPDs can be used to generate momentum-dissected transverse space images of the
proton~\cite{Burkardt:2000za}. A new class of experimental processes, deeply-virtual
exclusive processes (DVEP), including deeply-virtual Compton scattering (DVCS) in which the final state
is a diffractive real photon plus a recoiling proton, has been found to
measure them~\cite{Ji:1996ek,Ji:1996nm}.
The second concept is the transverse-momentum-dependent (TMD) PDFs (or TMDPDFs),
in which the parton's transverse momentum is explicit~\cite{Collins:1981uk,Collins:2011zzd}.
Much theoretical progress has been made in recent years regarding their proper definitions, factorizations,
and spin correlations~\cite{Collins:2012uy,Echevarria:2012js,Collins:2017oxh}.
TMDPDFs can be measured in experimental processes by
observing the transverse momentum of the final-state particles.

\begin{figure}[htb]	
	\centering
	\includegraphics[width=0.95\linewidth]{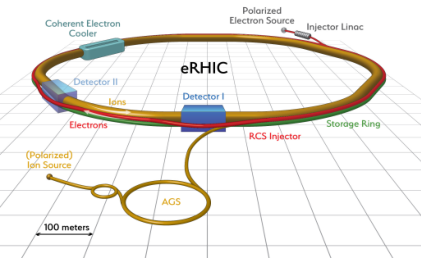}
	\caption{A realization of Electron-Ion Collider at BNL (figure credit to BNL),
		which can be used to probe the partonic landscape of the proton.}
	\label{fig:EIC.png}
\end{figure}

Over the years, it has gradually become clear that a dedicated experimental
facility to fully explore the partonic landscape of the proton is required. To meet this requirement,
the US nuclear science community has proposed, to build a high-energy, high-luminosity
Electron-Ion Collider (EIC)~\cite{Geesaman:2015fha}, which has been recently approved by the US
Department of Energy.
The new collider accelerates electrons to 10-30 GeV and ions--- including the proton and heavy
nuclei all the way up to Pb or U--- up to 100 GeV per nucleon, realizing the center-of-mass
collision energy $E_{\rm cm}$ from 40 to 170 GeV.
The corresponding electron energy in fixed-target experiments
would be 100 GeV to 10 TeV.
The beams are polarized, with high-luminosity up to $10^{33-34}$
collisions/(${\rm cm}^2\cdot{\rm s}$), which are critical
for studying exclusive processes such as DVCS.
The kinematic range of the collisions
covers the Bjorken $x_B$ (which coincides with the parton momentum fraction $x$ in the naive parton model to be discussed in the next section) down to sub-$10^{-4}$, and $Q^2$ as high as $10^4$
GeV$^2$. Much of the EIC science has been discussed in
a dedicated study~\cite{Accardi:2012qut}.

Of course, the EIC and lattice QCD efforts will not stop at the precision parton physics of
the proton. More importantly, we need to develop ways or languages to
describe the nucleon as a strongly-coupled relativistic quantum system, in much
the same way as we understand, for example, the quantum Hall effects in
condensed matter physics. Without a deep understanding of the mechanisms
of  strongly-coupled QCD physics, we cannot claim a fundamental
understanding of the structure of the proton and neutron, in particular,
the origin of their mass and spin. This is one of the most challenging
goals facing the standard model of particle and nuclear physics today.

This review is to systematically expose the idea, formalism, and results
of the LaMET approach to parton physics. We do not claim to be entirely complete
because the field is rapidly developing. References in the related fields
are not meant to be complete either, and we apologize for any important omissions.
Closely-related reviews on lattice parton physics can be found in~\cite{Cichy:2018mum,Zhao:2018fyu}.
There have been studies on the effectiveness of LaMET in various
models~\cite{Ji:2018waw,Gamberg:2014zwa,Broniowski:2017wbr,Xu:2018eii,
	Son:2019ghf,Ma:2019agv,Nam:2017gzm,Bhattacharya:2018zxi,Jia:2015pxx,Hobbs:2017xtq,Broniowski:2017gfp,
	Radyushkin:2017gjd,Bhattacharya:2019cme,Kock:2020frx,DelDebbio:2020cbz},
some of which we will mention in the following for illustrative purposes.
There have been also papers questioning the validity of LaMET method~\cite{Carlson:2017gpk,Rossi:2017muf,Rossi:2018zkn} and some got clarified later in the literature~\cite{Briceno:2017cpo,Ji:2017rah,Radyushkin:2018nbf},
We will not discuss them here and interested readers may refer to the above references.
We have used {\it proton} in most places in the text
to emphasize its importance in nuclear and particle physics. However,
the discussions apply equally to the neutron and other hadrons as well.

The plan for the presentation of this review is as follows. In the remainder of the Introduction, we
explain the nature of parton physics as an effective description
of the internal structure of the proton at large momentum, as well as other existing methods
in the literature for solving the parton structure.
In \sec{lamet}, we introduce the LaMET method starting from
momentum renormalization group equation (RGE) of physical observables in a moving hadron,
followed by the matching between momentum distributions and PDFs. We then formulate an EFT
expansion to compute parton physics from theoretical methods suitable for the structure of a
large-momentum proton. In \sec{renorm}, we discuss some important details for
collinear PDFs: renormalization of the nonlocal operators, particularly
power divergences in lattice regularization, and matching to all orders
in perturbation theory. \sec{gpo} is devoted to applications to general collinear parton observables
including GPDs, parton distribution amplitudes
and higher-order parton correlations. We also discuss applications
for the OAM of the partons in a polarized proton.
In \sec{tmd}, we consider the application to
TMDPDFs, a new class of parton observables. We study matching of the quasi-TMDPDFs to the physical ones,
and explore the lattice calculation of the soft function.
Finally, \sec{lattice} summarizes the recent lattice calculations relevant to
the LaMET applications, and the conclusion is given in \sec{conclusion}.
The review is completed with an Appendix
with a list of acronyms and glossaries, as well as notations and conventions.

\subsection{Partons through Infinite-Momentum States}
\label{sec:lc}
Although partons have become a ubiquitous language
for high-energy scattering, their role as
effective degrees of freedom of QCD for describing
the internal structure of the nucleon is less emphasized
in the literature.
In applications within QCD factorization theorems,
they are---following Feynman---objects arising
from the limit of infinite momentum, with the potential
UV divergences regulated and renormalized after the limit.
Thus,  the partons are an idealized concept,
referring to the quark and gluon Fock components
of the nucleon or other hadrons only in the context of
IMF and LF gauge $A^+=(A^0+A^z)/\sqrt 2=0$.
They are in the same category of concepts as
the {\it infinitely-heavy quark} in heavy-quark effective
field theory (HQET)~\cite{Manohar:2000dt}. To motivate LaMET,
it is important to understand this origin and nature of
partons.

Built from the knowledge of electron scattering in
non-relativistic systems (atoms and molecules)~\cite{West:1974ua},
Feynman introduced the {\it naive parton model} to
describe deep-inelastic scattering (DIS) on the proton, and to explain
the observed phenomenon of Bjorken scaling~\cite{Feynman:1969ej,Bjorken:1969ja,Feynman:1973xc}.

\begin{figure}[htb]	
	\includegraphics[width=0.6\linewidth]{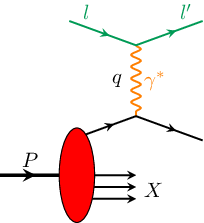}
	\caption{Deep-inelastic scattering in which partons are probed in the proton.}
	\label{fig:DIS}
\end{figure}

Shown in Fig. \ref{fig:DIS} is the DIS process in which a virtual photon with
large momentum $q^\mu$ is absorbed by a proton of momentum $P^\mu $ and mass $M$.
The invariant variables are $Q^2 = -q^\mu q_\mu $ and $P\cdot q = M\nu$, and
Bjorken $x_B=Q^2/(2P\cdot q)$ fixed in the scaling (or Bjorken) limit $Q^2\to \infty$, $P\cdot q\to \infty$. The inclusive DIS cross section can be factored into a product of leptonic and hadronic tensors, where the former is associated with the electromagnetic current of the lepton, while the latter contains all information about the electromagnetic interaction with the target proton.

To learn about the proton structure, it is best to consider
the scattering in the Breit frame where
\begin{align}\label{eq:diskin}
	q^\mu &= (0,0,0,-Q)  \,, \nonumber \\
	P^\mu &=\left(\sqrt{\frac{Q^2}{4x_B^2}+M^2}, 0, 0, \frac{Q}{2x_B}\right) \,,
\end{align}
and the virtual photon has zero energy. The probe is sensitive only to
the spatial structure as in non-relativistic electron scattering.
However, relativity now constrains the proton to move at a large
momentum $P^z= Q/(2x_B)$ with boost factor $\gamma = Q/(2x_BM)$,
which approaches $P^z=\infty$ in the Bjorken limit.

Feynman made intuitive assumptions about
the proton structure and scattering mechanism, without QFT subtleties~\cite{Feynman:1973xc}:
The proton structure at different large $P^z$ should be similar, and can be
approximated by that at $P^z=\infty$, or in the IMF. The interactions between
constituents (partons) are infinitely time-dilated, and
the wave function configurations are frozen. The proton in high-energy scattering
can be seen as being made of non-interacting partons, each with a longitudinal
momentum $xP^z$ with $0< x< 1$.

The internal structure of non-relativistic
systems is independent of their overall momentum. However, relativistic systems
is different as they least experience the Lorentz contraction.
The structures of such systems are inextricably mixed with
the overall motion, and their dependence on the external momentum is a
dynamical problem. On the other hand, if the internal structure
depends on a particular hadron scale $\Lambda_{\rm QCD}$, the protons at all large-momentum
with $P^z\gg \Lambda_{\rm QCD}$ have a similar structure, corresponding to the
$P^z\to \infty$ limit. This means that if $f(k^z,P^z)$
is the constituent momentum-$k^z$ distribution in a proton of momentum $P^z$,
it might be analytical at $P^z=\infty$ and admits Taylor series expansions
in $1/P^z$,
\begin{align}\label{Eq:expansion}
	f(k^z, P^z) = f(x) + f_2(x)(\Lambda_{\rm QCD}/P^z)^2 + ...\,,
\end{align}
where $x=k^z/P^z$. If so, one may find a large-momentum symmetry of
the proton properties up to power corrections ${\cal O}(1/P^z)$ (we omit the upper index $z$ sometimes
for simplicity), and $f(x)$ is the parton distribution.

The above picture can be shown to hold in certain simple
QFT models, where the dynamical frame dependence of wave functions
for composite systems can be studied straightforwardly.
There are many interesting examples of
two-dimensional systems, for which solutions can be found.
One of the much studied cases is the large $N_c$ QCD, also called
't Hooft model~\cite{tHooft:1974pnl}, in which
the bound states have a well-defined large-momentum limit.
The wave functions can be expanded in
$1/P$, with the corrections starting from $(1/P)^2$.
The momenta of the constituents, $k$ and $P-k$, scale in this
limit. When plotted as a function of $x=k/P$, the change in the
wave function with the magnitude of the momentum can be found
in Figs. 8--11 in~\cite{Jia:2017uul}. This is the type of example
in which Feynman's intuition applies.

However, such a intuition fails in many 3+1 dimensional QFTs, such as QCD.
When a bound state travels at increasingly
large momentum, more and more high-momentum modes of a field theory
are needed to build up its internal structure. Lorentz contraction
indicates that the range of constituent momentum important for the
structure also increases. If these high-momentum modes
do not decouple effectively from the low-momentum ones,
large logarithms of the form $\ln P$, will develop in the structural
quantities. Hence a singularity (cut) at $P=\infty$ can
exist in these theories, making $P\to \infty$ limit ill-defined
and the large momentum expansion impossible. This situation
is intimately related to UV properties of the theories, for which the limits of
taking the UV cut-off $\Lambda_{\rm UV}\to \infty$ and $P\to \infty$
do not commute. While the physically-relevant one is
$(\Lambda_{\rm UV}\gg P)\to \infty$, partons
in QCD factorizations
are obtained in the other limit $(\Lambda_{\rm UV}\ll P)\to \infty$ when
the UV divergences are ignored. Thus one can
formally write the parton distribution as
\begin{align}\label{eq:pd}
	f(x) = \int \frac{d\lambda}{2\pi} e^{ix\lambda}\langle P^z=\infty|\psi^\dagger(z) \psi(0)|P^z=\infty\rangle\,,
\end{align}
where $\lambda = \lim_{P^z\to\infty, z\to 0} (zP^z)$,
and $\psi$ is a quantum field.

Historically, the IMF limit
of field theories has been studied first at the level
of diagrammatic rules for perturbation
theory~\cite{Weinberg:1966jm}. It was found that
taking $P\to\infty$ by ignoring the UV divergences considerably simplifies the perturbation
theory rules: Many time-ordered diagrams vanish and only few
have finite contributions. Moreover, scattering in this limit resembles
that in non-relativistic quantum mechanics, and the wave function
description becomes useful.  The Fock states define
the partons which have the proper kinematic support ($0<x<1$).
After the limit is taken, all physical quantities are
now independent of $P$, and large-momentum symmetry is exact
before UV divergences are regulated. Therefore,
{\it it is the ``naive'' limit, $\Lambda_{\rm UV}\ll P\to \infty$,
	that corresponds to Feynman's naive parton model}.

In the standard QCD study of high-energy scattering,
the above concept of partons as effective degrees of freedom
has been used implicitly. The PDFs are defined in terms of the
naive $P=\infty$ limit, and are used to match the experimental cross sections,
resulting in QCD factorization theorems~\cite{Collins:2011zzd}.

\subsection{Partons through Light-Front Correlators}
\label{sec:lc-imf}

In the literature and textbooks, parton distributions are not traditionally represented
in terms of the Euclidean matrix elements as in \eq{pd}.
Rather, they are represented by the so-called LF correlators of
quantum fields (``operator formalism'')~\cite{Brodsky:1997de,Collins:2011zzd}.
A more explicit formulation in terms of collinear quantum fields
and effective lagrangian is made in the soft-collinear effective theory (SCET)~\cite{Bauer:2000yr,Bauer:2001ct,Bauer:2001yt}.

There is a physical way to see that the parton description of high-energy scattering
results in the light-front correlations. Consider DIS in the rest
frame of the proton, where the virtual photon has momentum
\begin{align}
	q^\mu = (\nu, 0, 0,  -\sqrt{\nu^2+2x_B M\nu}) \,.
\end{align}
In the Bjorken limit $\nu\to\infty$, although the invariant mass $Q$ of the photon
goes to infinity, the photon momentum becomes actually light-like
in the sense that it approaches the light front.
Therefore, in inclusive DIS cross section, the separation of the two electromagnetic currents in the hadronic tensor, which is Fourier conjugate to the photon momentum, also approaches the light-cone direction.

Thus, it appears natural that
all the structural physics of the proton in the IMF can also be
expressed in terms of time-dependent LF correlators or
correlations of quantum fields on the LF. Formally, this is simple to see
if one writes
\begin{align}
	|P\to\infty\rangle = U(\Lambda_\infty) |P=0\rangle \,.
\end{align}
The boost operator $U(\Lambda_\infty)$ can be applied to
the static nonlocal operators in the ordinary momentum distributions.
In doing so, all static correlations become light-cone ones.
The boost process is then similar to shifting the Hamiltonian evolution
in quantum mechanics from Schr\"odinger to Heisenberg
picture where time-dependence is now in the operators.

To express light-cone correlations, it is convenient to introduce
two conjugate light-like (or light-cone) vectors,
$p^\mu=(\Lambda,0,0,\Lambda)$ and $n^\mu=(1/2\Lambda,0,0,-1/2\Lambda)$, with the following properties,
$n^2=p^2=0$, and $n\cdot p=1$, where $\Lambda$ is a parameter. Then any four-vector can be expanded
as,
\begin{align}
	k^\mu = k\cdot n p^\mu + k\cdot p n^\mu  + k_\perp^\mu \ .
\end{align}
In particular, the momentum $P^\mu$ of a proton moving in the $z$-direction can be expressed as
\begin{align}
	P^\mu = p^\mu + (M^2/2)\, n^\mu \,,
\end{align}
where $M$ is the proton mass.

Using the above notation, one can express the unpolarized quark distribution
in the proton as~\cite{Collins:2011zzd},
\begin{align}
	q(x) =\frac{1}{2} \int \frac{d\lambda}{2\pi} e^{i\lambda x}
	\langle P|\overline{\psi}(0) \slashed{n} W(0,\lambda n)\psi(\lambda n)|P\rangle_c \ ,
	\label{eq:LCqPDF}
\end{align}
where $\psi$ is the quark field and $W$ is a gauge-link defined as
\begin{align}
	&W(x_2, x_1)= \nonumber\\
	&\hspace{0.5em}{\cal P}\exp[-ig\int_0^1 dt\,(x_2-x_1)_\mu A^\mu(x_1+(x_2-x_1)t)]
\end{align}
to ensure gauge invariance with ${\cal P}$ denoting the path ordering. $c$ indicates
the connected contributions only, and will be suppressed in the rest of this work.
It is a property of gauge theories in which the charge fields
are not gauge-invariant, and the physical distributions must include a beam of collinear gauge
particles. Note that the above expression is true
for any momentum $P$ (a residual momentum symmetry), in particular,
in the rest frame of the nucleon.
The $x$-support of the above distribution is $[-1,1]$. For negative $x$, one
defines the antiquark distribution with $-q(-x)\equiv \bar q(x)$.
The above expression has been more
familiar in the literature than Feynman's original formulation
of PDFs. In the single quark target, one finds
$q(x)=\delta(x-1)$.

To expose the partons in the above equation, one can follow the QCD light-front quantitzation
~\cite{Chang:1968bh,Kogut:1969xa,Drell:1970yt}, suggested by Dirac in 1949~\cite{Dirac:1949cp}.
In LFQ~\cite{Brodsky:1997de}, one defines the LF coordinates,
\begin{align}
	\xi^\pm = (\xi^0\pm \xi^3)/\sqrt{2} \ ,
\end{align}
where $\xi^+$ is the LF ``time'', and $\xi^-$ is the
LF ``spatial coordinate''. And any four-vector $A^\mu$ will be now written
as $(A^+,A^-,\vec{A}_\perp)$.
Dynamical degrees of freedom are defined on the $\xi^+=0$ plane with
arbitrary $\xi^-$ and $\vec{\xi}_\perp$, with conjugate momentum $k^+$
and $\vec{k}_\perp$. Dynamics is generated
by the light-cone Hamiltonian $H_{\rm LC}=P^-$. For a free particle with
three-momentum $(k^+, \vec{k}_\perp)$ and mass $m$, the {on-shell} LF
energy is $k^-=(\vec{k}_\perp^2+m^2)/(2k^+)$.

For QCD, one can define the Dirac matrices
$\gamma^\pm = (\gamma^0\pm\gamma^3)/\sqrt{2}$, and the
projection operators for the quark fields
as $P_\pm = (1/2)\gamma^\mp\gamma^\pm$, so that any $\psi$
can be decomposed into $\psi=\psi_++\psi_-$ with
$\psi_\pm = P_\pm \psi$, where $\psi_+$ is considered as a dynamical degree of freedom.
For the gauge field, $A^+$ is fixed by the LF gauge
$A^+=0$. $A_\perp$ are dynamical degrees of freedom. $\psi_-$ and $A^-$ are dependent variables, which can be expressed in
terms of $\psi_+$ and $A_\perp$ using equations of motion~\cite{Kogut:1969xa}.

The physics of the LF correlations becomes manifest
if one introduces the canonical expansion,
\begin{align}
	& \psi_+(\xi^+=0,\xi^-,\vec{\xi}_\perp)
	= \int \frac{d^2k_\perp}{ (2\pi)^3}
	\frac{dk^+}{ 2k^+}\sum_\sigma \Big[ b_\sigma(k) u({k,\sigma})
	\nonumber \\
	&  \times e^{-i(k^+\xi^--\vec{k}_\perp\cdot\vec{\xi}_\perp)}
	+ d_\sigma^\dagger(k) v({k,\sigma})e^{i(k^+\xi^--\vec{k}_\perp\cdot\vec{\xi}_\perp)}
	\Big],
\end{align}
where $b^\dagger(k)$ and $d^\dagger(k)$ ($b(k)$ and $d(k)$) are quark and antiquark creation (annihilation)
operators, respectively. {$\sigma$ is the light-cone
	helicity of the quarks which can take $+1/2$ or $-1/2$.} Covariant normalization is adopted for the particle
states and the creation and annihilation operators, i.e.,
\begin{align}
	& \big\{b_\sigma(k),b_{\sigma'}^\dagger(k')\big\}=
	\big\{d_\sigma(k),d_{\sigma'}^\dagger(k')\big\} \nonumber \\
	& =(2\pi)^3\delta_{\sigma\sigma'}2k^+\delta(k^+-k'^{+})
	\delta^{(2)}(\vec{k}_\perp-\vec{k}_\perp^\prime)\,.
\end{align}
Substituting the above expansion into \eq{LCqPDF}, one finds the quark distribution as
\begin{align}
	q(x) = \frac{1}{2x} \sum_\sigma \int \frac{d^2{\vec k}_\perp}{(2\pi)^3}
	\langle P|b_\sigma^\dagger(x,\vec{k}_\perp)b_\sigma(x,\vec{k}_\perp)|P\rangle/\langle P|P\rangle
	\label{eq:partonfock}
\end{align}
for $x>0$, and similarly for $x<0$ for which one gets the antiquark distribution.
The factor $1/x$ comes from the normalization of the creation and annihilation operators. The matrix element above should be interpreted as the matrix element in a wave packet state, in the limit of a state of definite momentum~\cite{Collins:2011zzd}.
This way, one recovers the physical meaning of PDFs in the LF correlator (operator) formalism.

\subsection{Other Approaches to Parton Structure}
\label{sec:intro_other}

Calculating the partonic structure of the hadrons from QCD has always
been an important goal in hadronic physics. There have been two main
approaches apart from various phenomenology and models: light-front quantization
and lattice QCD. Here the authors give a very brief review on LFQ and lattice
approaches that are different from the main subject of this review.

Although LFQ explicitly uses the parton degrees of freedom, it has not been very successful
in practical calculations. First of all, LF perturbation theory,
like the standard Hamiltonian perturbation theory, breaks Lorentz
symmetry manifestly and requires a sophisticated renormalization scheme to restore it.
A potential renormalization scheme must deal with the long-range correlations
in the $\xi^-$ direction which require functional dependence on
the renormalization counterterms~\cite{Wilson:1994fk}. Thus LF perturbation theory has not been used
for any calculations beyond one loop, except for the two-loop anomalous magnetic
moment in QED~\cite{Langnau:1992zj}. In fact, the common wisdom
of using dimensional regularization (DR) for the transverse integrals
and cut-off regularization for the longitudinal one has not been proven useful for
multi-loop calculations, although it has been successfully used to
derive the BFKL evolution by Mueller from the quarkonium wave
functions~\cite{Mueller:1993rr}.

The enthusiasm for using LFQ in QCD is not about perturbation theory, but to solve the hadron states.
Discretized LFQ was proposed in~\cite{Pauli:1985ps} to
make practical calculations for the
bound state problems. This non-perturbative method turns out to be successful
for models in 1+1 dimension, such as the Schwinger
model ~\cite{McCartor:1994im,Harada:1995au}, the 1+1 QCD~\cite{Burkardt:1989wy,Srivastava:2000cf},
the 1+1 $\phi^4$ theory~\cite{Harindranath:1987db} and the sine-Gordon model~\cite{Burkardt:1992sz}.
For 3+1 dimensional theories, simple approximations have been considered, like
the Tamm-Dancoff approximation~\cite{Perry:1990mz}. For QCD itself, one again has
to use severe truncations in the number of Fock states. Some recent works
of this type include~\cite{Vary:2009gt,Lan:2019vui,Jia:2019iyk}.
However, to derive a fully-renormalized hamiltonian is difficult and moreover,
there has been no demonstration so far that the Fock-space truncation actually converges~\cite{Wilson:1994fk}.
Therefore a systematic approximation for QCD bound states in LFQ has yet to be found.

Given the rapid development in lattice QCD, it is natural to use it
to compute parton physics. However, simulating real-time evolution directly
is numerically challenging, which runs into the so-called sign problem or more
generally NP-hard problem. Over the years, a number of methods
have been proposed previously to indirectly calculate the PDFs, which
includes well-studied moment methods, hadronic tensor and Compton amplitude method,
coordinate space factorization, etc. These approaches calculate lattice observables that can be related to the PDFs/structure functions through OPE or the dispersion relation, and thus can be used to probe certain information on the partonic structure of hadrons. However, their aims are
mainly to get the lower moments of PDFs and/or
segments of certain coordinate correlations, not directly in parton
degrees of freedom.

The most-adopted approach on the lattice has been to calculate the moments of PDFs
as the matrix elements of local operators~\cite{Kronfeld:1984zv,Martinelli:1987si}.
In the moments approach, one starts with the so-called twist-two operators~\cite{Christ:1972ms},
\begin{align}
	O^{\mu_1...\mu_n}= {\overline \psi}\gamma^{(\mu_1}iD^{\mu_2} ... iD^{\mu_n)} \psi
	- {\rm trace}
	\label{eq:t2opq}
\end{align}
in the quark case, where $(\mu_1...\mu_2)$ indicates that all the indices are symmetrized, the trace terms are those with at least one factor of the metric
tensor $g^{\mu_i \mu_j}$ multiplied by operators of dimension $(n+2)$ with $n-2$ Lorentz indices, etc.
Their matrix elements in the proton state are
\begin{align}
	\langle P|O^{\mu_1...\mu_n}(\mu)|P\rangle =  2a_n(\mu) (P^{\mu_1}\cdots P^{\mu_n} \!-\! {\rm trace}) \,,
	\label{eq:t2mom}
\end{align}
and the PDFs are related to the local matrix elements through
\begin{align}
	{a_n(\mu)} &= { \int^1_{-1} dx x^{n-1} q(x, \mu^2)}\nn\\
	&= \int^1_0 x^{n-1} \big[q(x, \mu^2)+(-1)^n \bar q(x, \mu^2)\big]
	\label{eq:t2sum}
\end{align}
with $n=1,2,...$. The time-dependent correlation for the PDF
in Eq. (\ref{eq:LCqPDF}) is recovered by taking all the
components as $+$ in Eq. (\ref{eq:t2mom}),
\begin{align}
	\langle P|O^{+...+}(\mu)|P\rangle =2 a_n(\mu) P^+\cdots P^+ \,,
\end{align}
and packaging all the moments into a distribution.
Likewise, for the gluon PDF, its moments are again related
to the matrix elements of local operators,
\begin{align}
	O_g^{\mu_1...\mu_n} = -F^{(\mu_1\alpha}iD^{\mu_2}\cdots .iD^{\mu_{n-1}}F^{\mu_n)}_{\ \alpha}\ ,
	\label{eq:t2opg}
\end{align}
with $n=2,4,6,...$.

A large number of lattice QCD calculations of PDF moments have been done
so far with various degrees of control in systematics~\cite{Lin:2017snn},
which include discretization errors,
physical pion mass, finite volume effects, excited state contaminations,
and proper renormalization. Most of the lattice calculations
have been focused on the first and second moments, $\langle x\rangle$
~\cite{Green:2012ud,Alexandrou:2017oeh,Bali:2014gha},
and $\langle x^2\rangle$\cite{Dolgov:2002zm,Deka:2008xr} for the unpolarized distributions,
and the zero-th and first moments, $\langle 1\rangle$~\cite{Gong:2015iir,Alexandrou:2017oeh,Chang:2018uxx,Alexandrou:2019brg},
and $\langle x\rangle$~\cite{Aoki:2010xg,Abdel-Rehim:2015owa} for the polarized distributions.
However, it has been difficult to
calculate higher moments, due to power divergences
and rapid decay in signals. Nonetheless, moment calculations can provide a useful
calibration for any comprehensive lattice approach to PDFs.

To get more information about the PDFs, it was proposed to
calculate the hadronic tensor of DIS in Euclidean space, and analytically continue the result
to Minkowski space~\cite{Liu:1993cv,Liu:1998um,
	Liu:1999ak,Liu:2016djw,Liu:2017lpe,Liu:2020okp}. Since numerical methods for
analytical continuation are known to be difficult for precision control (similar to NP-hard
or sign problem mentioned earlier), the approach
is useful mainly for the nucleon low-lying excitations. It is very challenging
to obtain parton physics this way.

A similar approach called ``operator product expansion (OPE) without OPE'' was suggested in~\cite{Aglietti:1998ur,Martinelli:1998hz}, see also~\cite{Dawson:1997ic,Capitani:1998fe,Capitani:1999fm}.
The point is that the Compton amplitude in the non-dispersive region
can be calculated in the Euclidean space~\cite{Ji:2001wha}. Through dispersion
relation and Taylor-expansion at $\nu = P\cdot q=0$, one can extract
the higher moments of structure functions from the lattice Compton amplitude.
The recent works and references for parton structure from this approach can be found in~\cite{Chambers:2017dov,Hannaford-Gunn:2020pvu,Horsley:2020ltc}.
A similar method has been adopted for Compton amplitude with heavy-light
currents~\cite{Detmold:2005gg}.
This approach has been used to calculate the second moment of pion distribution amplitude~\cite{Detmold:2018kwu,Detmold:2020lev}.

The current-current correlators can also be studied through OPE
in the coordinate space without momentum insertion into the
currents~\cite{Braun:2007wv}. The spatial correlation at small distances
can be used to calculate higher-moments of distribution amplitudes of the mesons.
A number of lattice studies have been performed in~\cite{Braun:2015axa,Bali:2018spj,Bali:2019dqc}.
Similar strategy has been suggested more recently by Qiu et al.~\cite{Ma:2017pxb}
for parton distributions, and has been used in lattice simulations~\cite{Sufian:2019bol,Sufian:2020vzb}.
The pseudo-PDF has been proposed based on the equal-time correlation---or the quasi-PDF in Fourier space---used
in LaMET~\cite{Radyushkin:2017cyf,Radyushkin:2019owq}, and uses a coordinate-space factorization or OPE at small distance
as in~\cite{Braun:2007wv}. Because of its close connection with the quasi-PDF, we
will discuss comparisons of the pseudo-PDF data analysis method with that for the quasi-PDF
in \sec{renorm-coord}.

There have been pioneering studies on moments of the ``quasi'' quark TMDPDFs on lattice ~\cite{Hagler:2009mb,Musch:2010ka,Musch:2011er,Engelhardt:2015xja,Yoon:2017qzo}.
The staple-shaped gauge link operators have been used to connect the quark fields
separated in the spatial direction to simulate the moments of TMDPDF. The ratios of these moments
are presumed independent of the unknown soft function and may be compared with experimental data.
However, a rigorous relation of these constructions to the physical moments of TMDPDFs
had not been investigated before LaMET, particularly the relationship between
large momentum limit and the rapidity cutoff which is an essential ingredient of
TMD physics. Comparison of this approach and LaMET will be made in \sec{quasitmd}.
\section{Large-Momentum Effective Theory}
\label{sec:lamet}
As has been explained in \sec{lc}, Feynman's partons
were motivated from describing the structure of a bound state
travelling at large momentum $P$. On the other hand, in QCD
factorizations, they appear as effective degrees of freedoms
arising in infinite momentum limit disregarding UV divergences.
Reconciling these two pictures results in large-momentum
effective theory (LaMET) for the parton structure of hadrons.

In this section, we start by considering the structure
of the proton at finite momentum. We define the
ordinary momentum distributions of the constituents,
and trying to illustrate their dependence on
the proton momentum. We demonstrate that the large-$P$ momentum-dependence
follow a RGE, similar to the
well-known RGE for partons. In \sec{lamet-fact},
we show that momentum distributions at large $P$,
are related to PDFs through a matching between different orders of
$P\to \infty$ and UV cut-off limits. This matching process has a standard EFT
explanation: Parton physics or observables
can be obtained from an effective theory in which
$P\ll \Lambda_{\rm UV}$ are calculated non-perturbatively in the so-called ${\cal P}$ space~\cite{Messiah:1979eg},
after ``integrating out'' degrees of freedom between $P$ and $\infty$  (or ${\cal Q}=1-{\cal P}$ space)
through perturbation theory. Therefore, the LaMET approach to partons
is in some sense similar to lattice QCD  as an EFT approach for continuum field theories,
in which all active degrees of freedom (${\cal P}$ space) are bounded by $|k| \le \pi/a$,
where $a$ is lattice spacing, whereas those at $|k| \ge \pi/a$ (${\cal Q}$ space) are taken into account through
perturbative coefficients and higher dimensional operators.

In \sec{lamet-pp}, we outline the formalism
of LaMET for a general parton observable. The method
can in principle be used also to calculate any LF
correlations in terms of large momentum external states (see in particular the application to
soft function in \sec{tmd}).
The strategy is also applicable for the components of the LF wave functions.
Thus, LaMET offers a practical and systematic way to carry out the program of
LFQ. Instead of working with the LF coordinates directly, one uses the
instant form of dynamics and large momentum or boost factor $\gamma$ as a
regulator for the LF divergences. In a certain sense, the
quantization using tilted light-cone coordinates~\cite{Lenz:1991sa} is
similar to the spirit of the LaMET approach.

At present, the only systematic approach to solve non-perturbative QCD
is lattice field theory~\cite{Wilson:1974sk}. Therefore,
a practical implementation of LaMET can be done
through lattice calculations.  It can also be done
with other bound-state methods using Euclidean approaches, such as the
instanton liquid model~\cite{Schafer:1996wv}. While LFQ may provide an attractive physical picture for the proton,
the Euclidean equal-time formulation is more practical for carrying out the
calculations, and LaMET serves to bridge them.

\subsection{Structure of the Proton at Finite Momentum}
\label{sec:lamet-frame}
In relativistic theories, the internal structure of
a composite system is frame-dependent (we always refer to the total momentum eigenstates),
and we are interested in the properties of the proton
at a momentum much larger than its rest mass.

We start from the quark momentum density in a fast-moving proton, assuming that it moves in the $z$-direction. A
straightforward definition is
\begin{align}
	N_P(\vec{k})=\sum_\sigma \langle P|b^\dagger_\sigma (\vec{k}) b_\sigma
	(\vec{k})|P\rangle/\langle P|P\rangle \ ,
\end{align}
where the quark helicity, color, and other implicit indices are summed over.
This equation should be compared with the parton density in Eq. (\ref{eq:partonfock}).
To make it gauge invariant, it is convenient to consider
the definition from a coordinate-space correlator,
\begin{align}
	N_{P,W}(\vec{k})=\int {d^3\xi\over {(2\pi)^3}} \ e^{-i\vec{k}\cdot \vec{\xi}}
	\langle P|\overline{\psi}(0)\gamma^0W(0,\vec{\xi})\psi(\vec{\xi})|P\rangle\,,
\end{align}
where the Dirac matrix $\gamma^0$ ensures that it is a number density. Clearly,
it is a static quantity without time-dependence and can be calculated
in Euclidean field theories, in contrast to Eq. (\ref{eq:LCqPDF}) for partons.
The gauge invariance is ensured by the Wilson line $W(0,\vec{\xi})$
between the quark fields separated by $\vec{\xi}$, which is defined in the fundamental representation of the color SU(3) group. There are infinitely many
choices for the Wilson line, generating infinitely many momentum densities.
For example, one can choose a straight-line link between $0$ and $\vec{\xi}$.
One can also let the Wilson line run from the fields along the $z$-direction for
a long distance (if not infinity) before joining them together along
the transverse direction (a staple).

For its obvious connection to the PDFs, we consider a transverse-momentum
integrated, longitudinal-momentum distribution,
\begin{align}
	N_P(k^z)&=\int  d^2\vec{k}_\perp\ N_{P,W}(\vec{k}) \nonumber \\
	&=\int {dz\over{2\pi}} e^{-ik^z z} \langle P|\overline{\psi}(0)\gamma^0 W(0,z)\psi(z)|P\rangle,
	\label{eq:momdis1}
\end{align}
where we ignore the question of convergence at large $|\vec{k}_\perp| $.
Now the gauge-link $W(0,z)$ is naturally taken as a straight-line,
\begin{align}
	W(0,z) &= \exp(-i\int^z_0 A^{z'}(z') dz') \nonumber \\
	&= \exp(i\int^0_\infty A^{z'}(z') dz')\exp(i\int^\infty_z A^{z'}(z') dz')\nonumber \\
	&= W^\dagger(\infty,0) W(\infty,z)\,,
\end{align}
where in the second line we have split the gauge link into two, going from $z$ to the infinity
and coming back from the infinity to zero. We can define a ``gauge-invariant''
quark field
\begin{align}
	\Psi(\vec{\xi}) = W(\infty,\vec{\xi})\psi(\vec{\xi}) \ ,
\end{align}
and the above density becomes,
\begin{align}
	N_P(k^z)= \int {dz\over {2\pi}} e^{-ik^z z}  \langle P|\overline{\Psi}(0)\gamma^0\Psi(z)|P\rangle \ ,
	\label{eq:momdis}
\end{align}
where again we have not considered UV divergences. The momentum
distribution defined above has been called {\it quasi-PDF}, but it is really
a physical momentum distribution in a proton of momentum $P$.

In the rest frame of the proton, $N_{{P=0}}(k^z)$ is symmetric in positive and negative $ k^z$,
probably peaks around $k^z=0$ and decays away as $k^z\to \pm\infty$.
Due to the perturbative QCD effects, it decays algebraically at large $k^z$, instead of exponentially.
Because of this property, the high moments of the distribution, $\int dk^z (k^z)^n N_0(k^z)$ with $n>0$,
have the standard QFT UV divergences.

As $P$ becomes non-zero and large, the peak $N_P(k_z)$ will be around $\alpha P^z$,
where $\alpha$ is a constant of order one. The density
at negative $k^z$ becomes smaller, but not zero. This is due to the so-called
backward-moving particles from the
large momentum kick in perturbation theory. For the same reason, the density at $k^z>P^z$ is not zero
either.

$N_P(k^z)$ has a renormalization scale dependence because the quark fields must be renormalized.
One can choose DR and modified minimal subtraction ($\overline{\rm MS}$)
scheme. Any other regularization scheme can be converted into this one perturbatively.
For $z\ne 0$, the only renormalization necessary is the quark wave
function (with anomalous dimension $\gamma_{F}$) in the $A^z=0$ gauge, because the linear divergence associated with the gauge link vanishes in the $\MS$ scheme. More extensive discussions on the renormalization
issue, particularly about non-perturbative renormalization,
will be made in the following section.

\begin{figure}[htb]	
	\centering
	\includegraphics[width=0.95\linewidth]{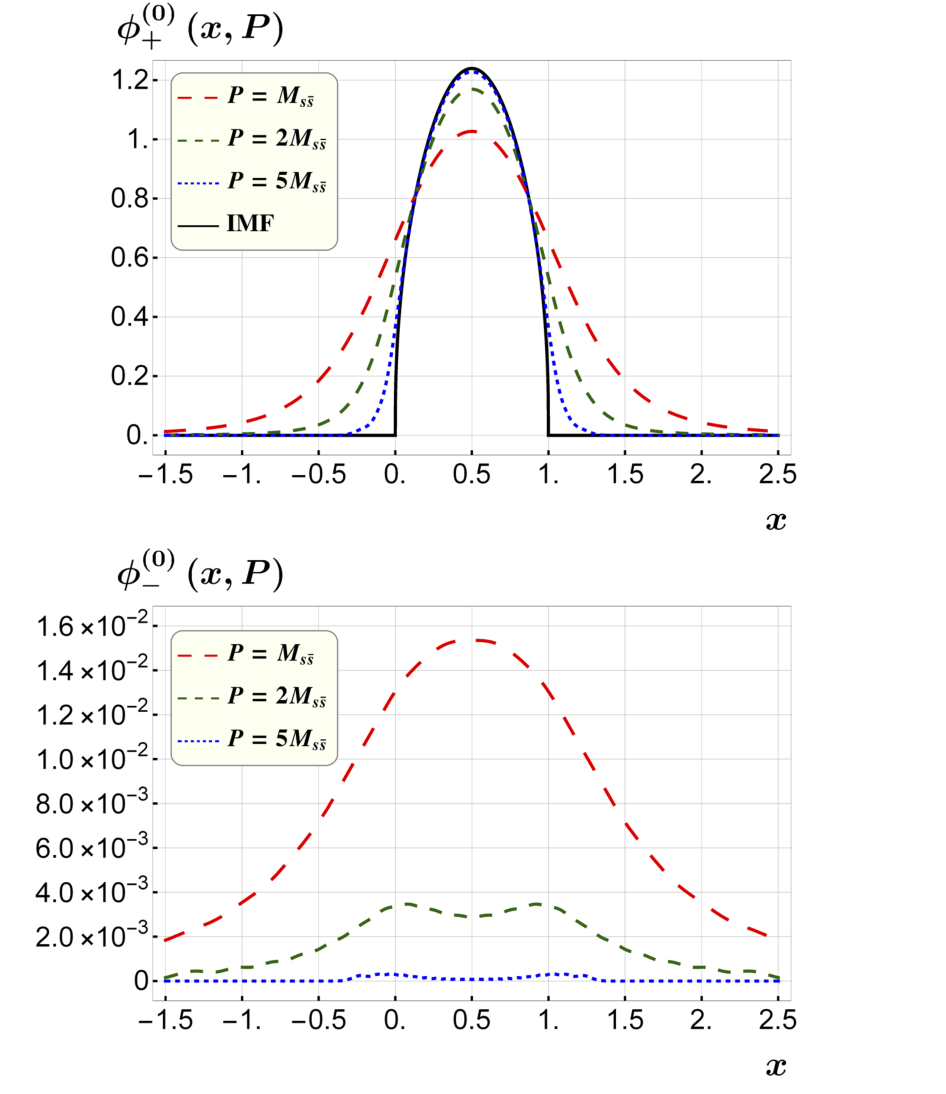}
	\caption{Wave function amplitudes of a meson in the 't Hooft model at different external momenta~\cite{Jia:2017uul}.}
	\label{fig:thooftwf}
\end{figure}

As an example showing how the parton momentum density depends on $P$, we depict in Fig.~\ref{fig:thooftwf} the quark wave function amplitude of a
meson in the 't Hooft model (1+1 dimensional QCD with $N_c\to\infty$)~\cite{tHooft:1974pnl} , the square of which yields the quark momentum density. In this model, a meson of momentum $P^\mu$ can be built as
\begin{align}
	|P_n^\mu\rangle & =
	\int \frac{dk}{2\pi|P|}
	\big[M(k-P,k)\phi_n^+(k,P)  \nonumber \\
	&\qquad\qquad + M^\dagger(k,k-P)\phi_n^-(k,P)\big]|0\rangle\,,
\end{align}
where $M(p,k) = \sum_i d^i_{-p}b_k^i/\sqrt{N_c}$, and $M^\dagger(p,k)
= \sum_i b_k^{i\dagger} d_{-p}^{i\dagger}/\sqrt{N_c}$ are annihilation
and creation operators for quark-antiquark pairs. The corresponding wave function amplitudes,
$\phi_n^+(k,P)$ and $\phi_n^-(k,P)$, satisfy a pair of equations
first derived in~\cite{Bars:1977ud}.

The meson bound state defined above has a well-defined large-momentum limit.
The wave functions can be expanded in
$1/P$, with the corrections starting from $(1/P)^2$.
The momenta of the constituents, $k$ and $P-k$, scale in this
limit. When plotted as a function of $x=k/P$, the change in the
wave function with the magnitude of the momentum is shown in Fig.~\ref{fig:thooftwf}.

\subsection{Momentum Renormalization Group}
\label{sec:lamet-rg}
In this subsection, we consider how to calculate the external momentum $P$
dependence of physical observables discussed in
the previous subsection.
Clearly, the dependence is related to the boost properties of
the operators under consideration, namely their commutation relations
with the boost generators, {$\hat{K}^i$}.
We argue that in the large momentum limit, one
has a {\it momentum RGE} which
is a differential equation relating
properties of the system at different momenta.
Momentum RGE will be, in the end, related to
the renormalization properties of the observables on
the LF.

Consider a generic operator $\hat O$, and its
matrix element in a state with momentum $P$,
\begin{align}
	O(P) = \langle P|\hat O|P\rangle \ .
\end{align}
We calculate the momentum dependence by writing $|P\rangle
= \exp(-i\omega(P) \hat K)|P=0\rangle$, where $\hat K$ is
the boost operator along the momentum direction and $\omega$ is
a boost parameter depending on $P$.
Taking a derivative with respect to the {boost} parameter gives
\begin{align}
	\frac{d O(P)}{dP} = i\frac{d\omega(P)}{dP}\langle P|[\hat O,\hat K]|P\rangle \ .
	\label{eq:momden}
\end{align}
The {r.h.s.} of the equation depends on the commutator
$[\hat O,\hat K]$, i.e., the boost properties of the operator.
For a scalar operator, the commutation relation vanishes, and $O(P)$
is frame independent.
For a vector operator, the commutation relation resembles that of an energy-momentum four-vector,
and the result is the standard Lorentz transformation of a four-vector.
For nonlocal operators, the commutation
relation requires the elementary formula,
\begin{align}
	[J^{\mu\nu}, \phi_i(x)] = i\left[l^{\mu\nu}\delta_{ij}+S^{\mu\nu}_{ij}\right]\phi_j(x)\ ,
\end{align}
where $l^{\mu\nu}\!=\!-i (x^{\mu}\partial^{\nu}\!-\!x^{\nu}\partial^{\mu})$ is the OAM operator and $S^{\mu\nu}$ is the intrinsic spin matrix.
Thus one of the fields is now { $\phi_i(t=\sinh\omega z, 0, 0, \cosh\omega z)$}
which generates a time-dependent correlation function.

In the large-momentum limit, because of asymptotic freedom,
the $P$-dependence is calculable in perturbation theory, and Eq. (\ref{eq:momden})
simplifies. One obtains the momentum or boost RGE~\cite{Ji:2014gla},
\begin{align}
	\frac{d O(P)}{dP} &= \lim_{\Delta P\to 0}[O(P+\Delta P) - O(P)]/\Delta P \\
	&\xrightarrow{P\gg M} C(\alpha_s(P)){\otimes} O(P) + {\cal O} (M^2/P^2) \ .
\end{align}
where $C(\alpha_s(P))$ is a perturbative expansion in the strong coupling $\alpha_s$. The symbol ``$\otimes$'' can be a simple multiplication or certain form of convolution, depending on the observable $O(P)$ studied.
The proof of the above equation is non-trivial, and it can be
analyzed on a case-by-case basis. There can be mixings among a set of independent operators with the same quantum numbers.
The momentum RGEs are very similar to those for scale transformation or that for the
coarse graining of a Hamiltonian. That the two are connected in some cases may
be traced to Lorentz symmetry.

\begin{figure}[t]	
	\centering
	\begin{subfigure}{.15\textwidth}
		\centering
		\includegraphics[width=\linewidth]{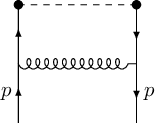}
		\caption{}
		\label{fig:qoneloop1}
	\end{subfigure}
	\begin{subfigure}{.15\textwidth}
		\centering
		\includegraphics[width=\linewidth]{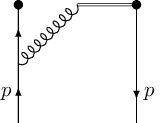}
		\caption{}
		\label{fig:qoneloop2a}
	\end{subfigure}
	\begin{subfigure}{.15\textwidth}
		\centering
		\includegraphics[width=\linewidth]{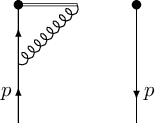}
		\caption{}
		\label{fig:qoneloop2b}
	\end{subfigure}
	\begin{subfigure}{.15\textwidth}
		\centering
		\includegraphics[width=\linewidth]{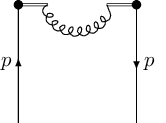}
		\caption{}
		\label{fig:qoneloop3a}
	\end{subfigure}
	\begin{subfigure}{.15\textwidth}
		\centering
		\includegraphics[width=\linewidth]{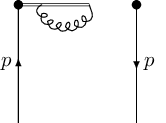}
		\caption{}
		\label{fig:qoneloop3b}
	\end{subfigure}
	\begin{subfigure}{.15\textwidth}
		\centering
		\includegraphics[width=\linewidth]{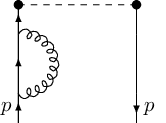}
		\caption{}
		\label{fig:qoneloop4}
	\end{subfigure}
	\caption{One-loop diagrams for the quasi-PDF in a free quark state in the Feynman gauge. The conjugate diagrams of (b), (c), (e), (f) do contribute but are not shown here.}
	\label{fig:qoneloop}
\end{figure}

As an example of the momentum RGE,
we calculate the quark momentum distribution
in a perturbative quark state using Eq.~(\ref{eq:momdis}).
Since it is gauge invariant, we can calculate it
in any gauge, for example, the Feynman gauge. The one-loop diagrams in QCD are shown in Fig.~\ref{fig:qoneloop}. There are two sources of UV
divergences, one is the logarithmic divergences from the vertex and self-energy diagrams, and the other
is the linear divergence in the self-energy of the Wilson line.
For the moment, we will use transverse momentum cut-off, $\Lambda_{\rm UV}$,  as the
UV regulator.  Using $y=k^z/P^z$, the one-loop result reads
for a large momentum quark~\cite{Xiong:2013bka},
\begin{align}\label{unpolvertex}
	& \tilde q^{(1)}(y,P^z,\Lambda_{\rm UV})=\frac{\alpha_s C_F}{2\pi} \nonumber
	\\
	& \times \left\{ \begin{array} {ll} \frac{1+y^2}{1-y}\ln \frac{y}{y-1}+1
		+\frac{\Lambda_{\rm UV}}{(1-y)^2P^z}\ , & y>1 \\
		\frac{1+y^2}{1-y}\ln \frac{(P^z)^2}{m^2}+\frac{1+y^2}{1-y}\ln\frac{4y}{1-y} \\
		\, \, \, \, -\frac{4y}{1-y}+1+\frac{\Lambda_{\rm UV}}{(1-y)^2P^z}\ , & 0<y<1 \\ \frac{1+y^2}{1-y}\ln \frac{y-1}{y}-1+\frac{\Lambda_{\rm UV}}{(1-y)^2P^z}\ , & y<0 \end{array} \right.
\end{align}
where we have ignored all power-suppressed contributions and keep the leading $P^z$ dependence only.
There is an additional contribution of the form $\delta Z_1(\Lambda_{\rm UV}/P^z) \delta(y-1)$.

The above result has several interesting features:
\begin{itemize}
	\item{The distribution does not vanish outside {$[0,1]$}.
		The radiative gluon can carry a large negative momentum fraction,
		resulting in a recoiling quark carrying larger momentum than the parent quark,
		and thus $y>1$. The same gluon can also carry a momentum larger than $P^z$,
		making the active quark have $y<0$.}
	
	\item{While the above effect is easy to understand perturbatively, it is
		surprising that a scaling contribution remains outside [0,1] in the IMF.
		As the proton travels faster,
		one might think any constituent has a momentum $k^z$ positive from Lorentz
		transformation. However, the order of limits matters because no matter
		how large the parent-quark momentum is, there are always quarks with much larger momentum, i.e.,
		$k^z\gg P^z \gg \Lambda_{\rm QCD}$. In this sense, Feynman's parton model does not describe
		the exact properties of the momentum distribution
		in a large-momentum nucleon.}
	
	\item{The contribution outside $[0,1]$ at one-loop is entirely perturbative because of the
		absence of any infrared (IR) divergence. This is no longer true at two-loop level, but
		the contribution depends only on the same one-loop IR physics in $[0,1]$. }
	
	\item{The distribution for $y$ in [0,1] has a term depending on $\ln P^z$. This dependence
		reflects that the quark substructure is resolved as a function of $P^z$, an interesting
		feature of boost.
		This dependence is perturbative in the sense that the derivative is
		IR safe,
		\begin{align}
			& P^z \frac{d \tilde q(y,P^z,\Lambda_{\rm UV})}{dP^z} \nonumber \\
			&= \frac{\alpha_s C_F}{\pi} \left[\left(\frac{1+y^2}{1-y}\right)_+ -\frac{3}{2}\delta(1-y) \right] \,.
		\end{align}
		Apart from the {$\delta$-function term}, the {r.h.s.} is similar to the one-loop
		quark splitting function in DGLAP evolution~\cite{Dokshitzer:1977sg,Gribov:1972ri,Altarelli:1977zs}. Therefore one might suspect that the momentum
		dependence is closely related to the familiar renormalization scale evolution in the
		PDFs. In fact, the physics is just the other way around: {\it It is the hadron-momentum dependence
			of the physical momentum distribution that generates
			the DGLAP evolution in the infinite-momentum limit.} One can derive an all-order momentum RGE for the momentum distribution function. Momentum RGE also provides a method
		to sum over the large logarithms of the momentum.}
	
	\item{There is a singularity at $y=1$. This singularity is generated
		from soft-gluon radiation. Fortunately, this singularity combined
		with the virtual contribution yields a finite result.}
	
	\item{There is a linear divergence in the cut-off regulator,
		leading to $\Lambda_{\rm UV}/P^z$ term, which is absent in DR. Thus, to keep $1/(P^z)^2$
		power counting, it is important
		to work in a renormalization scheme where this term does not exist.}
	
\end{itemize}

We can also move on to study the hadron momentum RGEs of other
structural properties considered in the previous subsection.
In particular, the RGE for TMD distributions will lead to
the familiar rapidity RGE in the literature. We reserve these discussions to \sec{tmd}.

\subsection{Effective Field Theory Matching to PDFs}
\label{sec:lamet-fact}
As we have seen in the previous subsection,
the momentum distributions of the constituents (now
called quasi-PDFs in the literature) in a proton at large $P$
are different from the PDFs or LF distributions in many ways.
In particular, the momentum fraction $y$ in a physical momentum distribution
is not limited to [0,1] due to backward moving particles,
which is the case even in the $P\to\infty$ limit. In fact, the
infinite-momentum limit is not analytical due to the presence of $\ln P$.

However, partons are effective objects arising
from a different limit $\Lambda_{\rm UV}\ll P\to \infty $.
There is also an important computational advantage in taking the naive limit $P\gg \Lambda_{\rm UV}$ in
perturbative calculations: Feynman integrals have one fewer four-momentum.
Therefore, this limit of QFTs serves as a
reference system where the structure of the bound states
is manifestly {independent of the hadron momentum}, and is similar to
scale-invariant critical points at which second-order phase transitions occur
in condensed matter systems. However, the theory in the naive IMF limit
introduces additional UV divergences.

Therefore, the partons in QCD are very similar
to the infinitely heavy quarks in HQET~\cite{Manohar:2000dt}. In certain
QCD systems, heavy quarks such as the bottom quark are present, and their
masses are much larger than the typical QCD scale $\Lambda_{\rm QCD}$.
In this case, one might study the dependence on the heavy quark mass
by expanding around $m_Q=\infty$. This expansion will generally
produce a power series in $1/m_Q$. However, the limits
of taking $\Lambda_{\rm UV}\to\infty$ and infinite heavy-quark
mass {limits} are not interchangeable, due to the presence
of the large logarithms $\ln m_Q$. In an EFT approach,
one takes the $m_Q\to\infty$ limit first, this will result in
a new theory with different UV behavior, but without the heavy-quark
mass, and symmetries among very different heavy-quark
systems become manifest. The renormalization
of the extra UV divergences yields a RGE
which can be used to resum large quark-mass logarithms.

Therefore, the momentum distribution at large-$P$ differs
from the parton distributions only in the order
of limits, their IR non-perturbative physics is
the same. In asymptotically free theories such as QCD, differences (or discontinuities) in taking the limits of $P\gg \Lambda_{\rm UV}$ and $\Lambda_{\rm UV} \gg P\to\infty$ are
perturbatively calculable, as only the high-momentum modes matter.
The differences are called {\it matching coefficients}.
Therefore, one is able to write down a power expansion for the momentum-dependent distribution (quasi-PDF) in terms of the PDF~\cite{Xiong:2013bka,Ma:2014jla,Ma:2017pxb,Izubuchi:2018srq},
\begin{align}\label{eq:fact_0}
	\tilde q(y, P^z, \mu)\! &=\! \int_{-1}^{1} \frac{dx}{|x|} C\left(\frac{y}{x}, \frac{\mu}{xP^z}\right) q(x,\mu)
	\nonumber \\
	&\qquad\qquad + {\cal O}\bigg(\frac{\Lambda_{\text{QCD}}^2}{(yP^z)^2},
	\frac{\Lambda_{\text{QCD}}^2}{((1-y)P^z)^2}
	\bigg)\,,
\end{align}
where the power correction is suppressed by the parton momentum $yP^z$ and  the spectator momentum $(1-y)P^z$~\cite{Ji:2020brr}.
This expansion {may be} also called
a factorization formula, as the quasi-PDF contains all the IR physics in the PDF, and $C$ involves only UV physics.
As we will discuss extensively in
the next section, this factorization formula is true to all orders in perturbation theory.
The above relation allows us to calculate the
LF parton physics from the momentum distribution at large
$P$. Since the expansion parameter is $\Lambda_{\rm QCD}^2/(yP^z)^2$ and $\Lambda_{\rm QCD}^2/((1-y)P^z)^2$ ,
for intermediate $y$ one might not need very large $P^z$ to neglect
the power corrections.

\begin{figure}[htb]	
	\centering
	\includegraphics[width=0.6\linewidth]{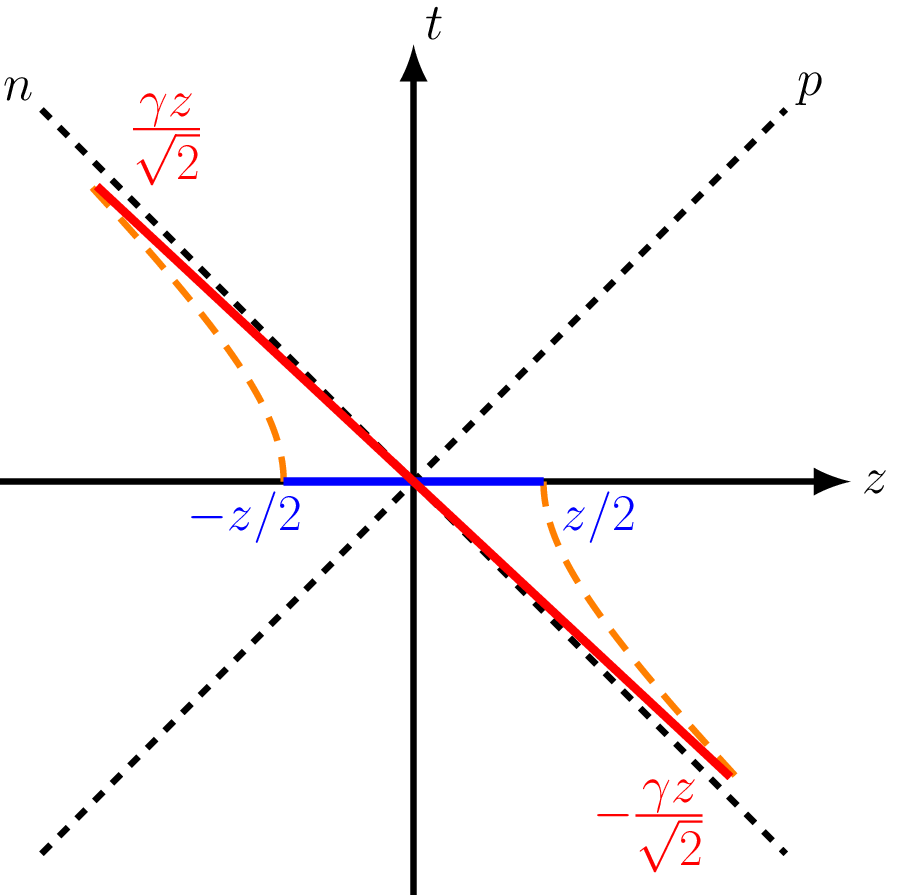}
	\caption{The line segment in the $z$-direction in the frame of a large-momentum hadron.
		Through Lorentz boost, it is equivalent to a line segment of length $\sim \gamma z$
		close to the light-one in the hadron state of zero momentum. Here $\gamma z/\sqrt{2}$ is the length of projection of the boosted line segment to the light-cone direction $n$. Thus, we call the dimensionless
		variable $\lambda=zP^z\sim \gamma zM$ as the quasi light-cone distance.}
	\label{fig:Boost}
\end{figure}

The above relation between the two quantities
has a simple explanation in terms of the Lorentz boost:
Consider the spatial correlation along $z$ shown
in Fig. \ref{fig:Boost} in a large momentum state.
It can be seen as approaching the LF one in the rest frame of the proton.
In other words, we are using a near-LF correlation to approximate a LF
correlation. Accordingly, we can invert the above equation recursively to express the PDF in terms of
quasi-PDF with their differences being taken care of
through the perturbative matching $\tilde C$ and power corrections,
\begin{align}\label{eq:fact_1}
	q(x, \mu)\! &=\! \int_{-\infty}^{\infty} \frac{dy}{|y|} \tilde C\left(\frac{x}{y}, \frac{\mu}{yP^z}\right) \tilde q(y,P^z,\mu)
	\nonumber \\
	&\qquad\qquad + {\cal O}\bigg(\frac{\Lambda_{\text{QCD}}^2}{(xP^z)^2},
	\frac{\Lambda_{\text{QCD}}^2}{((1-x)P^z)^2}
	\bigg)\,.
\end{align}
The above equation has an EFT interpretation:
The parton physics is calculated in an effective field theory
with physical momentum scale from $0$ to $P$, whereas the physics
from degrees of freedom from $P$ to $\infty$ can be integrated out to
generate the perturbative coefficients $\tilde C$ and
the high-order terms in $1/(P^z)^2$. In constrast to HQET, the full QCD degrees of freedom are used in LaMET calculations. In other words, the effective Lagrangian of LaMET
is the standard QCD one, while the large momentum $P$ for expansions
appears only in the external states.

\subsection{Recipe for Parton Physics in LaMET}
\label{sec:lamet-pp}

The principle of LaMET is to simulate the time-dependence of parton observables through external states at large momentum.
Thus, we can generalize the discussions in the previous subsection
to any type of physical observables for the large momentum
proton, which will be generally called quasi-parton observables.
Examples will be given in the later sections including
transverse-momentum dependent distributions and LF wave functions.

Consider any Euclidean quasi-observable $O$ which depends
on a large hadron momentum $P^z$
and UV cut-off $\Lambda_{\rm UV}\gg P^z$.
Using asymptotic freedom, we can systematically
expand the $P^z$ dependence,
\begin{align}
	O\Big(P^z,\Lambda_{\rm UV}\Big) = Z\Big({P^z\over\Lambda_{\rm UV}},
	{P^z\over \mu}\Big){\otimes}o(\mu) + {\cal O}\Big({\Lambda_{\rm QCD}^2\over (P^z)^2}\Big)+ ... \ ,
	\label{eq:Expansion}
\end{align}
where $\otimes$ refers to a convolution if appropriate, and $Z$ factorizes all the perturbative
dependence on $P^z$ and does not contain any IR divergence.
The quantity $o(\mu)$ is defined in a theory with $P^z\rightarrow \infty$, exactly
as in Feynman's parton model. In fact, $o(\mu)$ is a LF
correlation containing all the IR collinear (and soft)
singularities. The important point of the expansion is that it
may converge at moderately large $P^z$, say a few GeV, allowing
access to quantities needed for very large $P^z$ (a few TeV).
One can also use the large boost-factor $\gamma=P^z/M$ as the expansion
parameter $1/\gamma$.

Momentum dependence of the quasi-observables can be
studied through momentum RGEs. Defining the anomalous dimension through
\begin{align}
	\gamma_P(\alpha_s) = \frac{1}{Z}\frac{\partial Z}{\partial \ln P^z} \ ,
\end{align}
it follows{ that}
\begin{align}
	\frac{\partial O(P^z)}{\partial \ln P^z} = \gamma_P(\alpha_s) \otimes O(P^z) \,,
\end{align}
up to power corrections. One can resum large logarithms involving $P^z$ using the above equation.

When taking $P^z\rightarrow \infty$ first
in $O(P^z)$ before a UV regularization is imposed, one recovers from $\hat O$ the
light-cone operator $\hat o$, by construction. On the other hand, the physical matrix
element is calculated at a large $P^z$, with UV regularization such as the lattice cut-off
imposed first. Thus the difference between the matrix elements of $\hat o$ and $\hat O$ is a matter of
the order of limits. This is the standard set-up for an EFT. The different limits do
not change the IR physics. In fact, the factorization in terms of Feynman diagrams
can be proved order by order as in the renormalization program, as discussed in the
following section.

The parton physics can be calculated more directly by reversing
Eq. (\ref{eq:Expansion}) to produce an EFT expansion,
\begin{align}
	o(\mu)= \tilde Z\Big({P^z\over\Lambda_{\rm UV}},
	{P^z\over \mu}\Big){\otimes} O\Big(P^z,\Lambda_{\rm UV}\Big) + {\cal O}\Big({\Lambda_{\rm QCD}^2\over (P^z)^2}\Big)+ ... \ .
	\label{eq:Expansion1}
\end{align}
Thus, to compute any parton observable defined by an operator made of LF dynamical fields, $\hat o$, one constructs a time-independent version $\hat O$ which, under an infinite
Lorentz boost, approaches $\hat o$. Then, one calculates the matrix element
of $\hat O$ in a hadron with large momentum $P^z$ using
whatever approach (lattice QCD is an obvious choice for the time-independent operator $\hat O$)
and then uses Eq. (\ref{eq:Expansion1}) to systematically approximate the parton {observable}.
Usually the matrix element of $\hat O$ depends on $P^z$ as well as all the lattice UV artifacts.
In principle, the latter does not affect the EFT expansion and will be cancelled by the matching coefficient $\tilde Z$ and higher order terms in the expansion. However,
in practical applications such as the quasi-PDF calculations, a nonperturbative renormalization is still necessary to remove all the power divergences to ensure a continuum limit.

\subsection{Universality}
\label{sec:lamet-univ}
LaMET provides a framework to systematically
compute partonic observables on the LF from the properties of
a large-momentum proton.
However, the relationship is not one-to-one. There can be
infinitely many possible Euclidean operators in the large-momentum proton
which generate the same LF observable. This is because
the large-momentum physical states have built-in
collinear (as well as soft) parton modes, and upon acting on
a Euclidean operator, they help to {\it project out}
the leading LF physics. All operators projecting out
the same LF physics form a {\it universality class}.
Accordingly, in the operator formulation
for parton physics such as SCET, one uses
LF operators to project out parton physics off the external
states of any momentum, including $P=0$.

The concepts such as universality class have been used
in critical phenomena in condensed matter physics,
where systems with different microscopic Hamiltonians
can have the same scaling properties near their critical
points. Critical phenomena correspond to the IR fixed points of the scale transformation,
and are dominated by physics at long-distance scales.
In the present case, parton physics arises from the
infinite-momentum limit, $P=\infty$,
which is a UV fixed point of the momentum RGEs. It
is the longitudinal short distance (or large momentum)
physics that is relevant at the fixed point.
However, the short distance here does not mean everything
is perturbative. The part that is non-perturbative
characterizes the partonic structure of the proton. The critical region
near $P=\infty$ acts as a filter to select only the physics
that is relevant, so universality classes emerge.

In the case of unpolarized PDFs, the initial proposal in LaMET
starts from the matrix element of the following operator~\cite{Ji:2013dva},
\begin{align}
	O_1(z)=  \overline{\psi}(0)\gamma^zW(0,z)\psi(z) \ .
	\label{eq:eucbi1}
\end{align}
However, one can equally start from~\cite{Radyushkin:2016hsy,Constantinou:2017sej},
\begin{align}
	O_2(z)= \overline{\psi}(0)\gamma^0W(0,z)\psi(z) \ ,
	\label{eq:eucbi2}
\end{align}
and the leading contributions in the large-momentum expansion are the same. One can
also consider any linear combination of the two.
In~\cite{Jia:2018qee}, the calculations have been done with
these two operators in the 't Hooft model,
and the results have been compared at different
hadron momenta. For lattice simulations,
an important issue is about the operator mixing, which
depends on specific choices of the operators
in the universality class.

Another example of Euclidean operators for PDFs
is the current-current correlators
in a pure space separation,
\begin{align}
	O_3(z)=J^\mu(0)J^\nu(z) \ ,
\end{align}
where $J^\mu$ is, for example, an electromagnetic current.
This type of correlator was first considered in~\cite{Braun:2007wv,Bali:2017gfr}
for calculating pion DA, and recently
has been suggested to calculate PDFs with generalized
bilocal ``currents''~\cite{Ma:2017pxb}.
When the matrix elements are calculated in the large-momentum states,
$O(z)$ falls into the same universality class as the operators in Eqs. (\ref{eq:eucbi1}) and
(\ref{eq:eucbi2}). Instead of using light quarks as the intermediate propagator
in $O(z)$, one can have a number of other choices for LaMET applications, including
scalars~\cite{Aglietti:1998ur,Abada:2001if} and
heavy-quarks~\cite{Detmold:2005gg}. One can also similarly
work with quark bilinear operators in any physical gauge which become the light-front one in
the large momentum limit~\cite{Gupta:1993vp}.

Another important example of universality class is the gluon helicity
contribution to the spin of the proton, as we will
discuss in detail in \sec{spin}. The gluon
spin operator $\vec{E}\times \vec{A}$ is gauge-dependent. However, in physical gauges where the transverse degrees of
freedom are dynamical, its matrix element is the same in
the large-momentum limit. Therefore,
one can potentially choose different gauges to perform
calculations at finite momentum on lattice, such as Coulomb gauge $\vec{\nabla}\cdot \vec{A}=0$,
axial gauge $A^z=0$ or temporal gauge $A^0=0$.  Different gauge choices
will have different UV properties ($\ln P$) and hence different
matching conditions. However, the IR part of the matrix element
is the same~\cite{Hatta:2013gta}.

At a practical level, it is very useful to find which
operator has the fastest convergence in the LaMET expansion.
The current-current correlators use the light-quark propagator to simulate
the light-like Wilson line (sometimes called light-ray). The quasi-PDF approach not only starts from
a quantity with clear physical meaning (a momentum distribution),
but also generates the needed Wilson line simply by rotating a
space-like one, shown in Fig. \ref{fig:Boost}. Thus, it is plausible
that the quasi-PDF will provide mathematically the fastest large-$P$
convergence than the other choices.

\section{Renormalization and Matching for PDFs}
\label{sec:renorm}
In this section, we consider the LaMET application to calculating
the simplest collinear PDFs, which have been most extensively studied in the literature so far.
Although universality allows one to extract the collinear PDFs from the matrix elements of a wide class of operators evaluated at large momentum, we will focus on physical observables closely resembling the collinear PDFs, i.e.,
the quark and gluon momentum distributions or the {\it quasi-PDFs}. We also discuss
the coordinate-space factorization approach in which
the pseudo-PDF and current-current correlators have been studied.

We mainly review the technical progress made in renormalization and matching using the quasi-PDFs.
The matching can be done in principle at the bare matrix elements level,
since the factorization formula like \eq{fact_0} is valid for both bare and
renormalized momentum distributions. All the UV divergences in the bare quasi-PDF can be
factorized into the matching coefficient $C$, and the latter automatically
renormalizes the bare lattice matrix elements, so the continuum limit can be taken afterwards.
However,  such a matching coefficient then has to be calculated in lattice perturbation theory, which is computationally challenging and converges slowly~\cite{Lepage:1992xa}. More importantly, the quasi-PDF contains linear power divergence under UV cutoff regularization due to the Wilson line self-energy~\cite{Ji:2013dva,Xiong:2013bka}, which makes it impossible to take the continuum limit with fixed-order calculations in lattice perturbation theory. Though the latter problem can be improved by resumming the linear and possibly logarithmic divergences, it is usually preferred to nonperturbatively renormalize the quasi-PDFs on the lattice, after which a continuum limit can be taken and a perturbative matching can be done in the continuum theory. To this end, a thorough understanding of the renormalizability of Wilson-line operators defining the quasi-PDFs is required.
In addition to renormalization, the applications of LaMET rely on the validity of the large-momentum factorization formula~\eq{fact_0}, which can be proven in perturbation theory to all orders by showing that the collinear divergences
are the same in the momentum distributions and light-cone PDFs.

We begin in \sec{renorm-nl} with the proof of multiplicative renormalizability of the Wilson-line operators that define the quasi-PDFs. We first work in the continuum theory with $\MS$ scheme, and then generalize the conclusion to lattice theory.
Next, in \sec{factorization} we outline the factorization theorem for momentum distributions to all orders
in perturbation theory, and state the form of convolution between the matching coefficient and the PDF. In \sec{renorm-coord} we show that the factorization theorem has an equivalent form in coordinate space, which can be used as an alternate route to extract PDFs from lattice matrix elements. Finally, we discuss the nonperturbative renormalization of quasi-PDFs on the lattice and their matching to the $\MS$ PDF in \sec{renorm-npr}.

\subsection{Renormalization of Nonlocal Wilson-Line Operators}
\label{sec:renorm-nl}
The momentum distributions of the proton are defined from
equal-time nonlocal Wilson line operators of the form in Eq.~(\ref{eq:momdis1}).
In this subsection, we review the renormalization of these spacelike nonlocal operators (the renormalization of lightlike nonlocal operators defining the PDFs can be found in~\cite{Collins:1981uw,Collins:2011zzd}). We first discuss their renormalization in DR using an auxiliary field approach, followed by the discussion on similar
gluon operators. We then consider power divergences
in the momentum cutoff type of UV regularization. The conclusion is that they are
all multiplicatively renormalizable with a finite number of mixings with other operators.

\subsubsection{Renormalization of nonlocal quark operators}

We are interested in operators of the following kind,
\begin{equation}
	O_\Gamma(z)=\bar{\psi}\big({z\over2}\big)\Gamma W\big({z\over2},-{z\over2}\big)\psi\big(\!-\!{z\over2}\big) \ .
\end{equation}
Since the Wilson line $W(z_1,z_2)$ is a path-ordered integral of gauge fields,
it is not obvious that such operators are multiplicatively renormalizable. The renormalization of non-lightlike Wilson loops and Wilson lines has been studied in early literature~\cite{Dotsenko:1979wb,Craigie:1980qs}, and the all-order proof of their multiplicative renormalizability was first made using diagrammatic methods in~\cite{Dotsenko:1979wb,Craigie:1980qs} and then the functional formalism of gauge theories in~\cite{Dorn:1986dt}.
The same conclusion was conjectured to hold also for the quark bilinear operator $O_\Gamma(z)$, whose renormalization takes the following form~\cite{Musch:2010ka,Ishikawa:2016znu,Chen:2016fxx},
\begin{align} \label{eq:schemeren}
	O^B_\Gamma(z,\Lambda)=Z_{\psi,z}(\Lambda,\mu)e^{\delta m(\Lambda)|z|} O^R_\Gamma(z,\mu)\,,
\end{align}
where ``$B$'' and ``$R$'' stand for bare and renormalized operators respectively, and all the fields and couplings in $O^B_\Gamma(z,\Lambda)$ are bare ones which depend on the UV cutoff $\Lambda$. $\delta m(\Lambda)$ is the ``mass correction'' of the Wilson line, which includes all the linear power divergences of its self-energy. $Z_{\psi,z}(\Lambda,\mu)$ includes all the logarithmic divergences from wavefunction and vertex renormalizations.

An early two-loop study of the quasi-PDF in the $\MS$ scheme indeed indicated the multiplicative renormalizability of $O_\Gamma(z)$~\cite{Ji:2015jwa}. The first rigorous proof of \eq{schemeren} was given in the auxiliary ``heavy quark'' field formalism~\cite{Ji:2017oey,Green:2017xeu} which was used to prove the renormalizability of Wilson lines~\cite{Samuel:1978iy,Gervais:1979fv,Arefeva:1980zd,Dorn:1986dt}. This formalism is defined by extending the QCD Lagrangian to include the auxiliary ``heavy quark'' fields $Q,\bar{Q}$ and their gauge interaction,
\begin{align} \label{eq:hqet}
	{\cal L} = {\cal L}_{\rm QCD} + \bar{Q}_0in_z\cdot D_0 Q_0\,,
\end{align}
where the subscript ``0'' denotes bare quantities. $n_z^\mu=(0,0,0,1)$ is the direction vector of the spacelike Wilson line $W(z,0)$, $D_{0}^\mu = \partial^\mu + ig_0A_{0}^\mu$, and $Q_0$ is a color-triplet scalar Grassmann field in the fundamental representation of SU(3). Note that if we replace $n_z^\mu$ with the timelike vector $n_t^\mu=(1,0,0,0)$, then \eq{hqet} yields the leading order HQET Lagrangian.

In the theory defined by \eq{hqet}, the Wilson line can be expressed as the connected two-point function of the ``heavy-quark'' fields,
\begin{align}
	\langle Q_0(\xi)\bar{Q}_0(\eta) \rangle_Q = S_0^Q(\xi,\eta)\,,
\end{align}
where $\xi$ and $\eta$ are space-time coordinates, and
$\langle...\rangle_Q$ stands for integrating out the auxiliary fields. The above equation is valid up to the determinant of $in_z\cdot D_0$, which is a constant and can be absorbed into the normalization of the generating functional~\cite{Mannel:1991mc}. The Green's function $S_0^Q(\xi,\eta)$ satisfies
\begin{align}
	in_z\cdot D_0(\xi)\ S_0^Q(\xi,\eta) = \delta^{(4)}(\xi-\eta)\,,
\end{align}
which has the solution
\begin{align}
	S_0^Q(\xi,\eta)\! =\! W(\xi^3,\eta^3)\theta(\xi^3-\eta^3)\delta(\xi^0-\eta^0)\delta^{(2)}(\vec{\xi}_\perp -\vec{\eta}_\perp)
\end{align}
with a proper choice of boundary condition.
In this way, the Wilson-line operator $O^B_\Gamma(z)$ can be replaced by the product of two local composite operators averaged over all the ``heavy-quark'' field configurations~\cite{Dorn:1986dt},
\begin{align} \label{eq:product}
	O^B_{\Gamma}(z)& = \int d^4\xi\ \delta(\xi^3-z\big) \nn\\
	&\quad\times \langle \bar{\psi}_0\big({\xi\over2}\big) Q_0\big({\xi\over2}\big)\Gamma\bar{Q}_0\big(\!-\!{\xi\over2}\big)\psi_0\big(\!-\!{\xi\over2}\big)\rangle_Q\,,
\end{align}
where the UV regulator is suppressed.

Consequently, the renormalization of $O^B_{\Gamma}(z)$ is reduced to that of the two local ``heavy-to-light'' currents
\begin{align}
	J^B = \bar{Q}_0 \psi_0\,.
\end{align}
The renormalizability of this auxiliary field theory has been proven using the standard functional techniques for gauge theories~\cite{Dorn:1986dt}. After fixing the covariant gauge and introducing the ghost fields, the theory including the auxiliary ``heavy-quark'' has a residual BRST symmetry, from which one can derive the Ward-Takahashi identities to show that all the UV divergences of the Green's functions can be subtracted with a finite number of local counterterms. In analogy, the same method has also been used to prove the all-order renormalization of HQET in perturbation theory~\cite{Bagan:1993zv}.

According to~\cite{Dorn:1986dt}, the ``heavy-quark'' Lagrangian can be renormalized in a covariant gauge as
\begin{align}\label{eq:auxl}
	{\cal L} &= {\cal L}_{\rm QCD}[g_0,\psi_0,A_0,c_0] + \bar{Q}_0in_z\cdot D_0 Q_0\nn\\
	&={\cal L}_{\rm QCD}[g,\psi,A,c] +  {\cal L}_{\rm c.t.}[g,\psi,A,c] \nn\\
	& +  Z_Q\bar{Q}\left(in_z\cdot \partial - i\delta m\right) Q - g Z_{1}^{QQg}\bar{Q}n_z\cdot A_at^aQ\,,
\end{align}
where ${\cal L}_{\rm c.t.}[g,\psi,A,c]$ are the QCD counterterms, and the bare fields and coupling are related to the renormalized ones through
\begin{align}
	\psi_0=Z_\psi^{1\over2}\psi, \ A_0 = Z_A^{1\over2} A,\ Q_0 = Z_Q^{1\over2} Q,\ g_0=Z_gg\,.
\end{align}
The heavy-quark-gluon vertex renormalization constant $Z_1^{QQg}$ is related to $Z_g$ through the Slavnov-Taylor identities of the auxiliary field theory~\cite{Dorn:1986dt},
\beq
Z_g= Z_1^{QQg} Z_A^{-{1\over2}}Z_Q^{-1}	\,.
\eeq
The $i\delta m$ can be regarded as the mass correction of the ``heavy quark'' except that it is imaginary.
For Dirac fermions, the mass correction is logarithmically divergent and proportional to the bare mass, as a result of chiral symmetry; for HQET, the mass correction of the heavy quark is proportional to the UV cutoff, i.e. linearly divergent, which is also expected for the auxiliary field here. Since the proof of renormalizability for this auxiliary field theory is carried out in the $\MS$ scheme with DR ($d=4-2\epsilon$), all power divergences vanish, so does $\delta m$.
Nevertheless, $\delta m$ may include ${\cal O}(\Lambda_{\text{QCD}})$ contributions due to the renormalon ambiguities which are known to exist in HQET~\cite{Bigi:1994em,Beneke:1994sw}.

Since the auxiliary field theory is renormalizable, the renormalization of the operator product in \eq{product} amounts to the renormalizations of the two ``heavy-to-light'' currents. Using the standard techniques in quantum field theory~\cite{Collins:1984xc}, one can show recursively that the overall UV divergence of the insertion of $J^B$ into Green's functions is absorbed into a renormalization factor $Z_J$ to all orders in perturbation theory,
\begin{align}
	J^B&=Z_{J}J^R = Z_{\psi}^{1/2}Z_{Q}^{1/2}Z_{V}\ J^R\,,
\end{align}
where $Z_{V}$ is the vertex renormalization constant of the ``heavy-to-light'' current.
The renormalization of heavy-to-light currents in HQET has been calculated up to three-loop order in perturbative QCD~\cite{Shifman:1986sm,Politzer:1988wp,Ji:1991pr,Chetyrkin:2003vi,Broadhurst:1991fz}. More recently, it has been argued that the anomalous dimension of the ``heavy-to-light'' current is identical to that in HQET to all orders~\cite{Braun:2020ymy}, which is also the case for the ``heavy-to-gluon'' current that will be discussed below, so the renormalization factors for the spacelike and timelike Wilson line operators should be exactly the same.

Using the above results, we can show that
\begin{align} \label{eq:renproof}
	O^B_\Gamma(z)\!&= Z_J^2\!\int\! d^4\xi\ \delta(\xi^3\!-\!z)\big\langle \bar{J}^R\big({\xi\over2}\big) \Gamma  J^R\big(\!-\!{\xi\over2}\big)\big\rangle_Q\nn\\
	&= Z_J^2 e^{\delta m|z|} O^R_\Gamma(z)\,,
\end{align}
where $\delta m$ arises from the determinant of $(in_z\cdot \partial - \delta m)$ in \eq{auxl}. In this way, we identify that $Z_{\psi,z} =  Z_J^2 $ in \eq{schemeren} which is independent of $\Gamma$. At one-loop order~\cite{Stefanis:1983ke},
\beq \label{eq:zpsiz}
Z_{\psi,z} = 1+ \frac{\alpha_sC_F}{4\pi}{3\over\epsilon_{\mbox{\tiny UV}}}\,,
\eeq
where the UV regulator $\epsilon_{\mbox{\tiny UV}}$ is to be distinguished from the IR regulator $\epsilon_{\mbox{\tiny IR}}$ in DR.

The multiplicative renormalizability of $ O^B_\Gamma(z)$ has also been proven with a recursive analysis of all-order Feynman diagrams~\cite{Ishikawa:2017faj}. In addition to \eq{schemeren}, it was found that $O^B_\Gamma(z)$ does not mix with gluons or quarks of other flavors. This can also be easily understood within the auxiliary field formalism, as the flavor-changing ``heavy-to-light'' current does not mix with other operators~\cite{Green:2020xco}.

Finally, under lattice regularization we can still use the above techniques to prove \eq{renproof}, where the mass correction $\delta m$ is now nonvanishing and equal to the lattice UV cutoff $1/a$ multiplied by a perturbative series in the coupling constant $\alpha_s$.

\subsubsection{Renormalization of nonlocal gluon operators}

Using the same ``heavy-quark'' auxiliary field formalism, it has also been proven that the Wilson-line operators for the gluon quasi-PDF are multiplicatively renormalizable~\cite{Zhang:2018diq}, which is echoed by the diagrammatical proof in~\cite{Li:2018tpe}.

According to LaMET, the gluon quasi-PDF can be defined as~\cite{Ji:2013dva}
\begin{align} \label{eq:quasigluon}
	\tilde g(x,P^z) = N\int {d\lambda \over 4\pi x (P^z)^2}e^{i\lambda x} \langle P| O_g(z)|P\rangle\,,
\end{align}
where $N$ is a normalization factor, and
\begin{align}
	O^B_g(z) = g_{\perp,\mu\nu} F^{n_1\mu}_{0,a}\big({z\over2}\big) W^{ab}\big({z\over2},-{z\over2}\big) F^{n_2\nu}_{0,b}\big(\!-\!{z\over2}\big)
\end{align}
with $F^{n\mu}_{0,a}=n_\rho F^{\rho\mu}_{0,a}$ and $n_1^\mu$, $n_2^\mu$ being either $n_z^\mu$ or $n_t^\mu$. $a,b$ are color indices in the adjoint representation. The transverse metric tensor
\begin{align}
	g_\perp^{\mu \nu} = g^{\mu \nu}  - n_t^\mu n_t^\nu / n_t^2 - n_z^\mu n_z^\nu /n_z^2\,,
\end{align}
and $N = (n_z\cdot P/n_t\cdot P)^{(n_1+n_2)\cdot n_t}$.
For lattice implementation, $O^B_g(z)$ can also be defined as~\cite{Dorn:1986dt,Zhang:2018diq}
\begin{align}
	O^B_g(z)\!=\! 2g_{\perp}^{\mu\nu}\mbox{tr}\Big[F^{n_1}_{0,\mu}\big({z\over2}\big)W\!\big({z\over2},\!-\!{z\over2}\big)F^{n_2}_{0,\nu}\big(\!-\!{z\over2}\big)W\!\big(\!-\!{z\over2},\!{z\over2}\big)\Big]\,,
\end{align}
where $F^{\mu\nu}=F^{\mu\nu}_at^a$ and $W$ are in the fundamental representation.
Similar to \eq{product}, we can express $O^B_g(z)$ as a product of two local composite operators,
\begin{align} \label{eq:gluonprod}
	\tilde{O}_g^B(z)&=\int d^4\xi\ \delta(\xi^3\!-\!z)\\
	&\times g_{\perp,\mu\nu}\big\langle F^{n_1\mu}_{0,a}\big({\xi\over2}\big)Q^a_0\big({\xi\over2}\big)\bar{Q}^b_0\big(\!-\!{\xi\over2}\big)F^{n_2\nu}_{0,b}\big(\!-\!{\xi\over2}\big)\big\rangle_Q\nn\\
	&\equiv\!\int d^4\xi\ \delta(\xi^3\!-\!z) g_{\perp}^{\mu\nu}\big\langle J^{B}_{n_1\mu}\big({\xi\over2}\big)\bar{J}^{B}_{n_2\nu}\big(\!-\!{\xi\over2}\big)\big\rangle_Q\,,\nn
\end{align}
where the auxiliary ``heavy'' quark fields are in the adjoint representation, and
\beq
J^{\mu\nu}_B= F^{\mu\nu}_{0,a}Q_0^a,\ \ \ \ \bar{J}^{\mu\nu}_B= \bar{Q}_0^aF^{\mu\nu}_{0,a}\,.
\eeq

The renormalization of $J^{\mu\nu}_B$ and $\bar{J}^{\mu\nu}_B$ is more involved than the quark case, as they can mix with other composite operators of the same or less dimensions. In DR, BRST symmetry allows $J^{\mu\nu}_B$
to mix with~\cite{Dorn:1986dt,Zhang:2018diq}
\begin{align}
	J^{\mu\nu}_{2B} &\!=\! \left(n_z^\nu F^{\mu n_z}_{0,a} - n_z^\mu F^{\nu n_z}_{0,a} \right) Q_0^a/n_z^2\,,\\
	J^{\mu\nu}_{3B} &\!=\! (-in_z^\mu\! A^\nu_{0,a}\! +\! in_z^\nu A^\mu_{0,a})\big[(in_z\cdot\! D_0\!-\!i\delta m)Q_0\big]^a\!/n_z^2\,.
\end{align}
Their renormalization matrix is given by~\cite{Dorn:1986dt}
\begin{align}\label{eq:oprenormmix}
	\begin{pmatrix}
		J_{B}^{\mu\nu} \\  J_{2B}^{\mu\nu} \\ J_{3B}^{\mu\nu}
	\end{pmatrix}
	&=
	\begin{pmatrix}
		Z_{11} & Z_{12} & Z_{13}\\ 0 & Z_{22} & Z_{23} \\ 0 & 0 & Z_{33}
	\end{pmatrix}
	\begin{pmatrix}
		J_{R}^{\mu\nu}\\  J_{2R}^{\mu\nu} \\ J_{3R}^{\mu\nu}
	\end{pmatrix}
	,
\end{align}
where $J_{2B}^{\mu\nu}$ is gauge invariant while $J^{\mu\nu}_{3B}$ is gauge dependent and proportional to the equation of motion (EOM) for the auxiliary field. The Green's functions of the EOM operator will result in a $\delta$-function,
\begin{align}
	(in_z\cdot D_0(\xi)-i\delta m)\langle Q_0(\xi)\bar{Q}_0(0)\rangle_Q = \delta^{(4)}(\xi)\,,
\end{align}
which only contributes a contact term $\delta(z)$ after integrating over the auxiliary fields. As long as $z\neq0$, such mixing vanishes in all Green's functions of $O^B_g(z)$, so we can ignore the mixing between $J^{\mu\nu}_{B}$ and $J^{\mu\nu}_{3B}$ in the renormalization of $O^B_g(z)$. At $z=0$, $O^B_g(z)$ becomes a local operator and is known to mix with BRST-exact and EOM operators~\cite{Collins:1994ee}, whose renormalization can be performed in the standard way.

Note that when contracted with $n_z$,
\begin{align}
	J_{2B}^{n_z\mu}&\!=\! J_{B}^{n_z\mu} = F^{n_z\mu}_{0,a}Q_0^a\,,\\
	J_{3B}^{n_z\mu}&\!=\! i\big(\!-\!A^{\mu,a}_{0} + {n_z^\mu\over n_z^2} n_z\cdot A_0^a\big)\big[(in_z\cdot D_0-i\delta m)Q_0\big]_a\,,\nn
\end{align}
the $J_{B}^{n_z\mu}$ only mixes with the EOM operator $J_{3B}^{n_z\mu}$. As has been argued above, we can ignore such mixing for $z\neq0$.
Moreover, this degeneracy also leads to relations among elements in the renormalization matrix~\cite{Dorn:1986dt},
\begin{align}\label{eq:Zrelation}
	Z_{11}+Z_{12}=Z_{22}, \,\,\,  Z_{13}=Z_{23}\,.
\end{align}

When contracted with $n_t$,
\begin{align} \label{eq:gauxopt}
	J_{B}^{n_t\mu} &=F^{n_t\mu}_{0,a}Q_0^a\,,\nn\\
	J_{2B}^{n_t\mu} &=n_z^\mu F_{a,0}^{n_tn_z} Q_0^a/n_z^2\,,\nn\\
	J_{3B}^{n_t\mu}&= i{n_z^\mu\over n_z^2} n_t\cdot A_{0}^a \big[(in_z\cdot D_0-i\delta m)Q_0\big]_a\,.
\end{align}
As one can see, $J_{2B}^{n_t\mu}$ and $J_{3B}^{n_t\mu}$ vanish after contraction with $g_{\perp}^{\mu\nu}$, so $J_{B}^{n_t\mu}$ with transverse Lorentz index $\mu$ is multiplicatively renormalizable.

To summarize, for $z\neq0$ and transverse Lorentz index $\mu$, both $J_{B}^{n_z\mu}$ and $J_{B}^{n_t\mu}$ are multiplicatively renormalizable in coordinate space, thus proving the renormalizability of the gluon Wilson-line operator $O^B_g(z)$,
\begin{align} \label{eq:renproofgluon}
	O^B_g(z)&=Z_{J}Z_{\bar{J}}\!\int\! d^4\xi\ \delta(\xi^3\!-\!z)\ g_{\perp}^{\mu\nu}\big\langle J_{n_1\mu}^{R}\big({\xi\over2}\big)\bar{J}_{n_2\nu}^{R}\big(\!-\!{\xi\over2}\big)\big\rangle_Q\nn\\
	&=\ e^{\delta m|z|} Z_{J}Z_{\bar{J}}\ O^R_g(z)\,,
\end{align}
where
\begin{align}
	J_B^{n_1\mu}&=Z_{J}\ J_R^{n_1\mu}=(Z^g_Q)^{1\over 2}Z_A^{1\over 2}Z^g_V J_R^{n_1\mu}\,,\\
	\bar{J}_B^{n_2\nu}&=Z_{\bar{J}}\ J_R^{n_2\nu}=(Z^g_Q)^{1\over 2}Z_A^{1\over 2}Z^g_{\bar{V}} J_R^{n_2\mu}\,,
\end{align}
with $Z^g_V$ and $Z^g_{\bar{V}}$ being the renormalization constants for the vertex involving one gluon and one ``heavy quark'' fields. The wavefunction renormalization constant for the auxiliary ``heavy quark'', $Z^g_Q$, is different from the quark case as it is in the adjoint representation.

In addition, since $J_{B}^{n_z\mu}$ and $J_{B}^{n_t\mu}$ do not mix with ``heavy-to-light'' quark currents due to the mismatch of quantum numbers, it implies that the nonlocal gluon Wilson-line operator does not mix with the singlet quark one under renormalization.

\vspace{.3cm}
For the polarized gluon quasi-PDF, its definition is the same as \eq{quasigluon}, except that the gluon Wilson-line operator becomes
\begin{align}
	\Delta O^B_g(z) = \epsilon_{\perp,\mu\nu} F^{n_1\mu}_{0,a}(z) W^{ab}(z,0) F^{n_2\nu}_{0,b}(0)\,,
\end{align}
where $\epsilon_{\perp}^{\mu\nu}=\epsilon^{03\mu\nu}$. Since $\epsilon_{\perp}^{\mu\nu}$ only contracts with the transverse Lorentz indices, one can use the same proof for $O^B_g(z)$ to show that $\Delta O^B_g(z)$ is also multiplicatively renormalizable and does not mix with singlet quark case~\cite{Zhang:2018diq}.

Finally, one can also prove that \eq{renproofgluon} is valid under lattice regularization with $\delta m$ being linearly divergent~\cite{Zhang:2018diq}. This completes the proof of renormalizability of the gluon Wilson-line operators.

\paragraph*{One-loop renormalization.}
Now we demonstrate the above result by an explicit one-loop example. For the nonlocal Wilson-line operators to be multiplicatively renormalizable, it is important that all linear divergences associated with diagrams other than the Wilson line self-energy cancel out among themselves. To see this, a gauge symmetry preserving regularization is crucial. We use DR and keep poles around $d=3$ to identify the linear divergences~\cite{Zhang:2018diq,Wang:2019tgg}.

\begin{figure}[htb]	
	\centering
	\begin{subfigure}{.3\linewidth}
		\centering
		\includegraphics[width=\linewidth]{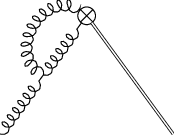}
		\caption{}
		\label{fig:Qgv1}
	\end{subfigure}
	\begin{subfigure}{.3\linewidth}
		\centering
		\includegraphics[width=\linewidth]{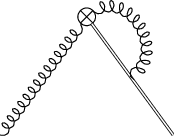}
		\caption{}
		\label{fig:Qgv2}
	\end{subfigure}	
	\begin{subfigure}{.3\linewidth}
		\centering
		\includegraphics[width=\linewidth]{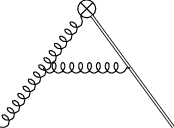}
		\caption{}
		\label{fig:Qgv3}
	\end{subfigure}
	\caption{One-loop vertex corrections to the ``heavy-to-gluon'' current.}
	\label{fig:ogmixing}
\end{figure}

The one-loop vertex correction to the ``heavy-to-gluon'' current is shown in \fig{ogmixing}. Each diagram contributes
\begin{align}\label{eq:pole_auxiliary}
	I_a^{\rho\nu}&= \frac{\alpha_s C_A}{\pi}\left[{1\over 4-d}{3\over 4} F_a^{\rho\nu}Q_a + {\rm finite\ terms}\right]\,,\nn\\
	I_b^{\rho\nu}&= \frac{\alpha_s C_A}{\pi}\Big[\frac{1}{d-4}(A_a^\nu n_z^\rho-A_a^\rho n_z^\nu) n_z\cdot \partial Q_a/n_z^2 \nn\\
	&+\frac{\pi\mu}{d-3}\big(n_z^\rho A_a^\nu-n_z^\nu A_a^\rho\big)Q_a+{\rm finite\ terms}\Big]\,,\nonumber\\
	I_c^{\rho\nu}&=\frac{\alpha_s C_A}{\pi}\Big\{\frac{1}{d-4}\Big[\frac{1}{2}\big(F_a^{\rho n_z}n_z^\nu -F_a^{\nu n_z}n_z^\rho \big) Q_a/n_z^2 \nn\\
	&+\frac{1}{4}F_a^{\rho\nu}Q_a+\frac 1 2 (A_a^\rho n_z^\nu-A_a^\nu n_z^\rho) n_z\cdot \partial Q_a/n_z^2\Big]\nonumber\\
	&-\frac{\pi\mu}{d-3}\big(n_z^\rho A_a^\nu-n_z^\nu A_a^\rho\big) Q_a+{\rm finite\ terms}\Big\}\,.
\end{align}

Both \fig{Qgv2} and \fig{Qgv3} include a linear divergence that is evident as the $\mu/(d-3)$ term, but they cancel among themselves. This guarantees that the overall UV divergence in the vertex correction is logarithmic, thus the renormalization of the ``heavy-to-gluon'' current is multiplicative. Combining the one-loop results in \eq{pole_auxiliary} and wavefunction renormalizations, we have
\begin{align}
	Z_{11} &= 1+ \frac{\alpha_s C_A}{4\pi}{1\over \epsilon_{\mbox{\tiny UV}}}\,,\qquad\qquad Z_{12} = 1- \frac{\alpha_s C_A}{4\pi}{1\over \epsilon_{\mbox{\tiny UV}}}\,,\nn\\
	Z_{13} &=Z_{23} = 1- \frac{\alpha_s C_A}{4\pi}{1\over \epsilon_{\mbox{\tiny UV}}}\,,\quad Z_{22} = 0\,,
\end{align}
where $C_A=N_c=3$ for QCD.
If we ignore the mixing to the EOM operator,
\begin{align}
	Z^{J^{n_z\nu}}_V &= Z^{J^{\nu n_z}}_V = 0 \,,\nn\\
	Z^{J^{n_ti}}_V &= Z^{J^{in_t}}_V = Z^{J^{ij}}_V = Z^{J^{ji}}_V = 1+ \frac{\alpha_s C_A}{4\pi}  {1\over \epsilon_{\mbox{\tiny UV}}}\,,
\end{align}
where $i,j=1,2$. As a result, the one-loop current renormalization constant is
\begin{align}
	Z_{J^{n_z\nu}} &= Z_{J^{\nu n_z}} =1 +{\alpha_s\over 4\pi}\left({1\over 6}C_A - {4\over 3}n_fT_F\right) {1\over\epsilon_{\mbox{\tiny UV}}} \,,\nn\\
	Z_{J^{n_ti}} &= Z_{J^{in_t}} = Z_{J^{ij}} =Z_{J^{ji}} \nn\\
	&=1 + {\alpha_s\over 4\pi}\left({7\over 6}C_A - {4\over 3}n_fT_F\right) {1\over\epsilon_{\mbox{\tiny UV}}}\,,
\end{align}
where $T_F=1/2$, and $n_f$ is the number of active quark flavors. The two-loop results can be found in~\cite{Braun:2020ymy}.

As one can see, the anomalous dimension of the ``heavy-to-gluon'' current is the same for $\mu,\nu=0,1,2$, which is due to $SO(2,1)$ (or  $SO(3)$ in Euclidean space) symmetry around the $z$-axis.

\subsection{Factorization of Quasi-PDFs}
\label{sec:factorization}

The key to LaMET applications for collinear parton physics
is the factorization formula that relates the quasi-PDFs
to light-cone PDFs~\cite{Ji:2013dva}. Here we use the perturbative properties
of the matching coefficients to write the factorization form
in the $\MS$ scheme in a way consistent with a direct EFT calculations
of PDFs at any given $x$~\cite{Izubuchi:2018srq,Wang:2019tgg}
\begin{align}
	q_i(x,  \mu)\! &=\! \int_{-\infty}^{\infty} \frac{dy}{|y|} \ \bigg[\!\sum_j \tilde C_{q_iq_j}\!\left(\frac{x}{y}, \frac{\mu}{yP^z}\right) \tilde{q}_j(y,P^z,\mu) \label{eq:factorization1} \\
	&\qquad + \tilde C_{qg}\left(\frac{x}{y}, \frac{\mu}{yP^z}\right) \tilde{g}(x,P^z,\mu)\bigg] + \cdots
	\ ,\nn\\
	g(x, \mu)\! &=\! \int_{-\infty}^{\infty} \frac{dy}{|y|} \ \bigg[\!\sum_j \tilde C_{gq}\left(\frac{x}{y}, \frac{\mu}{yP^z}\right) \tilde{q}_j(y,P^z,\mu) \label{eq:factorization2} \\
	&\qquad + \tilde C_{gg}\left(\frac{x}{y}, \frac{\mu}{yP^z}\right) \tilde{g}(y,P^z,\mu)\bigg] + \cdots\,,\nn
\end{align}
where $i,j$ runs over quark and anti-quark flavors.
The ``$\cdots$'' term includes mass corrections whose anayltical forms have been derived to all orders of $M^2/(P^z)^2$~\cite{Chen:2016utp}, and higher-twist contributions of order ${\cal O}\big(\Lambda_{\text{QCD}}^2/ (xP^z)^2,\Lambda_{\text{QCD}}^2/ ((1-x)P^z)^2\big)$ (see \eq{fact_0}). All $P^z$-dependence on the right hand side
cancels out, just like a renormalization scale.

As we have explained in the \sec{lamet}, the above factorization is guaranteed on the physics ground because
the difference between quasi-PDFs and light-cone PDFs is the order of limits in $P^z\to \infty$ and $\Lambda_{\rm UV}\to \infty$, and the IR physics in both quantities must be the same. An all-order
factorization proof for the quark quasi-PDF in perturbation theory was first given with a diagrammatical approach~\cite{Ma:2014jla}. The formula has also been derived using the operator product expansion (OPE) of nonlocal Wilson-line operators~\cite{Izubuchi:2018srq,Ma:2017pxb,Wang:2019tgg}.
Here we outline the diagrammatic proof similar to~\cite{Ma:2014jla}, showing
that the collinear divergences of the quasi-PDFs do factorize and
are equal to those of the light-cone PDFs.
Since the collinear divergence is a concept
in perturbation theory, we will show the factorization using a massless external quark state with lightlike momentum $P^\mu=(P^z,0,0,P^z)$.  While the proof is only for perturbative free quark states,
the factorization formulas are widely believed to be true nonperturbatively as well.
We use DR to regulate both UV and collinear divergences and only consider bare quantities, since UV renormalization does not change the leading collinear divergences.

Before the analysis, we should mention that all the soft divergences cancel between the real and virtual contributions to the quasi-PDFs, as discussed in \sec{lamet-rg}, thus we only need to focus on the collinear divergences. To obtain an intuitive understanding of the structure for collinear divergences, we start from the one-loop diagram in \fig{qoneloop1} in the Feynman gauge. The integral reads
\begin{align}
	\int \frac{d^{4-2\epsilon}k}{(2\pi)^{4-2\epsilon}}\frac{{\rm tr}\big[\slashed{P}\slashed{k}\gamma^z\slashed{k}\big]\delta(k^z-yP^z)}{(k^2+i0)^2((P-k)^2+i0)}\,.
\end{align}
The internal quark momentum is $k^\mu = (k^+,k^-,\vec{k}_\perp)$ and the gluon momentum is $P-k$. When $k^-$ and $k_\perp=|\vec{k}_\perp|$ are very small, the internal quark and gluon become collinear to the external quark, i.e. $k^{\mu}\sim (k^+,0,0_\perp)$ and $(P-k)^{\mu}\sim (P^+-k^+,0,0_\perp)$. In this case, the denominator of the quark and gluon propagators, $(k^2)^2$ and $(P-k)^2$, both vanish, which leads to collinear divergence. Conversely, for $k^2=(P-k)^2=0$, $k$ must be collinear to $P$ since the condition requires $k^2=k\cdot P=P^2=0$. For small $k^-$ and $k_\perp$, the $\delta$ function is dominated by the $k^+$ term of $k^z=(k^+-k^-)/\sqrt{2}$ and reduces to $\sqrt{2}\delta(k^+-yP^+)$. This is just the vertex which restricts $k^+=yP^+$ for the light-cone PDF, up to the factor $\sqrt{2}$. Furthermore, for collinear $k$ and $(P-k)$, the spinor trace in the numerator is dominated by the $\gamma^+$ part of $\gamma^z=(\gamma^+-\gamma^-)/\sqrt{2}$, ${\rm tr}\big[\slashed{P}\slashed{k}\gamma^z\slashed{k}\big] \sim {\rm tr}\big[\slashed{P}\slashed{k}\gamma^+\slashed{k}\big]/\sqrt{2}$. Thus in the collinear region $k^{\mu}\sim (k^+,0,0,0)$ the above integral reduces to that for the light-cone PDF:
\begin{align}
	\int_{\rm c} \frac{d^{4-2\epsilon}k}{(2\pi)^{4-2\epsilon}}\frac{{\rm tr}\big[\slashed{P}\slashed{k}\gamma^+\slashed{k}\big]\delta(k^+-yP^+) }{(k^2+i0)^2((P-k)^2+i0)}\,,
\end{align}
where the subscript ``c'' denotes the collinear region.

The above picture naturally arises in a highly boosted hadron state where the quark is approximately onshell.
Therefore, as explained in \sec{lamet-univ},  {\it although the operator contains no light-cone information, the large-momentum external hadron state can still generate collinear divergences equivalent to those in the light-cone PDFs.} By subtracting the full integral for light-cone PDF from that for the quasi-PDF, the logarithmic collinear divergence cancels, and the remaining difference is perturbative and can be absorbed into the matching kernel.

Similarly, for the vertex diagram in Fig.~\ref{fig:qoneloop2a}, the loop integral is proportional to
\begin{align}
	\int \frac{d^{4-2\epsilon}k}{(2\pi)^{4-2\epsilon}}\frac{1}{P^z-k^z}\frac{{\rm tr}\big[ \slashed{P} \gamma^z \slashed{k} \gamma^z\big] \delta(k^z-yP^z)}{(k^2+i0)((P-k)^2+i0)} \,.
\end{align}
The whole integral in the collinear region reduces to
\begin{align}
	\int_{\rm c} \frac{d^{4-2\epsilon}k}{(2\pi)^{4-2\epsilon}}\frac{1}{P^+-k^+}\frac{{\rm tr} \big[\slashed{P} \gamma^+ \slashed{k} \gamma^+\big]\delta(k^+-yP^+)}{(k^2+i0)((P-k)^2+i0)}  \ ,
\end{align}
which is the corresponding integral for the light-cone PDF. One key feature of the diagram is that while the gauge link probes the $z$-component of the gluon field $A^z =\left(A^+-A^-\right)/\sqrt{2}$, only the $A^+$ component (longitudinal polarization) contributes to the leading collinear divergence. While attaching a new collinear gluon to the gauge-link induces a power suppression from the link propagator of ${\cal O}(1/P^z)$, the $A^+$ component of the collinear gluon radiated from fast-moving color charges receives enhancement from Lorentz boost factor $\gamma$ that compensates for the suppression.

The above result can be generalized to all orders. Similar to the one-loop diagrams, in the leading region of collinear divergence there are an arbitrary number of longitudinally polarized $A^+$ gluons, which are emitted dynamically from the fast-moving state instead of being put in by hand using the lightlike gauge link, in contrast to the standard collinear PDF. The existence of the $A^+$ gluons clearly increases
the level of complication in showing the equivalence of collinear divergences between the quasi- and light-cone PDFs. For simplification, from now on we choose to work in the light-cone gauge $A^+=0$ to eliminate all the $A^+$ gluons. Therefore, the vertex diagrams
no longer contribute to the leading collinear divergence, thus making its structure much simpler.

\begin{figure}
	\includegraphics[width=0.8\columnwidth]{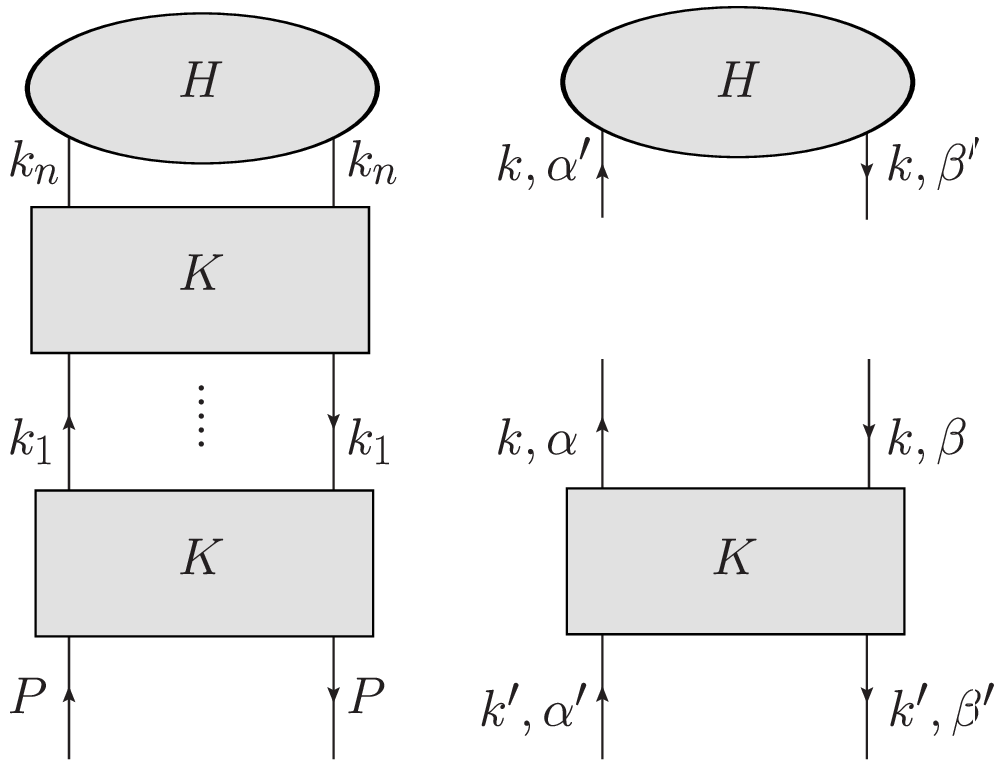}
	\caption{\label{fig:ladder} The ladder decomposition of the quasi-PDF (left). The upper 2PI kernel $H$ contains the operator defining the quasi-PDF, and external two legs at the bottom of the diagram is the external large $P^z$ state. The kernels $H$ and $K$ are shown on the right.}
\end{figure}

In a general diagram, we decompose the potential leading region of the quasi-PDF into the ladder structure shown in Fig.~\ref{fig:ladder}. The upper two-particle-irreducible (2PI) kernel that contains the nonlocal operator defining the quasi-PDF is $H$. The 2PI kernel in the ladder is $K$. $K$ contains the upper two external quark lines but not the lower ones. The momentum flowing out of the ladders are labeled as $k_1$ to $k_n$ from bottom to top when there are $n$ 2PI kernels. We write $H$ and $K$ as matrices in spinor and momentum space. $H=H_{\alpha'\beta'}(yP^z;k)$ where $k$ denotes the momentum flowing into $H$ and $K=K_{\alpha\beta;\alpha'\beta'}(k,k')$ where $k$, $k'$ are the momenta of the upper and lower external legs, respectively.
Here $\alpha\beta$ and $\alpha'\beta'$ are the spinor indices for the upper and lower two external legs, respectively.
Following the method in~\cite{Curci:1980uw,Collins:2011zzd}, we find that:
\begin{enumerate}
	\item There are no collinear divergences in the upper part $H$ in the light-cone gauge.
	\item If none of $k_1,\cdots,k_n$ is collinear, there will be no leading collinear divergence. More generally, for the $i$'th 2PI kernel, if either of $k_{i-1}$ and $k_{i}$ is not collinear, then the sub-integrals inside the kernel are finite and it does not contribute to leading collinear divergence.
	\item If $k_i$ is not collinear, then there are no collinear divergences for the upper part of the diagram above the $i$'th ladder.
\end{enumerate}
Therefore, the collinear divergences are generated in the momentum regions $R_i$ in which $k_1$ to $k_i$ are collinear while $k_{i+1}$ to $k_n$ are not. We can construct counter terms that subtract out the collinear divergences in each of the regions $R_i$.
For this we keep only the $+$ component of $k_i$ in the convergent upper part $HK^{n-i}$ as in the one-loop example, namely $k_i\rightarrow (k^+_i,0,0_\perp)$ in the upper part. This will clearly leave the collinear divergence unchanged.  Also notice that $[HK^{n-i}]_{\alpha\beta}=H_{\alpha'\beta'}K^{n-i}_{\alpha'\beta';\alpha\beta}$ should be understood as a 4 $\times$ 4 Dirac matrix with indices $\alpha\beta$, while the lower part is $[K^i \slashed P]_{\alpha\beta}=K^i_{\alpha\beta;\alpha'\beta'}\slashed P_{\alpha'\beta'}$. In the leading region of collinear divergence, $HK^{n-i}$ and $K^i\slashed P$ are proportional to $\gamma^+$ and $\gamma^-$ respectively.
Therefore, to obtain the leading collinear divergence, we can disentangle the spinor traces for the upper and lower parts by contracting them with $\gamma^-/2$ and $\gamma^+/2$ separately. The only communication between them is the $k^+$ integration. The collinear divergence is contained in the lower part
\begin{align}
	q^{i}(x,\epsilon_{\mbox{\tiny IR}})\!=\!\!\int \frac{dk^-\!d^{d-2}\!k_\perp}{2(2\pi)^{d}}{\rm tr} \big[\gamma^+\! K^i(xP^+\!,k^-\!,k_\perp;P)\slashed{P}\big],
\end{align}
where $d=4-2\epsilon$, $k^+=xP^+$, and the subtraction for the region $R_i$ can be written effectively as a convolution
\begin{align}
	\int \frac{dx}{x} \hat C^{n-i}(y,x,P^z)q^i(x,\epsilon_{\mbox{\tiny IR}}) \ ,
\end{align}
where
\begin{align}
	\hat C^{n-i}(y,x,\!P^z)=\frac{1}{2}{\rm tr}\left[HK^{n-i}(yP^z;xP^+,\!0,0_\perp)(xP^+)\gamma^-\right]
\end{align}
is the naive matching kernel. Here the $y$ dependence comes from the operator in $H$. However, the naive matching kernel still suffers from collinear sub-divergences that need to be subtracted. This can be achieved using the subtracted matching kernels $C^{n-i}(y,x)$ defined recursively in a way similar to the BPHZ relation for UV renormalization~\cite{Collins:2011zzd}. Summing over $n$ and $i$, the recursive relation leads to
\begin{align}\label{eq:subtraction}
	\tilde q(y,P^z,\epsilon_{\mbox{\tiny IR}})&=\sum_{n=0}^{\infty}\sum_{i=0}^n \int \frac{dx}{x} C^{n-i}(y,x,P^z)q^i(x,\epsilon_{\mbox{\tiny IR}})\nn\\
	&=\int \frac{dx}{x} C(y,x,P^z)q(x,\epsilon_{\mbox{\tiny IR}}) \ ,
\end{align}
where $\tilde q(y,P^z,\epsilon_{\mbox{\tiny IR}})$ is the quasi-PDF, $C(y,x,P^z)=\sum_{n=0}^{\infty} C^{n}(y,x,P^z)$ is the all-order matching kernel and $q(x,\epsilon_{\mbox{\tiny IR}})=\sum_{i=0} q^i(x,\epsilon_{\mbox{\tiny IR}})$. Based on the definition of $q^i(x,\epsilon_{\mbox{\tiny IR}})$, it is clear that $q^i$ equals the light-cone PDF with $i$ 2PI kernels and $q$ is the full light-cone PDF with natural support $0<x<1$. {\it The light-cone PDF $q(x)$ is independent of the operator defining the quasi-PDF}, as it is only sensitive to the explicit form of the collinear divergence. The r.h.s. of \eq{subtraction} contains all the collinear divergences from the quasi-PDF $\tilde q$. Thus the matching relation for bare quantities is established. A similar matching can be written down for the renormalized quantities, where the renormalization only affects the matching kernel $C(y,x,P^z)$. Note that the explicit solution for $C^{n-i}(y,x,P^z)$, which leads to \eq{fact_0}, can be given based on a subtraction operator defined similar to that in~\cite{Collins:2011zzd}. Besides, \eq{subtraction} can be inverted order by order in $\alpha_s$, thus proving \eq{fact_0}, which can also be generalized to \eqs{factorization1}{factorization2}.

Now we present the matching coefficient in the $\MS$ scheme at one-loop order.
The one-loop expansion of the $\MS$ quasi- and light-cone PDFs in a free massless quark state with momentum $p^\mu=(p^z,0,0,p^z)$ are
\begin{align}
	\tilde{q}(y,\mu/p^z,\epsilon_{\mbox{\tiny IR}}) &= \tilde{q}^{(0)}(y) + {\alpha_sC_F\over 2\pi}\tilde{q}^{(1)}(y,\mu/p^z,\epsilon_{\mbox{\tiny IR}})\,,\\
	q(x,\epsilon_{\mbox{\tiny IR}}) &= q^{(0)}(x) + {\alpha_sC_F\over 2\pi}q^{(1)}(x,\epsilon_{\mbox{\tiny IR}})\,.
\end{align}
At tree level, $\tilde{q}^{(0)}(y) = q^{(0)}(y) = \delta(1-y)$.
At one loop, the $\MS$ quasi-PDF and its counterterm are~\cite{Izubuchi:2018srq}
\begin{align}\label{eq:qPDFren}
	&\tilde{q}^{(1)}(y,\mu/p^z,\epsilon_{\mbox{\tiny IR}}) \nn\\
	=&\left\{
	\begin{array}{ll}
		\Big({1+y^2\over 1-y}\ln {y\over y-1} + 1 + {3\over 2y}\Big)^{[1,\infty]}_{+(1)}- {3\over 2y} & y>1\\
		\Big({1+y^2\over 1-y}\big[- {1\over \epsilon_{\mbox{\tiny IR}}} - \ln{\mu^2\over 4(p^z)^2} + \ln\big(y(1-y)\big)\big] &\\
		\  - {y(1+y)\over 1-y}+ 2\sigma (1-y)\Big)^{[0,1]}_{+(1)}  &0<y<1\\
		\Big(-{1+y^2\over 1-y}\ln {-y\over 1-y} - 1 + {3\over 2(1-y)}\Big)^{[-\infty,0]}_{+(1)}&\\
		\ - {3\over 2(1-y)} & y<0
	\end{array}\right.\nn\\
	&+ \delta(1-y)\left[{3\over2}\ln{\mu^2\over 4(p^z)^2} + {5 + 2\sigma\over2}\right]\,,
\end{align}
\begin{align}
	\delta \tilde{q}^{(1)}(y,\mu/p^z,\epsilon_{\mbox{\tiny UV}})
	= {3\over 2\epsilon_{\mbox{\tiny UV}}}
	\delta(1-y)\,,
\end{align}
where $\epsilon_{\mbox{\tiny IR}}$ regulates the collinear divergence, $\sigma=0$ for $\Gamma=\gamma^t$ and 1 for $\Gamma=\gamma^z$. The plus function at $y=y_0$ with support in a given domain $D$ is defined as
\begin{align}
	\int_D\! dy \big[ g(y)\big]_{+(y_0)}^D h(y)\! =\! \int_D\! dy\ g(y) \left[ h(y)\! -\! h(y_0)\right]
\end{align}
with arbitrary $g(y)$ and $h(y)$. Note that the $\MS$ renormalization of the quasi-PDF actually requires a subtle treatment of vector current conservation~\cite{Izubuchi:2018srq}. We only present results in the form that is sufficient for our discussion, which differs slightly from that in~\cite{Izubuchi:2018srq} by the $\delta$-functions at $y=\pm\infty$ and from the treatment in~\cite{Alexandrou:2019lfo}.

On the other hand,
\beq\label{eq:lcpdf}
q^{(1)}(x,\epsilon_{\mbox{\tiny IR}}) ={\alpha_sC_F\over 2\pi} {(-1)\over \epsilon_{\mbox{\tiny IR}}} \left({1+x^2\over 1-x}\right)^{[0,1]}_{+(1)}\,,
\eeq
which is limited to the physical region as expected.

By comparing the quasi- and light-cone PDFs in \eqs{qPDFren}{lcpdf}, we find that both of them have the same collinear divergence,  or in other words, they share the same IR physics, thus validating the factorization formula at one-loop order. Setting $p^z=xP^z$ and plugging the one-loop results into \eq{fact_0}, we extract the matching coefficient for the hadron matrix element which only depends on the perturbative scales $\mu$ and $P^z$,
\begin{align}\label{eq:msbarmatching}
	C^{\MS}\left(y, {\mu\over xP^z}\right) &= \delta\left(1-y\right) + {\alpha_sC_F\over 2\pi}\left[\tilde q^{(1)}\left(y,{\mu\over xP^z}, \epsilon_{\mbox{\tiny IR}}\right)\right.\nn\\
	&\qquad\left.-q^{(1)}(y,\epsilon_{\mbox{\tiny IR}})\right]\,.
\end{align}
The complete one-loop matching coefficients in \eq{fact_0} in the transverse-momentum cutoff and $\MS$ schemes can be found in \cite{Wang:2017qyg,Wang:2017eel,Wang:2019tgg}. The two-loop results were obtained recently in~\cite{Chen:2020arf,Chen:2020iqi,Chen:2020ody,Li:2020xml}.

\subsection{Coordinate-Space Factorization of Bilinear Operators}
\label{sec:renorm-coord}
Although the LaMET application to PDFs concerns the expansion of momentum densities
in the $P^z\to\infty$ limit, lattice QCD calculations actually start from
computing coordinate-space correlations, for example,
\beq
\tilde h(z, P^z) = {1\over N_\Gamma}\langle P^z| O_\Gamma(z) |P^z\rangle\,,
\eeq
at all $z$ and do Fourier transform with respect
to $\lambda=zP^z$ at a fixed $P^z$. Here the normalization factor $N_\Gamma=2P^z$ for $\Gamma=\gamma^z$ and $N_\Gamma=2P^t$ for $\Gamma=\gamma^t$.
The $\tilde h(z,P^z)$ is a function of two independent
variables $z$ and $P^z$, and in LaMET analysis the relevant
combinations are quasi-LF distance $\lambda$ (see \fig{Boost}) and $P^z$, hence
$\tilde h(\lambda, P^z)$ will be called quasi-LF correlation, which is distinguished from the LF correlation $h(\lambda,\mu)$ below.

The coordinate-space factorization approach in \cite{Braun:1994jq} has been
suggested as an alternative way to extract the PDFs from $\tilde h(z, P^z)$~\cite{Radyushkin:2017cyf,Orginos:2017kos,Radyushkin:2019mye}, which is closely related to the OPE.
Instead of working with variables $\lambda$ and $P^z$,
one may consider $\tilde h$ as a function of $\lambda$ and $z^2$, i.e.,
$\tilde h(\lambda, z^2)$. The Fourier transform
of $\tilde h(\lambda, z^2)$ with respect to $\lambda$ is no longer the momentum
distribution of the proton at a fixed momentum. Instead, it is called a pseudo-distribution~\cite{Radyushkin:2017cyf}.
At small $|z|\ll \Lambda^{-1}_{\rm QCD}$, $\tilde h(\lambda, z^2)$ can be factorized into the light-cone correlation~\cite{Radyushkin:2017lvu,Izubuchi:2018srq},
\begin{align}
	\label{eq:io-Q-fact}
	\tilde{h}(\lambda, z^2\mu^2)
	=\int_{-1}^1 d\alpha\ \mathcal{C}(\alpha, z^2\mu^2)\ h(\alpha\lambda,\mu) +...\,,
\end{align}
where $...$ are the power corrections in $z^2\Lambda^2_{\rm QCD}$,
and the matching coefficient $\mathcal{C}$
is related to $C$ in Eq.~(\ref{eq:fact_0}) by
\begin{align} \label{eq:ftc-C}
	C\left(\eta,{\mu\over xP^z}\right)
	\!= \!\int {d\lambda\over 2\pi}\ e^{i\eta\lambda}\int_{-1}^1\!\! d\alpha\: e^{-i\lambda\alpha}\: \mathcal{C}\left(\alpha,{\mu^2 \lambda^2\over (xP^z)^2}\right)\,.
\end{align}

To illustrate the connection between the above factorization in \eq{io-Q-fact} and OPE,
let us take the non-singlet quark case as an example~\cite{Izubuchi:2018srq,Wang:2019tgg}. In the $\MS$ scheme, the renormalized $O_{\gamma^{\mu_0}}(z,\mu)$ can be expanded in terms of local gauge-invariant twist-2 operators as $z^2\to 0$,
\begin{align}\label{eq:ope-tq}
	O_{\gamma^{\mu_0}}(z,\mu) &=\sum_{n=0}^\infty \Big[C_n ({\mu}^2 z^2)\frac{(iz)^n}{n!} (n_z)_{\mu_1}\cdots (n_z)_{\mu_n} \nn\\
	&\qquad \times O^{\mu_0\mu_1\cdots\mu_n}(\mu) +  \text{higher-twist}\Big]\,,
\end{align}
where $\mu_0\!=\!0$ or $3$, $C_n\!=\!1+{\cal O}(\alpha_s)$ is the Wilson coefficient, and $O^{\mu_0\mu_1\cdots\mu_n}(\mu)$
is the twist-two operator in \eq{t2opq}.

Using the hadron matrix elements in \eq{t2mom} and their relation to the light-cone PDF
in \eq{t2sum},
we write down the small-$|z|$ expansion of the hadron matrix element of $O_{\gamma^{\mu_0}}(z,\mu) $~\cite{Izubuchi:2018srq},
\begin{align}\label{eq:ope}
	\tilde{h}(\lambda,z^2\mu^2)&=\langle P| O_{\gamma^{\mu_0}}(z,\mu) |P\rangle/(2P^{\mu_0}) \nn\\
	&= \sum_{n=0}^\infty\ C_n(z^2\mu^2){(-i\lambda)^n\over n!}\left[1+{\cal O}\big({M^2\over (P^z)^2}\big)\right]\nn\\
	&\quad\times \int_{-1}^1 dx\ x^n q(x,\mu) + {\cal O}\left(z^2\Lambda_{\text{QCD}}^2\right)\,,
\end{align}
where the ${\cal O}\left(M^2/(P^z)^2\right)$ term comes from the kinematic trace contribution and
the ${\cal O}\left(z^2\Lambda_{\text{QCD}}^2\right)$ term from higher-twist.
The Wilson coefficients $C_n(z^2\mu^2)$ have been calculated at one-loop~\cite{Izubuchi:2018srq} and two-loop~\cite{Li:2020xml} orders.
Comparing the above equation with Eq.~(\ref{eq:io-Q-fact}),
we identify
\begin{align}  \label{eq:fac-C}
	\mathcal{C} (\alpha, {\mu}^2 z^2) \equiv \int\! \frac{d \lambda}{2 \pi} \,
	e^{i  \lambda \alpha}  \sum_n C_n ({\mu}^2 z^2)  \frac{(-i \lambda)^n}{n!}
	\,.
\end{align}
Since $z^2$ is fixed in $\mathcal{C} (\alpha, {\mu}^2 z^2)$, the integration in \eq{fac-C} is actually over $P^z$ from $-\infty$ to $+\infty$.
$\mathcal{C}(\alpha,z^2\mu^2)$ has support $-1 \le \alpha \le 1$,
and its one-loop result is
\begin{align} \label{eq:ps-c}
	&{\cal C}(\alpha,z^2\mu^2)\\
	=& \left[ 1+ {\alpha_sC_F\over 2\pi}\left({3\over2}\ln{z^2\mu^2e^{2\gamma_E}\over 4}+{3\over2}\right)\right]\delta(1-\alpha)\nn
	\\
	&+ {\alpha_sC_F\over 2\pi}\left\{\left({1+\alpha^2\over 1-\alpha}\right)^{[0,1]}_{+(1)}\left[-\ln{z^2\mu^2e^{2\gamma_E}\over 4}-1\right]\right.\nn\\
	&\left. - \left(4\ln(1-\alpha)\over 1-\alpha\right)^{[0,1]}_{+(1)}\!  +\!2(1+\sigma)(1-\alpha)\right\}\!\theta(\alpha)\theta(1-\alpha)
	\,,\nn
\end{align}
which was also calculated and further studied in~\cite{Ji:2017rah,Radyushkin:2017lvu,Radyushkin:2018cvn,Zhang:2018ggy,Li:2020xml}.
One can check that the above result is indeed related to one-loop momentum-space
matching by Eq.~(\ref{eq:ftc-C}).
Since we are interested in the relation between the
quasi-LF correlation with the matrix element of the light-ray operator
$O_{\gamma^+}(\lambda n)$, Eq.~(\ref{eq:io-Q-fact})
can also be obtained by using the light-ray operator
expansion in~\cite{Balitsky:1987bk,Braun:1994jq,Braun:2007wv}.

Using OPE or short-distance expansion, the exact factorization formula for the gluon and singlet quark quasi-PDFs, which includes their mixings, has also been derived in coordinate space and studied at one-loop order~\cite{Wang:2019tgg,Balitsky:2019krf}.

It is easy to see that the limits $P^z\to\infty$ in LaMET expansion and $z\to 0$
in coordinate-space factorization, keeping finite $\lambda=zP^z$, are equivalent.  However, in practical
lattice QCD calculations, one is limited by the largest
momentum $P^z_{\rm max}$ in a specific setup, and the two approaches start to differ.

In LaMET systematic approximation, one should calculate $\tilde h (z, P^z_{\rm max})$ with
all possible $z$ or $\lambda$, but in practice the largest $\lambda_{\rm max} = z_{\rm max}P^z_{\rm max}$
is limited by the lattice volume as well as data quality at large $z$. Due to QCD confinement, $\tilde h (z, P^z_{\rm max})$ has a correlation length $\sim 1/\Lambda_{\rm QCD}$, leading to an exponential decay
at large $z$~\cite{Ji:2020brr}. Therefore, if $z_{\rm max}$ is sufficiently large (e.g., the proton size $\sim1$ fm) for $\tilde h (z, P^z_{\rm max})$ to fall to almost zero, then the truncated Fourier transform of $\tilde h (z, P^z_{\rm max})$
should converge quickly, and the truncation effects mainly affect results at small $x \lesssim 1/\lambda_{\rm max}$. If $\tilde h (z, P^z_{\rm max})$ exhibits exponential decay but still has a nonzero value at $z_{\rm max}$, then one can perform a physically motivated extrapolation beyond $z_{\rm max}$~\cite{Ji:2020brr} to do the Fourier transform, which removes the unphysical oscillation from truncation and only affects the small-$x$ region. In the momentum space, one can use LaMET expansion to calculate the PDF point by point in $x$ with systematic error controlled by $\Lambda_{\rm QCD}^2/(xP^z_{\rm max})^2$ and $\Lambda_{\rm QCD}^2/((1-x)P^z_{\rm max})^2$, which gives the prediction for a cetain region of $x$, $[x_{\rm min}, x_{\rm max}]$, with a target error.

In coordinate-space factorization, one expands $\tilde h(\lambda, z^2)$ in $z^2\Lambda^2_{\rm QCD}$. For the factorization formula to be valid,  $z$ must remain in the perturbative region. For example, an estimate in~\cite{Ji:2020brr} gives $z_{\rm max}\sim 0.3$--$0.4$ fm. Although there have been observations that forming ratios of $\tilde h (\lambda, z^2)$ may cancel the higher-twist contributions at $z>0.4$ fm~\cite{Orginos:2017kos}, this cancellation needs be quantified
for precision calculations. With a finite range of quasi-LF correlations, the PDFs can be extracted through modelling
the $x$-dependence or more advanced techniques such as Bayesian analysis~\cite{Bringewatt:2020ixn} or neural network~\cite{Karpie:2019eiq,Cichy:2019ebf,DelDebbio:2020rgv}, which is similar to extracting the PDFs from experimental data~\cite{Ma:2017pxb}, although quantifying the
systematic error from fitting can be challenging. The coordinate-space factorization can also provide the extraction of the lowest moments of PDFs~\cite{Karpie:2018zaz,Shugert:2020tgq,Joo:2020spy,Gao:2020ito}, where the main systematic error is controlled by $z^2\Lambda_{\rm QCD}^2$.

So far, there have been very limited studies about the comparison between quasi- and pseudo-PDF analysis
~\cite{Bhat:2020ktg,Alexandrou:2020qtt}. It remains to be seen how systematic errors
in the two strategies are compared and contrasted.

\subsection{Nonperturbative Renormalization and Matching}
\label{sec:renorm-npr}
The multiplicative renormalizability of the nonlocal Wilson-line operators for quasi-PDFs allows a nonperturbative renormalization on the lattice, after which the continuum limit can be taken.
This is an important step in the application of LaMET.
One way of doing so is to perform a mass
subtraction of the Wilson line first~\cite{Musch:2010ka,Ishikawa:2016znu,Chen:2016fxx,Zhang:2017bzy,Green:2017xeu,Green:2020xco}, and then renormalize the remnant UV divergences with lattice perturbation theory or other nonperturbative schemes. Another scheme which has gained more popularity in recent years is the regularization-independent momentum subtraction (RI/MOM) scheme~\cite{Constantinou:2017sej,Stewart:2017tvs,Alexandrou:2017huk,Chen:2017mzz,Liu:2018uuj}.
In the coordinate space approach where $|z|\ll \Lambda_{\rm QCD}^{-1}$, the ratios of quasi-LF correlations in
different states~\cite{Radyushkin:2017cyf,Orginos:2017kos,Braun:2018brg,Li:2020xml} have also been proposed as a renormalization scheme. At large $z$, the RI/MOM and ratio schemes introduce extra nonperturbative effects at different levels, which may distort the IR property of the original quasi-LF correlations. Due to the suppression  of long-range contributions by large $P^z$ in the Fourier transform, this nonperturbative contamination mainly affects the end-point region in $x$-space, while the existing LaMET calculations with RI/MOM scheme at moderate $x$, for example in~\cite{Alexandrou:2018eet,Lin:2018qky}, suffers less from such systematics. Nevertheless, the above complication can be avoided by switching to the hybrid scheme~\cite{Ji:2020brr} where one utilizes the advantages of different schemes at short and large distances.
In the following, we discuss the above schemes in order, with a particular focus on the hybrid renormalization scheme.

Before we proceed, it should be noted that the current-current correlators in~\cite{Detmold:2005gg,Braun:2007wv,Ma:2017pxb} do not need or have simple renormalization on the lattice, though it might be more costly to simulate them. Besides, there is another distinct method based on a redefinition of the quasi-PDF with smeared fermion and gauge fields via the gradient flow~\cite{Monahan:2016bvm}. The smeared quasi-PDF is free from UV divergences and remains finite in the continuum limit, which can be perturbatively matched onto the PDF~\cite{Monahan:2017hpu}. Nevertheless, this method awaits to be implemented on the lattice.

\subsubsection{Wilson-line mass-subtraction scheme}
\label{sec:renorm-npr-mass}

Since the mass correction $\delta m$ includes all the linear UV divergences, it is highly favored to nonperturbatively subtract it from the quasi-PDFs. It is well known that the Wilson line renormalization is related to the additive renormalization of the static quark-antiquark potential, i.e., $\delta m$, especially in the context of finite temperature field theory. For a rectangle-shaped Wilson loop of dimension $L\times T$ in the spatial and temporal directions, its vacuum expectation value for large $T$ scales as
\begin{align}
	\lim_{T\to\infty}W(L,T) = c(L) e^{-V(L)T}\,.
\end{align}
The renormalized static potential is
\begin{align}\label{eq:deltam1}
	V^{R}(L) = V(L) + 2\delta m\,,
\end{align}
and $\delta m$ can be fixed by imposing the condition $V^{R}(L_0)=0$ for a particular value of $L_0$. Alternatively, one can also fit $\delta m$ from the famous string potential model,
\begin{align}\label{eq:deltam2}
	V(L) = \sigma L- {\pi \over 12 L} -2\delta m\,.
\end{align}

Apart from using the static potential to determine $\delta m$, it was also proposed to calculate this quantity in the auxiliary ``heavy quark'' field theory with the following condition~\cite{Green:2017xeu},
\begin{align}\label{eq:deltam3}
	\delta m = {d\over dz} \ln{\rm Tr}\big\langle Q(x+zn_z)\bar{Q}(x)\big\rangle_{{\rm QCD}+Q} \Big|_{z=z_0}\,.
\end{align}

Other suggestions have also been made for a nonperturbative calculation of $\delta m$~\cite{Ji:2020brr}. For example, one can investigate the asymptotic large-$z$ behavior of the hadron matrix element or the single quark Green's function, of the vacuum expectation value of $O_{\Gamma}(z,a)$ in a fixed gauge. The $\delta m$ calculated from all these matrix elements will have the following dependence
on the lattice spacing $a$,
\begin{equation}
	\delta m = m_{-1}(a)/a + m_0 \ ,
\end{equation}
where $m_{-1}$(a) is the coefficient of the power divergence which is independent of the specific matrix element, while $m_0~\sim O(\Lambda_{\rm QCD})$ is finite and depends on the external state.
The determination of $m_0$ can be rather nontrivial, and in practical calculations one could adopt a fine-tuning method, such as that for the Wilson-fermion mass, to find the critical value of $m_0$ at which the final result converges fastest in the large $P^z$ limit.

After the Wilson-line mass subtraction, there are still logarithmic UV divergences in $O_\Gamma(z,a)$. One can use lattice perturbation theory to match $\delta m$-subtracted $O_\Gamma(z,a)$ to the $\MS$ scheme~\cite{Ishikawa:2016znu,Xiong:2017jtn,Constantinou:2017sej}, but the convergence still needs to be examined at higher orders. In~\cite{Green:2017xeu,Green:2020xco}, the logarithmic divergences were nonperturbatively renormalized with RI/MOM-like schemes.

The Wilson-line mass-subtraction has been implemented on the lattice in~\cite{Musch:2010ka,Zhang:2017bzy,Green:2017xeu,Chen:2017gck,Alexandrou:2020qtt}.

\subsubsection{RI/MOM scheme}
\label{sec:renorm-npr-rimom}

The RI/MOM scheme has been widely used in lattice QCD for the renormalization of local composite quark operators that are free from power-divergent mixings ~\cite{Martinelli:1994ty}. It is essentially a momentum subtraction scheme in QFT and can be nonperturbatively implemented on the lattice. For an arbitrary composite quark bilinear operator $O^B$ that is multiplicatively renormalized as $O^B=Z_OO^R$, the RI/MOM scheme is defined by imposing the following condition on its off-shell quark matrix element at a subtraction scale $\mu_{\tiny R}$,
\begin{align} \label{eq:rimom}
	Z^{-1}_O \langle p|O^B|p\rangle \Big|_{p^2=-\mu_{\tiny R}^2}= \langle p|O|p\rangle_{\rm tree}\,.
\end{align}
where the subscript ``tree'' means the tree-level matrix element in perturbation theory. If $\mu_{\tiny R} \gg \Lambda_{\rm QCD}$, $Z_O$ defined in \eq{rimom} is in the perturbative region, and we can convert it to the $\MS$ scheme order by order in perturbation theory. In this sense, $Z_O$ is not literally nonperturbative, but an all-order calculable quantity.

Since the nonlocal quark bilinear operator $O_\Gamma(z)$ has been proven to be multiplicatively renormalizable in the coordinate space, one can also renormalize it in the RI/MOM scheme and then match the result to PDF in the $\MS$ scheme~\cite{Constantinou:2017sej,Stewart:2017tvs}.
On the lattice, the off-shell matrix element of an operator is defined from its amputated Green's function, or vertex function, with off-shell quarks. For the nonlocal Wilson-line operator, the latter is
\begin{align} \label{eq:Lambda}
	&\Lambda^{\Gamma}_0(z, a, p) \equiv
	\left[S^{-1}_0(p,a)\right]^\dagger \sum_{x,y}e^{ip\cdot (x-y)}
	\nn\\
	&\ \ \  \times\! \big\langle 0 \big|T\bigl[\psi_0(x,a) O^B_\Gamma(z,a) \bar{\psi}_0(y,a)\bigr]\big|0\big\rangle S^{-1}_0(p,a)\,,
\end{align}
where $S_0(p,a)$ is the bare quark propagator, and the external momentum $p$ is Euclidean on the lattice.
Since Green's functions are not gauge invariant, one needs to fix a gauge (usually Landau gauge $\partial\cdot  A=0$ is chosen), and the gauge dependence is expected to be canceled by the matching or scheme conversion order by order in perturbation theory.

After including the quark wavefunction renormalization $Z_q$, which can be determined independently on the lattice~\cite{Martinelli:1994ty}, \eq{rimom} is revised as
\begin{align} \label{eq:rimomvertex}
	Z_q Z^{-1}_{O_\Gamma} \Lambda^{\Gamma}_0(z,a,p)\Big|_{p=p_{\tiny R}}= \Lambda^{\Gamma}_{\rm tree}(z,a,p) = \Gamma e^{ip_{\tiny R}\cdot z}\,.
\end{align}
Since $O_\Gamma(z,a)$ is not $O(4)$ covariant, one needs to define the RI/MOM scheme with two scales, one is $\mu_{\tiny R}=|p_R|$, and the other $p_{\tiny R}^z$. For convenience we simply denote them as $p=p_{\tiny R}$. To work in the perturbative region and control the lattice discretization effects that are of order ${\cal O}\big(a^2\mu_{\tiny R}^2, a^2(p_{\tiny R}^z)^2\big)$, one must work in the window $\Lambda_{\rm QCD}\ll \mu_{\tiny R}\ll a^{-1}$, $p_{\tiny R}^z \ll a^{-1}$, which is attainable if the lattice spacing is small enough.

Since the quarks are off-shell, also finite mixings with the EOM operators can appear. As a result, \eq{rimomvertex} in general cannot be satisfied as a matrix equation. Instead, one usually needs a projection operator ${\cal P}$ to define the off-shell matrix elements, i.e.
\begin{align}
	\langle p| O^B_\Gamma |p\rangle = {\rm tr}\big[\Lambda^{\Gamma}_0(z,a,p) {\cal P}\big]\,,
\end{align}
so as to calculate the renormalization factor $Z_{O_\Gamma}$.

Then, the bare hadron matrix element $\tilde{h}_B(z,P^z,a)$ can be renormalized in coordinate space as
\begin{align} \label{eq:rimomh}
	\tilde{h}_R(z,P^z, p_{\tiny R}^z,\mu_{\tiny R},a)=Z^{-1}_{O}(z,p_{\tiny R}^z,\mu_{\tiny R},a)\tilde{h}_B(z,P^z,a) \,,
\end{align}
In the continuum limit, the renormalized matrix element is independent of the UV regulator, so we should obtain the same result in DR under RI/MOM scheme, i.e.,
\begin{align}\label{eq:continuum}
	&\tilde{h}_R(z,P^z, p_{\tiny R}^z,\mu_{\tiny R})=\lim_{a\to0}\tilde{h}_R(z,P^z, p_{\tiny R}^z,\mu_{\tiny R},a) \nn\\
	&\qquad = \lim_{\epsilon\to0}Z^{-1}_{O}(z,p_{\tiny R}^z,\mu_{\tiny R},\epsilon)\tilde{h}_B(z,P^z,\epsilon)\,,
\end{align}
which allows us to compute the matching coefficients in continuum perturbation theory. Note that $\delta m$ vanishes in $Z_{O}$ due to the use of DR.

By Fourier transforming the above renormalized matrix element to momentum space, one can then work out the RI/MOM matching coefficient for the quasi-PDFs~\cite{Stewart:2017tvs}. The one-loop matching coefficient for different spin structures has been obtained in \cite{Stewart:2017tvs,Liu:2018uuj,Liu:2018hxv}, and the two-loop result for the unpolarized case can be found in~\cite{Chen:2020ody}. Alternatively, one can also first convert the RI/MOM matrix element to the $\MS$ or modified $\MS$ schemes~\cite{Constantinou:2017sej,Alexandrou:2019lfo}, and then do the Fourier transform and momentum-space matching.

\subsubsection{Ratio scheme}
\label{sec:renorm-npr-ratio}

In the coordinate-space factorization, $|z|\ll \Lambda^{-1}_{\rm QCD}$ must be small, whereas
$P^z$ can be of any value. In this case, the ratio scheme in~\cite{Radyushkin:2017cyf,Orginos:2017kos} can be an effective choice for lattice renormalization.
Consider the ratio
\beq \label{eq:ratio}
\tilde{h}(\lambda,z^2,a)/\tilde{h}(0,z^2,a)\,,
\eeq
where the denominator is a nonperturbative matrix element at $P^z=0$. Since $\tilde{h}(\lambda,z^2,a)$ and $\tilde{h}(0,z^2,a)$ calculated from the same lattice ensemble are correlated with each other, the error in the ratio can be reduced.
Besides, the ratio does not need further renormalization on the lattice, so one can directly take the continuum limit
\begin{align} \label{eq:ratio1}
	\lim_{a\to0}\frac{\tilde{h}(\lambda,z^2,a)}{\tilde{h}(0,z^2,a)} = \frac{\tilde{h}(\lambda,z^2)}{\tilde{h}(0,z^2)}\,,
\end{align}
which has referred to as the ``reduced Ioffe-time pseudo'' distribution in~\cite{Radyushkin:2017cyf,Orginos:2017kos}.
In the $\MS$ scheme, $\tilde{h}(0,z^2\mu^2)$ has a small-$z$ expansion,
\begin{align}
	\tilde{h}(0,z^2\mu^2) = C_0(z^2\mu^2) + {\cal O}(z^2M^2, z^2\Lambda_{\rm QCD}^2)\,,
\end{align}
where the lowest moment of the iso-vector quark PDF $a_0$ is trivially one.
If we ignore all the power corrections, then $\tilde{h}(0,z^2\mu^2)$ is perturbative and can be regarded as a renormalization factor. Therefore, the ratio in \eq{ratio1} still satisfies a similar OPE or factorization formula to \eqs{ope}{io-Q-fact}, except that the matching coefficient must be modified correspondingly~\cite{Radyushkin:2017lvu,Izubuchi:2018srq},
\begin{align}
	{\cal C}^{\rm ratio}(\alpha,z^2\mu^2)= {\cal C}(\alpha,z^2\mu^2) - \delta(1-\alpha)C_0(z^2\mu^2)\,.
\end{align}

In other variants of the ratio scheme, it has also been suggested that one replaces $\tilde{h}(0,z^2,a)$ by the vacuum matrix element of the nonlocal Wilson line operator~\cite{Braun:2018brg,Li:2020xml}, as the UV divergence does not depend on the external state.

\subsubsection{Hybrid scheme}
\label{sec:hybrid}

Since the factorization formula for the quasi-PDF is only proven in the $\MS$ scheme, it is not legitimate to use momentum-space factorization for any other scheme that differ from $\MS$ nonperturbatively. The RI/MOM and ratio schemes fall into this category as the conversion factors that match them to $\MS$ includes logarithms of $z^2$~\cite{Constantinou:2017sej,Izubuchi:2018srq}, which requires running $\alpha_s$ to the IR region when $z\sim \Lambda_{\rm QCD}^{-1}$. In contrast, the Wilson-line mass-subtraction scheme with wavefunction renormalizations is essentially the same as $\MS$, so it will not introduce extra IR effects.

However, the Wilson-line mass-subtraction scheme also has disadvantages. On the lattice, due to discretization effects at $z\sim a$, the lattice scheme cannot reproduce the short-distance $\ln z^2$ behavior of the $\MS$ matrix elements of the nonlocal operator. Such discretization effects, however, are cancelled in the RI/MOM or ratio scheme. To take advantages of both types of schemes, the hybrid scheme was proposed in~\cite{Ji:2020brr} which provides a viable approach to renormalize the quasi-LF correlations at all $z$.

To begin with,
for $|z|\le z_{\rm S}$ where $z_{\rm S}$ is smaller than the distance at which the leading-twist approximation in the OPE becomes unreliable, one renormalizes the quasi-LF correlation as
\begin{align}\label{eq:ratio2}
	\frac{\tilde h(z,a,P^z)}{Z_X(z,a)}\,,
\end{align}
where ``$X$'' can be the RI/MOM or ratio scheme.

For $|z|>z_{\rm S}$, one applies the Wilson-line mass subtraction
\begin{align}
	{\tilde h(z,a,P^z)}e^{-\delta m |z|}Z_{\rm hybrid}\,,
\end{align}
where $Z_{\rm hybrid}$ denotes the wavefunction and vertex renormalizations, which can be nonperturbatively determined by imposing a continuity condition at $z=z_{\rm S}$,
\begin{align}
	Z_{\rm hybrid} e^{-\delta m |z_{\rm S}|}{\tilde h(z,a,P^z)}  = \frac{{\tilde h(z,a,P^z)}}{Z_X(z_{\rm S},a) }\,,
\end{align}
leading to
\begin{align}
	Z_{\rm hybrid}(z_{\rm S},a) = e^{\delta m |z_{\rm S}|}/{Z_X(z_{\rm S},a)  }\,.
\end{align}
In this way, one only has to calculate $\delta m$. Note that the final result should be independent of $z_{\rm S}$, so one should try multiple values and find the optimal one around which the result changes the most slightly.

The perturbative matching for the hybrid renormalized quasi-PDF can be derived accordingly. Taking $Z_X$ being the zero-momentum matrix element in the ratio scheme as an example, the $O(\alpha_s)$ matching has been derived as~\cite{Ji:2020brr}
\begin{align}\label{eq:hybridm2}
	&C_{\rm hybrid}(\xi, \mu^2/(p^z)^2, z_{\rm S}^2\mu^2) = C_{\rm ratio}(\xi, \mu^2/(p^z)^2) \nn\\
	&\quad+{\alpha_sC_F\over 2\pi}{3\over 2} \left[-{1\over |1-\xi|_+}+ \frac{2 \text{Si}((1-\xi) \lambda_{\rm S})}{\pi  (1-\xi)} \right]\,,
\end{align}
where $C_{\rm ratio}$ can be found in~\cite{Izubuchi:2018srq}, $\xi=y/x$, and $\lambda_{\rm S}=z_{\rm S}p^z$ with $p^z=xP^z$ being the parton momentum. The plus function is defined as
\begin{align}
	{1\over |1-\xi|_+} &\equiv \lim_{\beta\to0^+}\left[ {\theta(|1-\xi|-\beta)\over |1-\xi|} + 2\delta(1-\xi)\ln\beta\right] \,.
\end{align}

Due to finite lattice volume and deteriorating signal-to-noise ratios at large $z$, the available lattice data have to be truncated at  $z_{\rm L}$.
As we have discussed in \sec{renorm-coord}, the quasi-LF correlation has a correlation length $\xi_z\sim \Lambda_{\rm QCD}^{-1}$ and exhibits an exponential decay at large $z$ ($\sim 1$ fm). If $z_{\rm L}$ is not sufficiently large and the quasi-LF correlation still has a considerable nonzero value, then a direct Fourier transform truncated at $z_{\rm L}$ will lead to unphysical oscillations and other systematics in the quasi-PDF.

To improve this situation, it is suggested in the hybrid scheme to perform an extrapolation to $z\to\infty$~\cite{Ji:2020brr}. When $P^z$ is not very large and the lattice matrix elements exhibit the exponential behavior near $z_{\rm L}$, one can use the form $\sim e^{-z/\xi_z}$ to do the extrapolation, although some algebraic behavior can be added on top to better reflect the $z$-dependence. If $P^z$ is very large, then the signal-to-noise ratio gets worse, so $z_{\rm L}$ is smaller. In this case, the quasi-LF correlation is yet to show exponential decay and dominated by the leading-twist contributions, so one can use the algebraic decay form to do the extrapolation. Since $\lambda_{\rm L} =z_{\rm L}P^z$ can reach reasonably large values with contemporary computing resources, the extrapolation will only affect very small-$x$ region, for which the LaMET expansion is not well under control after all.

To summarize, the hybrid scheme provides a proper renormalization of the quasi-LF correlations at all $z$, which allows for a controlled calculation of the PDF for $x\in[x_{\rm min}, x_{\rm max}]$ through LaMET expansion in momentum space. Therefore,  we expect it to play a dominant role in the LaMET calculation of PDFs in the future.

\subsection{Total Gluon Helicity $\Delta G$ and Transversity PDF}
\label{sec:deltaG}

Apart from the collinear PDFs, the first application of LaMET is the gluon helicity contribution $\Delta G$ to the proton spin~\cite{Ji:2013fga}. In the naive sum rule for the proton spin~\cite{Jaffe:1989jz}, $\Delta G$ is related to the matrix element of a nonlocal light-cone correlation operator~\cite{Manohar:1990jx},
\begin{align}\label{eq:sginv}
	\Delta G &\!=\! \langle PS\big|i\!\int\! \frac{dxd\lambda}{2\pi xP^+} e^{i\lambda x} \! F^{+\alpha}(0) W(0,\!\lambda n)\tilde F_{\alpha}^{~+} (\lambda n)\big|PS\rangle\,,
\end{align}
which in the light-cone gauge $A^+=0$ reduces to
\begin{align}
	\Delta G &= \langle PS| \big(\vec{E}\times \vec{A}\big)^z|PS\rangle /(2P^+)\,.
\end{align}

Within the LaMET framework, one can start from  a static ``gluon spin'' operator, which is defined as $\vec{E}\times \vec{A}$ fixed in a time-independent gauge which maintains
the transverse polarizations of the gluon field in the IMF limit. For example, the Coulomb gauge $\vec{\nabla}\cdot \vec{A}=0$, axial gauges $A^z=0$ and $A^0=0$ are viable options~\cite{Hatta:2013gta}.

In the Coulomb gauge and $\MS$ scheme, the static ``gluon spin'' $\Delta \widetilde G$ in a massive on-shell quark state at one-loop order is~\cite{Chen:2011gn,Ji:2013fga}
\begin{align}  \label{eq:coulomb}
	&\Delta \widetilde G(P^z, \mu)  (2S^z)=\left.\langle PS| \epsilon^{ij}_\perp F^{i0}A^j|PS\rangle_q \right\arrowvert_{\vec{\nabla}\cdot \vec{A}=0}\\
	&\qquad\qquad = \frac{\alpha_sC_F}{4\pi} \left[ {5\over 3}\ln{\mu^2\over m^2} -\frac{1}{9} + \frac{4}{3}\ln \frac{(2P^z)^2}{m^2}\right](2S^z) \,,\nn
\end{align}
where the subscript $q$ denotes a quark, and $S^\mu$ is the spin vector. The collinear divergence is regulated by the finite quark mass $m$.

If we follow the procedure in~\cite{Weinberg:1966jm} and take $P^z\to\infty$ limit before UV regularization~\cite{Ji:2013fga}, then
\begin{align} \label{eq:coulombimf}
	\Delta \widetilde{G}(\infty, \mu) (2S^z)&= \left. \langle PS| \epsilon^{ij}_\perp F^{i0}A^j|PS\rangle_q\right\arrowvert_{\vec{\nabla}\cdot \vec{A}=0}\nn\\
	&=\frac{\alpha_sC_F}{4\pi}\left(3\ln{\mu^2\over m^2}+7\right) (2S^z)\,,
\end{align}
which is exactly the same as the light-cone gluon helicity $\Delta G(\mu)$~\cite{Hoodbhoy:1998bt}. Therefore,
despite the difference in the UV divergence, the collinear divergences of $\Delta \widetilde G(P^z,\mu)$ and $\Delta G(\mu)$ are exactly the same,
which allows for a perturbative matching between them.

The complete factorization formula that relates $\Delta\widetilde G(P^z,\mu)$ to $\Delta G$ and $\Delta \Sigma$ is
\begin{align} \label{eq:deltaGmatching}
	\Delta\widetilde G(P^z,\mu) &= Z_{gg}(P^z/\mu)\Delta G(\mu) \nn\\
	&\qquad\qquad+ Z_{gq}(P^z/\mu)\Delta\Sigma(\mu) +...\,,
\end{align}
where $...$ are power corrections suppressed by $1/P^z$, and the matching coefficients $Z_{gg}$ and $Z_{gg}$ have been calculated for the Coulomb gauge at one-loop~\cite{Ji:2014lra}.

Besides, one can also calculate the gluon helicity PDF $\Delta g(x)$
according to the factorization formula in \sec{renorm}, and then integrate it over $x$ to obtain $\Delta G$.\\

At leading-twist, apart from the unpolarized
and helicity PDFs that we have discussed before, there is also the transversity
PDF defined as~\cite{Jaffe:1991kp,Jaffe:1991ra}
\begin{equation}
	h_1(x)\! =\! \frac{1}{2P^+}\!\int \frac{d\lambda}{2\pi}
	e^{-i\lambda x} \langle PS_\perp|\overline{\psi}(0)\gamma^+\gamma_\perp\gamma_5\psi(\lambda n)|PS_\perp\rangle \,.
\end{equation}
The $h_1(x)$ simply counts the number of transversely polarized quarks carrying
the momentum fraction $x$ in a transversely polarized proton.
The first moment of this distribution corresponds to the
so-called tensor charge $\delta q$, which is the matrix element of a chiral-odd operator. $h_1(x)$ can be accessed through the transverse-transverse spin asymmetry in Drell-Yan processes~\cite{Ralston:1979ys,Jaffe:1991kp,Jaffe:1991ra} or the Collins single-spin asymmetry in SIDIS where the transversity TMDPDF couples to a chiral-odd TMD fragmentation function~\cite{Collins:1992kk}.
At present, experimental results on the transversity PDF are very
limited~\cite{Barone:2001sp,Kang:2015msa,Lin:2017stx,Radici:2018iag,Cammarota:2020qcw}, especially for the sea quark contributions~\cite{Chang:2014jba}, so this is one scenario where lattice
QCD calculation can make an important difference.

The LaMET calculation of $h_1(x)$ is straightforward as the nonlocal operator has the same renormalization as the unpolarized case, and its one-loop matching has been calculated in the $\MS$ and RI/MOM schemes at one-loop order~\cite{Alexandrou:2018eet,Liu:2018hxv}. First lattice calculations of $h_1(x)$ have been done in~\cite{Chen:2016utp,Alexandrou:2018eet,Liu:2018hxv}, which will be discussed with more details in \sec{lattice}.

\section{Generalized Collinear Parton Observables}
\label{sec:gpo}
In the previous section, we have extensively discussed the leading-twist collinear PDFs
that characterize the 1D structure of the proton in longitudinal momentum space.
There exist various other parton observables that provide complementary information.
In this section, we focus on observables defined by collinear parton correlators, in the sense
that only the collinear quark and gluon mode contribute, corresponding to the so-called
{\it collinear expansion} in QCD factorizations~\cite{Sterman:1994ce,Collins:2011zzd}.
We call them ``generalized collinear parton observables'' (GCPOs), and discuss their
calculations through LaMET framework. For observables defined by parton correlators
involving transverse separations, in particular, the TMDPDFs, Wigner functions,
and LFWFs, we will consider them in the following sections.

One of the important GCPOs is the GPDs introduced in~\cite{Mueller:1998fv}, and
rediscovered~\cite{Ji:1996ek} from their connection to the spin structure of the proton.
They describe the correlation between the transverse position and longitudinal momentum of partons inside the proton, and thus provide important information for 3D imaging of the proton. A proton spin sum rule was derived in terms of the moments of the GPDs, which has
stimulated considerable general interest in the GPDs. It was also found that
in the so-called zero skewness limit or when the longitudinal momentum transfer
vanishes, the GPD has a probability interpretation in the impact parameter space~\cite{Burkardt:2000za}.
In general case, it is related to the quantum phase-space distributions or Wigner functions~\cite{Ji:2003ak,Belitsky:2003nz}.
Experimentally, the GPDs can be measured through hard exclusive processes such as deeply virtual Compton scattering (DVCS) or meson production (DVMP) that were first proposed in~\cite{Ji:1996ek,Ji:1996nm}. Much effort has been devoted to measuring such processes at completed and ongoing experiments, including HERA, COMPASS and JLab. For a more comprehensive discussion on the GPDs, we refer the readers to the review articles~\cite{Ji:1998pc,Ji:2004gf,Diehl:2003ny,Belitsky:2005qn}.
Despite that the GPDs have more complicated kinematic dependence and relation to experimental observables,
various fitting methods have been proposed in the literature to fit available DVCS and DVMP data~\cite{Kumericki:2016ehc,Favart:2015umi}. In parallel, one can also extract certain information on the GPDs from lattice calculations of their moments~\cite{Hagler:2007xi,Gockeler:2003jfa,Alexandrou:2019ali}, which, however, is again very limited due to the same difficulties existing in lattice calculations of the PDF moments. For JLab 12 GeV program and future EIC, it
is critically important to have first-principle calculations of GPDs with much better understanding
of the physical landscape in different kinematic variables.

A simpler but closely related GCPO is the parton distribution amplitudes (DAs), which are collinear matrix elements of light-cone operators between a hadron state and the QCD vacuum, representing the probability amplitude of finding a given Fock state in the hadron. They can be probed in certain exclusive processes~\cite{Brodsky:2002st}, and are crucial inputs for processes relevant to measuring fundamental parameters of the Standard Model and probing new physics. There exists a vast amount of literature on this subject, particularly about the pion DA. For a review see e.g. \cite{Brodsky:1989pv,Grozin:2005iz,Braun:2006hn}.

Another type of GCPO is the higher-twist parton distributions.
They are defined by multi-parton correlation functions, and quantify the proton structure
in terms of longitudinal momentum correlations~\cite{Jaffe:1982pm,Ellis:1982cd,Jaffe:1991ra}.
Although physically interesting, they are hard to
separate theoretically due to mixing with the
leading-twist ones~\cite{Mueller:1984vh,Ji:1994md}, and difficult to
extract experimentally because they are power-suppressed~\cite{Ji:1993ey}.
Higher-twist effects can become important in kinematic regions where the
suppression is relaxed. Moreover, some twist-three distributions,
$g_T$ and $h_L$, are different; they have no leading-twist to mix with and are dominant
in spin-related observables~\cite{Jaffe:1991ra}.
Twist-three GPDs are also relevant for studying parton OAM in the
proton~\cite{Hatta:2012cs,Ji:2012ba,Courtoy:2013oaa} and can be
accessed through DVCS process~\cite{Penttinen:2000dg,Kiptily:2002nx}.

In principle, all the GCPOs discussed above can be computed within LaMET.
In addition, an accurate LaMET expansion for the leading-twist PDFs requires
calculations of quasi higher-twist matrix elements.
In the following, we begin with the flavor non-singlet quark GPDs and hadronic DAs for which the computational procedure has been well established, and then give some generic discussions on higher-twist distributions, followed by the discussion on power-suppressed contributions required to extract the leading-twist quark PDFs, which have been investigated using different approaches though not yet implemented in numerical computations.

\subsection{Generalized Parton Distributions}
\label{sec:others-gpd}

The operators defining the GPDs are the same as those defining the PDFs. Thus, the LaMET calculation of PDFs can be rather straightforwardly generalized to the GPDs by taking into account the non-forward kinematics~\cite{Liu:2018tox}. To illustrate how it works, let us take the nonsinglet unpolarized quark GPDs in the nucleon as an example.

\begin{figure}
	\centering
	\includegraphics[width=.9\linewidth]{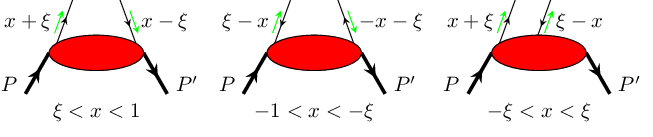}
	\caption{Parton interpretation of the GPDs in different kinematic regions.}
	\label{fig:gpdkin}
\end{figure}

The unpolarized quark GPDs are defined through the following matrix element~\cite{Ji:2004gf}
\begin{align}\label{eq:GPD}
	F&=\frac{1}{2{\bar P}^+}\int \frac{d\lambda}{2\pi}e^{-i x\lambda}\langle P'S'|O_{\gamma^+}(\lambda n)|PS\rangle\nn\\
	&=\frac{1}{2{\bar P}^+}\bar{u}(P'S')\bigg[H\gamma^+
	+E\frac{i\sigma^{+\mu}\Delta_\mu}{2M}\bigg]u(PS)\,,
\end{align}
where we have suppressed the arguments $(x,\xi,t, \mu)$ of $F$, $H$ and $E$ for simplicity. The operator
\begin{align}
	O_{\gamma^+}(\lambda n)=\bar\psi(\frac{\lambda n} 2)\gamma^+ W(\frac{\lambda n}{2},-\frac{\lambda n}{2})\psi(-\frac{\lambda n} 2)
\end{align}
with $n^\mu=1/\sqrt 2(1/{\bar P}^+, 0,0, -1/{\bar P}^+)$ is the same operator used to define the unpolarized quark PDF, $M$ is the nucleon mass.
The momentum fraction $x\in [-1,1]$, and
\begin{align}\label{eq:kinematic_variable}
	\Delta\equiv P'-P,\;\; t\equiv \Delta^2,\;\; \xi\equiv -\frac{P'^+-P^+}{P'^++P^+}=-\frac{\Delta^+}{2{\bar P}^+}\,,
\end{align}
where without loss of generality we have chosen a Lorentz frame in which the average momentum takes the following form
\begin{align}
	{\bar P}^\mu\equiv\frac{P'^\mu+P^\mu}{2}=({\bar P}^0,0,0,{\bar P}^z)\,.
\end{align}
The skewness parameter $\xi\in[-1,1]$ since $P^+, P'^+\ge0$. Besides, there exists another kinematic constraint on $\xi$, which follows from $\vec \Delta^2_\perp\ge 0$,
\beq\label{eq:sklimit}
\xi\le\xi_{\rm max}(t)=\sqrt{\frac{-t}{-t+4M^2}}\,.
\eeq
In the following, we will also assume $\xi>0$ without loss of generality. With these kinematic constraints, the GPDs can be divided into several kinematic regions that have different physical interpretations. As shown in \fig{gpdkin}, in the region $\xi<x<1\, (-1<x<-\xi)$ the distribution describes the emission and reabsorption of a quark (antiquark), while in the region $-\xi<x<\xi$ it represents the creation of a quark and antiquark pair. The first region is similar to that present in usual PDFs and referred to as the DGLAP region, whereas the second is similar to that in a meson DA, which will be discussed later in this section, and referred to as the Efremov-Radyushkin-Brodsky-Lepage (ERBL) region. The easiest way to see this is in light-cone quantization and light-cone gauge where the matrix element defining the GPDs can be rewritten in terms of parton creation and annihilation operators, for details see e.g.~\cite{Ji:2004gf}.

The quark GPDs defined above have a number of remarkable properties, see, e.g.,~\cite{Ji:1998pc,Ji:2004gf,Diehl:2003ny,Belitsky:2005qn}
, which either hold or have similar counterparts for the quark quasi-GPDs to be defined below. Apart from their physical significance, these properties also serve as useful checks on calculations related to GPDs.

According to LaMET, the unpolarized quark GPDs defined above can be determined by calculating the following quasi-GPDs
\begin{align}\label{eq:quasiGPD}
	\tilde F&=\frac{1}{2{\bar P}^0}\int \frac{d\lambda}{2\pi}e^{i y\lambda }\langle P'S'|{O}_{\gamma^0}(z)|PS\rangle\nn\\
	&=\frac{1}{2{\bar P}^0}\bar{u}(P'S')\bigg\{\tilde H\gamma^0
	+\tilde E\frac{i\sigma^{0\mu}\Delta_\mu}{2M}\bigg\}u(PS)\,,
\end{align}
where we have again suppressed the arguments $(y,\tilde{\xi},t,{\bar P}^z, \mu)$ of $\tilde F$, $\tilde H$, and $\tilde E$. The operator ${O}_{\gamma^0}(z)=\bar\psi(\frac{z}{2})\gamma^0 W(\frac z 2,-\frac z 2)\psi(-\frac z 2)$ is the same operator defining the unpolarized quark quasi-PDF, and $\lambda=z {\bar P}^z$.
As in the quasi-PDF case, the momentum fraction $y$ extends from $-\infty$ to $\infty$. The skewness parameter for the quasi-GPD
\begin{align}
	\tilde{\xi}=-\frac{P'^z-P^z}{P'^z+P^z}=-\frac{\Delta^z}{2{\bar P}^z} =  \xi +\mathcal{O}\left({M^2\over ({\bar P}^z)^2},{t\over ({\bar P}^z)^2}\right)
\end{align}
differs from the light-cone skewness $\xi$ by power suppressed corrections. Moreover, the constraint from $\vec\Delta_\perp^2\ge0$ becomes~\cite{Ji:2015qla}
\begin{align}
	{\tilde \xi}\le\frac{1}{2{\bar P}^{z}}\sqrt{\frac{-t\left[({\bar P}^{z})^{2}+M^{2}-t/4\right]}{M^{2}-t/4}},
\end{align}
which differs from the constraint in \eq{sklimit} by corrections of $\mathcal O(M^2/({\bar P}^z)^2, t/({\bar P}^z)^2)$. We can replace $\tilde \xi$ with $\xi$ and attribute the difference to generic power suppressed contributions.

The quasi-GPDs defined above can be renormalized by observing that their UV divergence depends only on the operators defining them, but not on the external states. Since ${O}_{\gamma^0}(z)$ is multiplicatively renormalized, we can choose the same renormalization factor as that for the quasi-PDF~\cite{Stewart:2017tvs,Liu:2018uuj} to renormalize the quasi-GPD.
After renormalization, the quasi-GPD can then be matched to the usual GPD through a factorization formula.

The factorization of quasi-GPDs was first proposed and verified at one-loop order in~\cite{Ji:2015qla,Xiong:2015nua}, where a transverse momentum cutoff and a quark mass were used as the UV and IR regulator, respectively. Later on, a detailed derivation based on OPE was given in~\cite{Liu:2019urm}. In contrast with the OPE for the quasi-PDF, a crucial difference here is that the total derivative of operators can come into play, as it simply gives momentum transfer factors when sandwiched between non-forward external states, and therefore is non-vanishing. In other words, the local twist-two operators as those in Eq.~(\ref{eq:ope-tq}) will mix under renormalization with operators with total derivatives.
The RGE that governs the mixing reads~\cite{Braun:2003rp},
\begin{align} \label{eq:rge}
	&\mu^2 {d\over d\mu^2}O^{{\mu}_0 {\mu}_1 \ldots {\mu}_n}(\mu) = \sum_{m=0}^{[n/2]}\Gamma_{nm}\\
	&\times\left[i\partial^{(\mu_1}\cdots i\partial^{\mu_{2m}} \bar{\psi} \gamma^{\mu_0}i\overleftrightarrow{D}^{\mu_{2m+1}}\cdots i\overleftrightarrow{D}^{\mu_n)}\psi-{\rm trace}\right]\,,\nn
\end{align}
where $\Gamma_{nm}$ is the anomalous dimension of the associated operators, $\overleftrightarrow{D}=(\overrightarrow{D}-\overleftarrow{D})/2$ with $\overrightarrow{D} (\overleftarrow{D})$ denoting the covariant derivative acting to the right (left). The above equation can be diagonalized by choosing an appropriate operator basis. Such an operator basis has been studied in the literature and known as the ``renormalization group improved'' conformal operators~\cite{Braun:2003rp,Mueller:1993hg}. In terms of the matrix elements of these operators, we have
\begin{align}\label{eq:ope-twist2}
	&\langle P'|{O}_{\gamma^0}(z)|P\rangle  =  2P^0\sum_{n=0}^\infty C_n ({\mu}^2 z^2){\cal F}_n(-\lambda)\sum_{m=0}^n{\cal B}_{nm}(\mu)\nn\\
	&\hspace{3em}\times\xi^n\int_{-1}^1 dx\ C_m^{3/2}\left({x\over \xi}\right) F(x,\xi,t,\mu)+\ldots\,,
\end{align}
where ${\cal F}_n(-\lambda)$
are partial wave polynomials whose explicit forms are known in the conformal OPE of current-current correlators for the hadronic light-cone DAs~\cite{Braun:2007wv}, ${\cal B}_{nm}$ can be found in~\cite{Braun:2003rp,Mueller:1993hg}, and $\ldots$ denotes the higher-twist contributions $\mathcal{O}\left({M^2/ ({\bar P}^z)^2},{t/ ({\bar P}^z)^2},z^2\Lambda_{\rm QCD}^2\right)$.

Fourier transforming the l.h.s of the above equation to momentum space and invert it order by order in $\alpha_s$, we then obtain the following EFT expansion of the unpolarized quark GPD,
\begin{align}\label{eq:gpdfact}
	&F(x,\xi,t,\mu) \\
	&\hspace{0.5em}=  \int_{-\infty}^\infty {dy\over |\xi|}\bar{C}\left({x\over \xi}, {y\over \xi}, {\mu \over \xi {\bar P}^z }\right){\tilde F}(y,\xi,t, {\bar P}^z,\mu) +\ldots\nn\\
	&\hspace{0.5em}= \int_{-\infty}^\infty {dy\over |y|}C\left({x\over y}, {\xi\over y}, {\mu \over y{\bar P}^z }\right){\tilde F}(y,\xi,t,{\bar P}^z,\mu)+ \ldots\,,\nn
\end{align}
which has been organized following the same spirit as the factorization of PDFs in previous sections. Both forms have been used in the literature~\cite{Ji:2015qla,Xiong:2015nua,Liu:2019urm} with the matching coefficients being related by
\beq
C\left({x\over y}, {\xi\over y}, {\mu \over y{\bar P}^z }\right) = \left|{y\over \xi}\right|\bar{C}\left({x\over \xi}, {y\over \xi}, {\mu \over \xi {\bar P}^z }\right)\,,
\eeq
and $\ldots$ denotes the higher-twist contributions which have the same power-counting as in Eq.~(\ref{eq:ope-twist2}) except that $z^2$ is replaced by $1/(x{\bar P}^z)^2$. For the helicity and transversity quark quasi-GPDs, the factorization formula has the same form as \eq{gpdfact}~\cite{Liu:2019urm}.

The matching coefficient can be obtained by replacing the hadron states in Eqs. (\ref{eq:GPD}) and (\ref{eq:quasiGPD}) with the quark states carrying momentum $p+\Delta/2$ and $p-\Delta/2$ with $p^\mu=(p^0,0,0,p^z)$, and calculating the quark matrix element in perturbation theory.
The explicit expression for the $\mathcal O(\alpha_s)$ matching coefficients can be found in~\cite{Liu:2019urm}.
An important feature of the result is: The quasi-GPDs do not vanish in all $y$ range, but the collinear singularities only show up in DGLAP and ERBL regions at one-loop. They are exactly the same as those in light-cone GPDs, and thus cancel in the matching coefficient. Moreover, one can derive momentum RGEs for the quasi-GPDs,
which are turned into RGE for the scale dependence of the GPDs by
the matching procedure.

To conclude this subsection, let us make some remarks on the EFT formula for the quark GPD above. First, at zero skewness $\xi=0$, we have
\begin{align} \label{eq:zeroskew}
	F(x,0,t,\mu)&=\!\! \int_{-\infty}^\infty\! {dy\over |y|}C\!\left({x\over y}, 0, {\mu \over y{P}^z }\right)\! {\tilde F}(y,0,t,P^z,\mu)\nn\\
	&\qquad +\ldots\,,
\end{align}
where the matching kernel $C(x/y,0,\mu/(yP^z))$ is exactly the same as the matching coefficient for the quasi-PDF~\cite{Izubuchi:2018srq}, even when $t\neq0$. This can be understood as follows: At zero skewness, both the longitudinal momentum transfer and the energy transfer vanish, the momentum transfer is purely transverse and thus is not affected by Lorentz boost along the longitudinal $z$ direction. As a result, no extra matching related to $t$ is required in the large $P^z$ limit, and the matching remains the same as in the quasi-PDF case. If we take the forward limit $\Delta \to0$, then Eq.~(\ref{eq:zeroskew}) reduces exactly to the EFT expansion formula for the PDF~\cite{Ji:2015qla,Izubuchi:2018srq}.

Second, in the limit $\xi\to1$ and $t\to0$, the quasi-GPD reduces to the quasi-DA that will be discussed in the next subsection, and the corresponding matching kernel also reduces to that for the quasi-DA.

\subsection{Hadronic Distribution Amplitudes}
\label{sec:others-da}

Within LaMET, the DAs of protons as well as other hadrons can also be extracted from lattice simulations of appropriately chosen quasi-DAs. In this subsection, we show how this can be done in practice. For illustration, we take the leading-twist pion DA as an example. The application to other hadrons~\cite{Chen:2017gck,Wang:2019msf} is analogous.

The leading-twist DA of the pion is the simplest and most extensively studied hadronic DA. It represents the probability amplitude of finding the valence $q\bar q$ Fock state in the pion with the quark carrying a fraction $x$ of the total pion momentum, and is defined as
\begin{align}\label{eq:pionDA}
	\phi_{\pi}(x)&={\frac{1}{if_{\pi}}}\int \frac{d\lambda}{2\pi P^+}e^{-i (x-\frac 1 2)\lambda}\langle 0|O_{\gamma^+\gamma_5}(\lambda n)|\pi(P)\rangle\,,
\end{align}
with normalization $\int_0^1 dx\, \phi_\pi(x)=1$. Here $f_\pi$ denotes the decay constant, and
$O_{\gamma^+\gamma_5}(\lambda n)$ has the same structure as that used in Eq.~(\ref{eq:GPD}) with $\gamma^+$ replaced by $\gamma^+\gamma_5$. The pion DA can be constrained from experimental measurements of, e.g.,
$\gamma \gamma* \rightarrow \pi^0$ from BaBar and Belle~\cite{Aubert:2009mc,Uehara:2012ag},
and then used as an input to test QCD in other measurements such as the pion form factor~\cite{Efremov:1978rn,Farrar:1979aw}.
In the asymptotic limit, it is well known that the pion DA takes the form
$6 x(1-x)$~\cite{Efremov:1978rn,Lepage:1979zb}.
However, how it behaves at lower scales remains under debate (see e.g.~\cite{Chernyak:1981zz}).
Calculating the pion DA with controllable systematics in LaMET will be able to shed new lights on its shape and thus on our understanding of pion structure.

Following the same strategy as before, we can access the $x$-dependence of the pion DA by studying the following quasi-DA~\cite{Ji:2015qla,Zhang:2017bzy}
\begin{align}\label{eq:quasiDA}
	\tilde\phi_{\pi}(y, P^z)&={\frac{1}{if_{\pi}}}\int \frac{d\lambda}{2\pi P^z}e^{i (y-\frac 1 2) \lambda}\langle 0|{O}_{\gamma^z\gamma_5}(z)|\pi(P)\rangle\,,
\end{align}
The longitudinally and transversely polarized vector meson quasi-DAs can be defined analogously by replacing $\gamma^z\gamma_5$ in the above equation with $\gamma^0$, $\gamma^z\gamma_\perp$, respectively~\cite{Liu:2018tox}.

The quark bilinear operators defining quasi-DAs follow the same renormalization pattern as those defining the quasi-PDFs or quasi-GPDs.
In the literature, the Wilson-line mass-subtraction scheme was used in the first LaMET calculations of the meson DAs~\cite{Zhang:2017bzy,Chen:2017gck}, and the RI/MOM scheme has been adopted in more recent works~\cite{Zhang:2020gaj}.

The LaMET expression for DAs takes the following form in the $\MS$ scheme~\cite{Ji:2015qla,Liu:2018tox}
\begin{align}\label{eq:matching2}
	\phi_\pi(x,\mu) &=\int_{-\infty}^\infty dy\, C_{\pi}\left(x,y,P^z/\mu\right) \tilde\phi_\pi(y,P^z,\mu)+\ldots\,.
\end{align}
The matching coefficient  for the quasi-DAs can be obtained by replacing the meson state $| \pi(P)\rangle$ in Eqs. (\ref{eq:pionDA}) and (\ref{eq:quasiDA}) with the lowest Fock state $| q(y P) \bar q((1-y)P)\rangle$ and calculating the quark matrix elements, where $y P$ and $(1-y)P$ are the momenta of the quark $q$ and anti-quark $\bar q$, respectively. Its one-loop results have been calculated in both $\MS$ and RI/MOM schemes~\cite{Liu:2018tox}, which agrees with matching coefficient for the quasi-GPDs~\cite{Ji:2015qla,Xiong:2015nua,Liu:2019urm} in \eq{gpdfact} with the replacement of $\xi\to 1/(2y-1)$, $x/\xi\to 2x-1$, and the external momentum $p^z$ to $p^z/2$.

Apart from the LaMET approach in momentum space,
the shape of the pion DA has also been studied using equal-time current-current correlation in coordinate space approach~\cite{Bali:2017gfr,Bali:2018spj},
\begin{align}
	&\langle 0 | T\Big\{J_\mu\Big(\frac{z}{2}\Big) J_\nu\Big(-\frac{z}{2}\Big)\Big\}| \pi^0(P)\rangle \nn\\
	& \hspace{3em}= \frac{2i\, f_\pi}{3 \pi^2 z^4} \epsilon_{\mu\nu\alpha\beta} P^\alpha z^\beta \Phi_\pi(\lambda, z^2)\,,
	\label{eq:pigg-in-z-space}
\end{align}
where $\Phi_\pi(\lambda, z^2)$ can be factorized as
\begin{align} \label{twist_decomposition_heuristic}
	\Phi_\pi(\lambda, z^2) &= C_2 (\lambda,z^2\mu^2, x)\otimes \phi_\pi(x,\mu)+\cdots\,.
\end{align}
Here the matching coefficient $C_2$ depends  on the choice of the currents. Its explicit expression can be found in~\cite{Bali:2018spj}.
The above factorization is controlled by ${\mathcal O}(z^2\Lambda_{\rm QCD}^2)$, with power corrections denoted by ``$\cdots$''. In~\cite{Bali:2018spj}, a combined analysis of several current-current correlations has been performed where twist-four contributions were also included using the model estimate in~\cite{Braun:1989iv,Ball:2006wn}. {The leading-twist pion DA was then extracted from a global fit to the data, and the second moment of the pion DA has been fitted with controlled precision}, both of which favor a considerably broader shape than the asymptotic DA at a scale of $2$ GeV.
A large pion momentum is required to access information at large $\lambda$ so that we can extract wider range of $x$ or higher moments of the pion DA~\cite{Bali:2017gfr}.

\subsection{Higher-Twist Distributions}
\label{sec:others-ht}

Higher-twist distributions are quantities of great interest because they describe the coherent quark-gluon correlations in
the proton. In contrast with the leading-twist distributions, our understanding of the higher-twist ones is rather poor.
On one hand, they often depend on more than one parton momentum fractions; on the other hand, there is no physical intuition about what they may look like, in particular, about how they behave asymptotically at small and large $x$~\cite{Braun:2011aw}. There have been studies on the higher-twist distributions in the context of their connection to the DIS structure function, the transverse single-spin asymmetries in various hadron productions, GPDs related to quark and gluon OAM, parton DAs, etc. LaMET will be able to shed new lights by providing a possibility to access them from the lattice.

Higher-twist contributions also appear in LaMET expansion, where
the suppression is provided by powers of the hadron momentum squared.
In all factorizations presented in previous sections, only the leading-twist terms that capture the logarithmic dependence on hadron momentum are taken into account. The higher-twist contributions have been assumed to be small. If the hadron momentum is not sufficiently large compared and/or one is close to the endpoint region ($x\to0$ and $x\to1$), the higher-twist contributions can become non-negligible, whose structure and impact require understanding.

\subsubsection{Higher-twist collinear-parton observables}

Beyond leading-twist, there exist three simplest twist-three quark distributions
$e(x)$, $g_T(x)$ and $h_L(x)$ related to the unpolarized, transversely and longitudinally polarized
proton~\cite{Jaffe:1991ra},
\begin{align}
	e(x) &= \frac{1}{2M}\int \frac{d\lambda}{2\pi}e^{ix\lambda} \\
	&\quad\times \langle PS|\psi^\dagger_+(0)\gamma_0\psi_-(\lambda n)|PS\rangle + {\rm h.c.}  \,, \nn\\
	g_T(x)& = \frac{1}{2M}\int \frac{d\lambda}{2\pi}e^{ix\lambda}  \\
	&\quad\times \langle PS_\perp|\psi^\dagger_+(0)\gamma_0\gamma_\perp\gamma_5\psi_-(\lambda n)|PS_\perp\rangle + {\rm h.c.} \,,\nn\\
	h_L(x)&  =\frac{1}{2M}\int \frac{d\lambda}{2\pi} e^{ix\lambda} \\
	&\quad\times \langle PS_z|\psi^\dagger_+(0)\gamma_0\gamma_5\psi_-(\lambda n)|PS_z\rangle + {\rm h.c.}  \,,\nn
	\label{eq:twist3}
\end{align}
where we have again employed the decomposition of quark fields $\psi=\psi_+ +\psi_-$ in Sec.~\ref{sec:lc} and the light-cone gauge $A^+=0$, and ``h.c.'' stands for Hermitian conjugate.

The twist-three distributions can contribute as leading effects in certain experimental observables. For example, $g_T(x)$ and $h_L(x)$ can be measured as the leading effects in the longitudinal-transverse spin asymmetry in polarized Drell-Yan process.

Since $\psi_-$ is a non-dynamical component depending on $\psi_+$, all the above distributions can be shown to be related to more complicated quark-gluon correlation functions~\cite{Ji:1990br,Balitsky:1996uh}. A complete set of such correlation functions has been given in~\cite{Qiu:1991pp,Ji:1992eu,Ji:2001bm,Kang:2008ey}, where the quark-gluon correlations
in a transversely-polarized proton take the following form
\begin{align}
	&T_q(x_1, x_2)=\frac{1}{(P^+)^2}\int\frac{d\lambda d\zeta}{(2\pi)^2}e^{i\lambda x_1+i\zeta(x_2-x_1)}\\
	&\hspace{2em}\times\langle PS_\perp|\bar\psi(0)\gamma^+ \epsilon^{+-S_\perp i}g F^{+i}(\zeta n)\psi(\lambda n)|PS_\perp\rangle,\nn\\
	&T_{\Delta q}(x_1, x_2)=\frac{1}{(P^+)^2}\int\frac{d\lambda d\zeta}{(2\pi)^2}e^{i\lambda x_1+i\zeta(x_2-x_1)}\\
	&\hspace{2em}\times\langle PS_\perp|\bar\psi(0)i\gamma^+\gamma_5 S^i_\perp g F^{+i}(\zeta n)\psi(\lambda n)|PS_\perp\rangle\,.\nn
\end{align}
There are also ones in an unpolarized and longitudinally-polarized proton.
Generalizing to off-forward kinematics, the resulting twist-three GPDs are also related to quark
and gluon OAM contribution to the proton spin~\cite{Hatta:2012cs,Ji:2012ba}.

One can also define twist-four distributions in a similar way as in Eq. (\ref{eq:twist3})
by using $\psi_-$ for both quark fields. More general twist-four distributions
will involve three light-cone variables, which will contribute to, e.g.,
$1/Q^2$ term in DIS~\cite{Jaffe:1982pm,Ellis:1982cd,Jaffe:1991ra,Ji:1993ey}.

In principle, all the above higher-twist distributions, as well as others that have not been listed here, can be
computed using the LaMET approach by choosing appropriate quasi-LF correlations. For example, the first exploratory lattice calculation of $g_T(x)$ has been done in~\cite{Bhattacharya:2020cen}, which will be discussed in \sec{lattice-other_distributions}. However, extra complications are expected due to their complex structure. For instance, the lightcone zero modes that do not enter in dealing with leading-twist distributions come into play here. Recently, one of the authors has shown how to study the properties of these zero modes from lattice simulations in LaMET~\cite{Ji:2020baz}. In addition, the higher-twist distributions will have a more complex mixing pattern~\cite{Ji:1990br,Balitsky:1996uh}. Thus,
their matching from the corresponding quasi distributions must take into account such
mixings, making them more challenging than calculating the twist-two PDFs.  One-loop studies of the matching for twist-three disributions have been carried out in~\cite{Bhattacharya:2020xlt,Bhattacharya:2020jfj}.

\subsubsection{Higher-twist contributions to quasi-PDFs}

Let us turn to the power suppressed higher-twist contributions appearing in the extraction of leading-twist quark PDFs using LaMET. Such contributions have two distinct origins. To understand them, let us recall the OPE for the quasi-LF correlation in \eq{ope}. For simplicity, we ignore the renormalization here.
Recovering the leading-twist quark PDF requires removing the contributions of both trace terms in that equation.
The trace terms on the r.h.s. of \eq{ope}, which lead to contributions suppressed by powers of $M^2/(P^z)^2$, are known as kinematic power contributions or target mass corrections. In DIS, they can be accounted for by changing the scaling variable $x$ to the Nachtmann variable~\cite{Nachtmann:1973mr}. In the case of LaMET, it behaves slightly differently, as shown in the following. The second type of power corrections come from the trace terms in the operators on the r.h.s. of \eq{ope-tq}, and in general leads to contributions of $\mathcal O(\Lambda^2_{\rm{QCD}}/(P^z)^2)$ . These are genuine higher-twist contributions that involve multi-parton correlations, sometimes also known as dynamical higher-twist contributions. The target mass corrections have been computed to all orders in $M^2/(P^z)^2$ for the quark quasi-PDFs in~\cite{Chen:2016utp,Radyushkin:2017ffo}. The genuine higher-twist contributions have been investigated using two different approaches~\cite{Chen:2016utp,Braun:2018brg}.

According to~\cite{Chen:2016utp}, the $M^2/(P^z)^2$ corrections can be obtained from the ratio
\begin{align}
	K_m  &\equiv
	\frac{n_{(\mu_1}\cdots n_{\mu_m)}
		P^{\mu_1}\cdots P^{\mu_m}}{n_{\mu_1}\cdots n_{\mu_m}
		P^{\mu_1}\cdots P^{\mu_m}} = \sum_{i=0}^{i_\text{max}} C_{m-i}^i c^i   \label{Kn}\,,
\end{align}
where $i_\text{max}=(m-\text{Mod}[m,2])/2$,
$C$ is the binomial function and
$c=-n^2 M^2/4\left(n\cdot P\right)^2 = M^2/4 (P^z)^2$
with $n^\mu = (0,0,0,-1)$ and $n \cdot P=P^z$.

Plugged into the tree-level OPE formula in \eq{ope}, the above factors can then be converted to the following relation between unpolarized PDF and quasi-PDF~\cite{Chen:2016utp}
\begin{align}
	q(x)&=\sqrt{1+4c}\sum_{n=0}^\infty \frac{(4c)^n}{f_+^{2n+1}}\Big[(1+(-1)^n)\tilde q\Big(\frac{f_+^{2n+1}x}{2(4c)^n}\Big)\nn\\
	&\qquad+(1-(-1)^n)\tilde q\Big(\frac{-f_+^{2n+1}x}{2(4c)^n}\Big)\Big],
\end{align}
where $f_{+}=\sqrt{1+4c}+ 1$. It is worth noting that quark number conservation is preserved in the above result. The target mass corrections for the longitudinally and transversely polarized quasi-PDFs can be derived analogously.

The trace part on the r.h.s. of \eq{ope-tq} is a genuine higher-twist effect. One may try to construct a non-local form of the higher-twist operators from OPE. The leading trace term, which is a twist-four effect, has been studied in~\cite{Chen:2016utp} (see also~\cite{Balitsky:1987bk}) and shown to give rise to a twist-four PDF
\begin{align}\label{tildeq-twist4}
	&{q}_\text{4}(x,P^z)
	=  \int_{-\infty}^\infty \!\frac{d\lambda}{8\pi P^z}\,
	\Gamma_0\left(-ix\lambda\right)
	\left\langle P\left\vert  O_\text{tr}(z)\right\vert P\right\rangle ,
\end{align}
with
\begin{align}
	&O_\text{tr}(z) =
	\int_0^z \!dz_1\, \bar{\psi}(0) \Big[
	\Gamma^\nu W\left(0,z_1\right) D_\nu W\left(z_1,z\right)\\
	&\hspace{-.5em}+ \int_0^{z_1} \!\! \! dz_2\,  n \cdot  \Gamma\,
	W\left(0,z_2\right) D^\nu W\left(z_2,z_1\right) D_\nu
	W\left(z_1,z\right) \Big] \psi(z n)\,,\nn
\end{align}
where one has $\Gamma^{\mu}=\gamma^{\mu}, \gamma^{\mu}\gamma^5, \gamma^\perp\gamma^{\mu}\gamma^5$ for the unpolarized, helicity and transversity PDFs, respectively. $\Gamma_0$ is the incomplete Gamma function
\begin{equation}
	\Gamma_0\left(-ix\right) = \int_0^1 \frac{dt}{t} e^{ix/t} \,.
\end{equation}
The above twist-four contribution needs to be removed from the quasi-PDF to recover the leading-twist PDF. It also provides a possibility for practical computations on the lattice. However, as a multi-parton correlation involving more gauge links and covariant derivatives, its lattice computation is rather challenging and has not been carried out in any existing work yet.

Another approach that has been used to estimate power corrections related to quark quasi-PDFs is the
renormalon model (see~\cite{Beneke:1998ui} for a comprehensive review).
It is based on the observation that the perturbative expansion of the matching coefficient for the quasi-PDF diverges factorially with the loop order, implying that it is only well defined up to a power accuracy. This is known as the renormalon ambiguity, which must be cancelled by terms in the higher-twist contributions.

In~\cite{Braun:2018brg}, it was shown that the cancellation of renormalon ambiguity requires that the leading higher-twist or twist-four contribution takes the following form
\begin{align}
	\label{eq:Q4-cutoff}
	{q}_4(y,P^z,\mu) = \mu^2 \int_{-1}^1 \!\frac{dx}{|x|} D\Big(\frac{y}{x}\Big) q(x,\mu) + q'_4(y,P^z,\mu)\,,
\end{align}
where the first term on the r.h.s. cancels the renormalon ambiguity from the leading-twist matching coefficient, and $q'_4$ depends on $\mu$ at most logarithmically. Since the first term is to merely cancel similar contributions in the matching coefficient, it does not contribute to any physical observable. The renormalon model of power corrections \cite{Beneke:1995pq,Dokshitzer:1995qm,Dasgupta:1996hh,Dasgupta:1996ki,Beneke:1997sr,Braun:2004bu} is based on the assumption that, by replacing $\mu$ with a suitable nonperturbative scale, this contribution reflects the order and the functional form of actual power-suppressed contributions. This was known as ``ultraviolet dominance'' in~\cite{Braun:1995tp,Beneke:1998ui,Beneke:2000kc}. Under this assumption, we obtain the following estimate,
\begin{equation}
	{q}_4(y,P^z,\mu) = \kappa
	\Lambda_{\rm QCD}^2 \int_{-1}^1 \!\frac{dx}{|x|} D\Big(\frac{y}{x}\Big)\, q(x,\mu)\,,
	\label{UV_Lambda_int}
\end{equation}
where $\kappa$ is a dimensionless coefficient of $\mathcal O(1)$ that cannot be fixed within theory and remains a free parameter.

A detailed analysis~\cite{Braun:2018brg} showed that for the quasi-PDF we have
\begin{align}
	\label{t4-Qparallel}
	& {q}_4(y,P^z) =\frac{\kappa  \Lambda_{\rm QCD}^2}{y^2(1-y) (P^z)^2}\\
	&\hspace{-.2em}\times \hspace{-.2em}(1-y)\!\Big[\!\!
	\int_{|y|}^1 \!\!\frac{dx}{x} \Big[\frac{x^2}{(1-x)_+}\!-\!2x^2\Big]q\Big(\frac{y}{x}\Big)\!
	+\! 2q(y)\! -\! |y|q'(y) \Big]\,,\nn
\end{align}
where the first term in the integral was reproduced in a recent analysis of the renormalon effects in the quasi-PDF~\cite{Liu:2020rqi}. As one can see, the second row vanishes as $q(y)$ when $y\to1$ if $\lim_{y\to1}q(y)\sim (1-y)^a$ with $a>0$.
This gives an estimate of the twist-four contribution on the r.h.s. of Eq.~(\ref{eq:fact_0}), which  implies that the higher-twist contributions are enhanced as $1/y^2$ and $1/(1-y)$ for $y\sim 0$ and $y\sim 1$, respectively. Similar analysis can also be done for the pseudo-PDF.
The above result can be used as a way to model the twist-four contribution with $\kappa$ as the only parameter.

\subsection{Orbital Angular Momemntum of Partons in the Proton}
\label{sec:spin}

Over the past three decades, much experimental and theoretical
work has been done on the origin and structure of proton spin, which has been covered in depth in the review articles
~\cite{Filippone:2001ux,Bass:2004xa,Aidala:2012mv,Leader:2013jra,Ji:2016djn,Deur:2018roz,Ji:2020ena}.

In addition to the spin-dependent PDFs and TMDs, the GCPOs---in particular the GPDs---also play an important role in understanding the spin structure of the proton. Since GPDs describe the correlation between the transverse position and longitudinal momentum of quarks and gluons inside the proton, they offer a unique channel to study the orbital angular momentum (OAM) from experiments.

There are two widely known definitions of OAM in literature. One is the kinetic OAM in the gauge-invariant and frame-independent  sum rule for the proton spin~\cite{Ji:1996ek,Ji:1996nm}, which is related to the first moment of twist-two GPDs and can be calculated from the form factors of the symmetric QCD energy-momentum tensor. A review of the lattice calculations of kinetic OAM can be found in~\cite{Ji:2020ena}. The other definition, which has a clear partonic interpretation in comparison to the kinetic OAM, is the canonical OAM in the naive partonic sum rule~\cite{Jaffe:1989jz} based on the free-field form of the QCD angular momentum,
\begin{align}\label{eq:jmam}
	\vec{J} &= \int d^3\xi\ \psi^\dagger \frac{\vec{\Sigma}}{2} \psi + \int d^3\xi\ \psi^\dagger\left[\vec{\xi}\times (-i\vec{\nabla} )\right]\psi \nn\\
	&\ \ \ \ \ + \int d^3\xi\ \vec{E}\times\vec{A} + \int d^3\xi\ E^i\ \left(\vec{\xi}\times\vec{\nabla}\right) A^{i} \ ,
\end{align}
where $i$ is the spatial Lorentz index. Except for the first one,
the other three operators are gauge dependent, and their matrix elements are
generally frame dependent. In high-energy scattering, there is one frame and
gauge that are special: the IMF and light-front gauge, $A^+=0$.
Therefore, the naive partonic sum rule for proton spin can be expressed as~\cite{Jaffe:1989jz}
\beq\label{eq:jm}
\frac{1}{2}= \frac{1}{2}\Delta \Sigma(\mu)  + l^z_q(\mu) + \Delta G(\mu) + l^z_g(\mu) \,,
\eeq
where $l^z_q(\mu)$ and $l^z_g(\mu)$ are the canonical
OAM of the quark and gluon partons, respectively.
Both $l_q^z$ and $l_g^z$ can be related to twist-three GPDs~\cite{Hatta:2011ku,Ji:2012ba,Hatta:2012cs}, which can be accessed through spin-asymmetries in hard exclusive processes~\cite{Ji:2016jgn,Hatta:2016aoc,Bhattacharya:2017bvs,Bhattacharya:2018lgm} (see the recent review~\cite{Ji:2020ena}).

To fully understand the partonic spin structure of the proton,
one also needs to determine the quark and gluon canonical OAM, $l^z_q$ and $l^z_g$.
LaMET allows extraction of $l_q^z$ and $l_g^z$ from lattice calculation in the same way as the gluon helicity that was reviewed in \sec{deltaG}.

The quasi-partonic OAM operators can be chosen as the free-field operators fixed in gauges that
belong to the universality class~\cite{Hatta:2013gta}.
Their matrix elements $\tilde l^z_q$ and $\tilde l^z_g$ can be calculated from the off-forward matrix elements of the relevant energy-momentum tensors~\cite{Zhao:2015kca}, for example,
\begin{align}
	\tilde l^z_q(2S^z) &= \lim_{\Delta\to0} \epsilon^{ij}{\partial \over \partial i\Delta^i}\langle P'S| \psi^\dagger (0) i\partial^j \psi(0)|PS\rangle \Big|_{\vec{\nabla}\cdot \vec{A}=0}\,.
\end{align}
where the kinematics is the same as \eq{kinematic_variable}.

Along with $\Delta G$,
$\tilde l^z_q$ and $\tilde l^z_g$ can be matched to the partonic
quantities defined in the Jaffe-Manohar sum rule through the factorization formulas,
\begin{align}\label{eq:oammatching}
	\tilde l^z_q(P^z,\mu)  &= P_{qq}  l^z_q(\mu) + P_{gq} l^z_g(\mu) \nn\\
	&\qquad+ p_{qq} \Delta \Sigma(\mu) + p_{gq}\Delta G(\mu) + ...\, ,\\
	\tilde l^z_g(P^z,\mu)  &= P_{qg}  l^z_q(\mu) + P_{gg} l^z_g(\mu) \nn\\
	&\qquad+ p_{qg} \Delta \Sigma(\mu) + p_{gg}\Delta G(\mu) + ...\, ,
\end{align}
where $\cdots$ are power corrections suppressed by the momentum $P^z$, and the one-loop matching coefficients in front of each term on the r.h.s. have been calculated in the Coulomb gauge~\cite{Ji:2014lra}. Since the quasi-partonic operators are gauge-variant and need to be fixed in a particular gauge, they can mix with new operators that are not allowed by
Lorentz or gauge symmetries. For example,
the gauge-dependent potential angular momentum $\psi^\dagger(\vec{r}\times \vec{A})\psi$
comes into play~\cite{Wakamatsu:2014zza,Ji:2015sio}. Such mixings must be taken into account in lattice renormalization  to have a controlled calculation of the canonical OAM.

Apart from the above approach, it has also been proposed to calculate the ratio of $l_q^z$ and the valence quark number from the derivatives of off-forward matrix elements of staple-shaped quark Wilson line operators~\cite{Engelhardt:2017miy}, whose definition can be found in \eq{quasi_TMD} below. The first lattice calculations with this approach have been carried out in~\cite{Engelhardt:2017miy,Engelhardt:2019lyy}, which shows different size of effects between the kinetic and canonical OAM. {For systematic improvement in this calculation, one should include the matching of such matrix elements to the physical $l_q^z$ in the limit when the transverse separation of the quark fields approaches zero.}

For the transverse polarization, it is natural to define a twist-two
partonic transverse angular momentum density of quarks~\cite{Hoodbhoy:1998yb,Ji:2012sj,Ji:2020hii,Ji:2020ena},
\begin{equation}
	J^{q}_\perp (x) = x\left[q(x)+ E_q(x)\right]/2,
\end{equation}
and similarly for the gluons, where $q(x)$ is the unpolarized quark/antiquark distributions,
and $E_{q,g}(x)$ is the GPDs defined earlier in this section.
Thus to get a simple partonic picture of the
proton transverse spin from the first principles,
it is important to calculate the GPD $E(x)$ using LaMET.

\section{Transverse-Momentum Dependent PDFs}
\label{sec:tmd}
The transverse-momentum-dependent (TMD) parton distribution functions (TMDPDFs) are a natural generalization
of the collinear PDFs to include both longitudinal and transverse momentum of partons.
They are in principle probability distributions $f_i(x,\vec{k}_\perp,\sigma)$ of finding a parton of given species $i$,
longitudinal and transverse momentum $(xP^+,\vec{k}_{\perp})$, and polarization $\sigma$
inside the hadron state. TMDPDFs are playing an increasingly important role in
understanding the partonic structure of hadrons and high-energy scattering.

The TMD parton densities were firstly introduced by Collins and Soper in 1980s~\cite{Collins:1981uk,Collins:1981va,Collins:1982wa,Collins:1984kg,Collins:1985ue,Bodwin:1984hc} to understand the Drell-Yan (DY) and $e^{+}e^{-}$ annihilation process, and generalized in~\cite{Ji:2004wu,Ji:2004xq} to semi-inclusive deep-inelastic scattering(SIDIS) process. The TMD factorization has been reanalyzed in the framework of SCET in which modes are made manifest by effective fields ~\cite{Bauer:2000yr,Bauer:2001yt,Manohar:2006nz,Becher:2010tm,GarciaEchevarria:2011rb,Echevarria:2012js,Chiu:2012ir}. Various TMD factorization formalisms finally converged to the standard one where a scheme-independent TMDPDF can be defined~\cite{Echevarria:2012js,Collins:2012uy,Collins:2017oxh}.

The TMD parton densities are important in understanding the experimental processes where the transverse momenta of final state particles are measured. For example, in DY pair and $W, Z$ production it is known that the differential cross section $d\sigma/dQ_T^2$ normally peaks at relatively small transverse momentum.  For $Q\sim 10 $ GeV, the peak is typically located at $Q_\perp\sim 1 $ GeV where nonperturbative effects are important~\cite{Collins:1984kg}. A good knowledge of TMD parton densities is therefore crucial for the determination of the
cross sections and precision test of perturbative QCD predictions.

Besides their importance in understanding the high-energy experimental data, the TMD parton densities are also important by themselves for their crucial role in describing hadron structures. With them, one can simultaneously study the fast-moving collinear physics through the longitudinal $x$-dependencies, and the nonperturbative effect from the transverse $\vec{k}_\perp$-dependencies. Moreover, the TMDPDFs are sensitive to effects such as soft radiations. Therefore, the physics in the presence of transverse degrees of freedom is rather rich. This is particularly true in studies of spin-dependent phenomena where one can define various TMDPDFs through Lorentz decompositions (see Sec. V.B). One example is the Sivers function for a transversely polarized proton, $\epsilon_{ij}k^i_\perp S^i_\perp f^{\perp}_{1T}(x,k_\perp)$, which is naive-time-reversal odd and is predicted to change sign between the DY and SIDIS processes~\cite{Collins:2011zzd}. Similar properties also exist in the Boer-Mulders function~\cite{Boer:1997nt} concerning a transversely-polarized parton
distribution in an unpolarized hadron. These two functions are related to the single transverse spin asymmetry. If we generalize the TMDPDFs to include the impact parameter dependence, we can further define the Wigner function, the parton orbital angular momentum distributions, etc~\cite{Belitsky:2003nz, Lorce:2011ni}. Therefore, the TMDPDFs allow for a more complete and refined 3D description (or tomography) of the hadron structure~\cite{Burkardt:2000za, Boer:2011fh}.
The 3D tomography of the proton is a major physical goal of the EIC program. The TMDPDFs are also important
in understanding small-$x$ physics~\cite{Kuraev:1977fs,Balitsky:1978ic,Balitsky:1995ub,Kovchegov:1999yj,Kovchegov:2012mbw}.

Our current knowledge on TMDPDFs mainly comes from fitting to the experimental data~\cite{Landry:1999an,Konychev:2005iy,Su:2014wpa,Echevarria:2014xaa,Kang:2015msa,Bacchetta:2017gcc,Scimemi:2017etj,Bertone:2019nxa,Scimemi:2019cmh,Bacchetta:2019sam}. This is, however, rather primitive due to the paucity of data. Although the future EIC will make up the gap and produce more data for TMD measurements, it is still important to develop first-principle methods for the determination of nonperturbative TMDPDFs, which can serve as a test or provide useful inputs to constrain the global fits.
LaMET provides a systematic way to extract TMDPDFs from the lattice calculations. Early studies~\cite{Ji:2014hxa,Ji:2018hvs,Ebert:2018gzl,Ebert:2019okf} have tried to construct a quasi-TMDPDF on the lattice, but its relation to the physical TMDPDF is expected to be nonperturbative due to complications in the soft function~\cite{Ebert:2019okf}. The recent works in~\cite{Ji:2019sxk,Ji:2019ewn} provide a formulation
to calculate the soft function so that a perturbative matching formula can be established between the quasi- and physical TMDPDFs, allowing for a complete determination of the latter from lattice QCD. In this section we review the application of LaMET to the nonperturbative TMDPDFs. The investigation is still in its early stage and a lot remains
to be explored, particularly in lattice calculations and matching.

In the first subsection we introduce the TMDPDFs and discuss the associated rapidity divergences. In the following subsections, we define the quasi-TMDPDFs or TMD momentum distributions in a proton of finite momentum,
and study their momentum RGEs and UV renormalization properties. In the process,
we introduce the off-light-cone soft functions. We then present the factorization of the quasi-TMDPDFs into the light-cone TMDPDFs and the off-light-cone soft function, where various one-loop results and the relevant RGEs are also given.
The properties of the off-light-cone soft function are discussed in the last subsection, where it is shown to be related to
the form factor of a pair of charged color sources, which paves the way for its calculation on a Euclidean lattice.

\subsection{Introduction to TMDPDFs and Rapidity Divergence}
\label{sec:lightconetmd}

As explained in \sec{lamet}, we can define various TMDPDFs by choosing different
gauge-links between the quark or gluon bilinears. The one relevant to high-energy phenomena is defined with
light-like Wilson lines. The links represent the propagation of high-energy color-charged particles, and are crucial in forming gauge-invariant nonlocal operators~\cite{Belitsky:2002sm}. As argued in previous sections, such
operators are the result of an EFT description (more explicitly so in SCET) arising from
taking the infinite-momentum limit of the proton. Thus, it is natural to expect that they require additional regularization and renormalization.

\begin{figure}[htb]
	\centering
	\includegraphics[width=0.9\columnwidth]{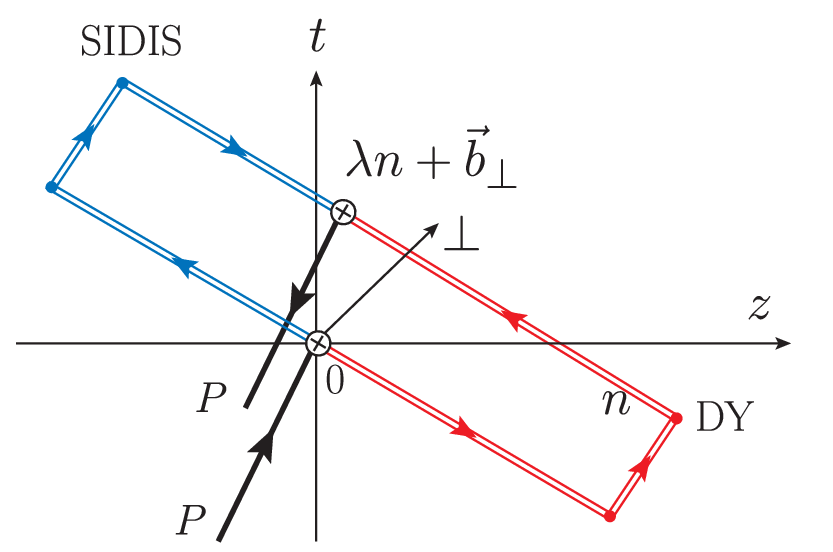}
	\caption{The space-time picture of TMDPDF for DY and SIDIS process. The circled crosses denote the quark-link vertices. Notice that the vertices are placed at $\lambda n+\vec{b}_\perp$ and $0$ which gives the same result as the symmetric choice in Eq.~(\ref{eq:beam}). }
	\label{Fig:tmdpdf}
\end{figure}

Let us take the non-singlet quark unpolarized TMDPDF as an example.
Without the field theoretic subtleties, the distribution is
\begin{align}\label{eq:beam}
	& f(x,\vec{k}_\perp) = \frac{1}{2P^+}\int\frac{d\lambda}{2\pi}\frac{d^2\vec{b}_\perp}{(2\pi)^2}
	e^{-i\lambda x+i\vec{k}_\perp \cdot \vec{b}_\perp}\\
	& \times \langle P| \bar \psi(\lambda n/2 +\vec{b}_\perp)\gamma^+{\cal W}_n(\lambda n/2+\vec{b}_\perp)\psi(-\lambda n/2)|P\rangle  \ ,\nn
\end{align}
where ${\cal W}_n(\lambda n+\vec{b}_\perp)$ is the staple-shaped gauge-link of the form
\begin{align}
	&{\cal W}_n(\xi)=W^{\dagger}_{n}(\xi)W_{\perp}W_{n}(-\xi \cdot p n) \label{eq:staplen} \ , \\
	&W_{n}(\xi)= {\cal P}\exp\left[-ig\int_{0}^{-\infty} d\lambda n\cdot A(\xi+\lambda n)\right] \ ,
\end{align}
along the light-cone direction $n^\mu$, as shown in Fig.~\ref{Fig:tmdpdf}.The $W_\perp$ is a transverse gauge-link at light-cone infinity to maintain gauge-invariance..
If one uses LFQ and ignores the transverse gauge-link, the above distribution
is just $\langle P|b^\dagger (x,\vec{k}_\perp)b(x,\vec{k}_\perp)|P\rangle$ for $x>0$, as expected.

However, there are a number of qualifications in the above definition.
First, the light-like gauge-links ${\cal W}_{n}$ are chosen to be past-pointing in accordance with the
DY kinematics, but for SIDIS  they should be chosen as future-pointing, as shown in Fig.~\ref{Fig:tmdpdf}.
For unpolarized TMDPDFs there is no distinction between the two choices,
but for spin-dependent TMDPDFs there are physical consequences associated with
the direction of gauge-links.

Second, there exists a new type of divergence associated with the infinitely-long light-like gauge-links. These divergences are due to radiation of gluons
collinear to the light-like gauge-link and cannot be regularized by the standard UV regulators. An example is
the following integral in dimensional regularization (DR)~\cite{Ebert:2019okf},
\begin{equation}\label{eq:RD}
	I = \int dk^+dk^- \frac{f(k^+k^-)}{(k^+k^-)^{1+\epsilon}} = \frac{1}{2}
	\int \frac{dy}{y} \int dm^2 \frac{f(m^2)}{m^{2+2\epsilon}}\,,
\end{equation}
where $m^2=k^+k^-$ and $y = k^+/k^-$ is the rapidity-related variable. The divergences
in $y$ arise from large and small $y$ where the integral is unregulated. The contribution
from $k^+=0$ is called the light-zero mode in LFQ, where it is also called light-cone divergence
which causes considerable problems.

To regulate the light-cone or rapidity divergences, a number of methods have been introduced in the literature (for a review see~\cite{Ebert:2019okf}).
They can be put into two classes: on-light-cone regulators and off-light-cone regulators.
In the former case, the gauge-links are kept along the light-cone direction $n^\mu$ after regularization. For example, the so-called {\it $\delta$ regulator}~\cite{Echevarria:2015usa,Echevarria:2015byo} regularizes the gauge-link as:
\begin{align}
	&W_n(\xi)\rightarrow W_{n}(\xi)|_{\delta^-}\nonumber \\
	&={\cal P}{\rm exp}\left[-ig\int_0^{-\infty} d\lambda A^+(\xi+\lambda n)e^{-\frac{\delta^-}{2p^+} |\lambda|}\right] \,,
\end{align} and similarly for the conjugate direction. The $\delta$ regulator breaks gauge-invariance, but preserves the boost invariance $\delta^{\pm}\rightarrow e^{\pm Y}\delta^{\pm}$ where $Y$ is the rapidity of the Lorentz boost. Other on-light-cone regulators include the exponential regulator~\cite{Li:2016axz}, $\eta$ regulator~\cite{Chiu:2012ir}, analytical regulator~\cite{Becher:2010tm}, etc. In the remainder of this section, we will use the $\delta$ regulator as a representative whenever we need an on-light-cone regulator.

The off-light-cone regulator was introduced in ~\cite{Collins:1981uk,Ji:2004wu,Ji:2004xq,Collins:2011zzd}, and also used in ~\cite{Ji:2004wu}. This type of regulator chooses off-light-cone directions to avoid the rapidity divergence. One can choose, for instance, to deform the gauge-links into the space-like region:
\begin{align}
	n\rightarrow n_Y=n-e^{-2Y}\frac{p}{(p^+)^2}\,.
\end{align}
Here $Y$ plays the role of a rapidity regulator, as when $Y \rightarrow \infty$, $n_Y \rightarrow n$. In certain cases one can also deform $n_Y$ into time-like region~\cite{Collins:2004nx}.

The on-light-cone regulators are consistent with the spirit of parton physics, and therefore are useful to define COM-momentum-independent parton densities. The off-light-cone regulators, on the other hand, follow a similar spirit as LaMET, and therefore can be exploited for practical lattice QCD calculations, as we shall see in the next subsection.

To avoid light-cone divergences, from now on we include the rapidity regulator in the definition of the
light-cone TMDPDFs. Using the same label $f$ for the TMDPDFs in both momentum and
coordinate spaces, we have
\begin{align}
	&f(\lambda,b_\perp,\mu,\delta^-/P^+)\\
	&= \langle P| \bar \psi(\lambda n/2+\vec{b}_\perp)\slashed{n}{\cal W}_{n}(\lambda n/2+\vec{b}_\perp)|_{\delta^-}\psi(-\lambda n/2) |P\rangle  \,,\nn
\end{align}
where $\mu$ is the $\MS$ scale for UV renormalization.
Due to rotational invariance, the bare TMDPDF defined above is a function of $b_\perp=|\vec{b}_\perp|$, so we have omitted the vector arrow for $\vec{b}_\perp$ in $f$ and will do so throughout the discussion for the soft functions, quasi-TMDPDFs, etc. The subscript $\delta^-$ denotes that the staple-shaped gauge-link ${\cal W}$ is regulated by the $\delta$ regulator in the light-cone minus direction. $f$ diverges logarithmically as $\delta^- \rightarrow 0$, and the finite part also depends on the rapidity
regulator. To define the physical TMDPDF, we need to remove all divergences and rapidity regularization scheme dependencies in $f$, in a way similar to removing UV divergences in physical quantities.
\begin{figure}[htb]
	\centering
	\includegraphics[width=0.9\columnwidth]{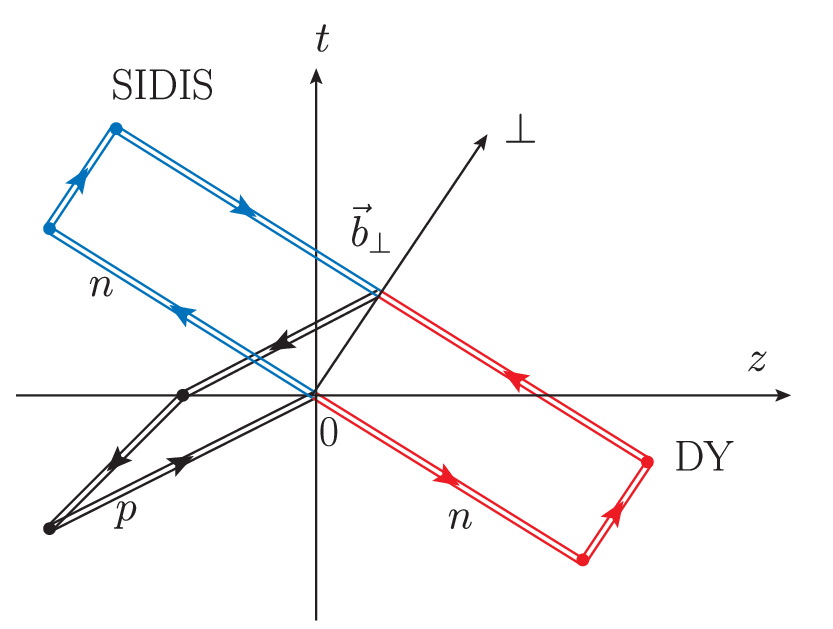}
	\caption{The soft function $S(b_\perp,\mu,\delta^+,\delta^-)$ as space-time Wilson-loop arising in the
		factorization of DY and SIDIS process.}
	\label{Fig:soft1}
\end{figure}

The rapidity divergence for TMDPDFs can be removed by the
soft function, which also plays an important role in TMD factorization.
Intuitively, the soft function represents a cross section for fast-moving
charged particles emitting soft gluons into final states. It has rapidity divergence associated with the light-cone direction, which
is ultimately related to the mass singularity. The TMD soft function corresponding to Drell-Yan process is defined~\cite{Collins:2011ca,Echevarria:2015byo} as
\begin{align}\label{eq:soft}
	& S(b_\perp,\mu,\delta^+,\delta^-)\nonumber \\
	&=\frac{{\rm Tr}\langle 0|{\cal \bar T}W_p(\vec{b}_\perp)|_{\delta^+}W_n^{\dagger}(\vec{b}_\perp)|_{\delta^-}{\cal T} W_n(0)|_{\delta^-}W^{\dagger}_p(0)|_{\delta^+}|0\rangle }{N_c} \nonumber \\
	&=\frac{{\rm tr}\langle 0|{\cal W}_{n}(\vec{b}_\perp)|_{\delta^+} {\cal W}^{\dagger}_{p}(\vec{b}_\perp)|_{\delta^-}|0\rangle}{N_c} \,,
\end{align}
where ${\cal T}/\bar{\cal T}$ stands for time/anti-time ordering. The first equality defines the soft function in terms of cut-diagrams as an amplitude square. Since the soft function for DY process is independent of time ordering, one can also define it with a single time ordering or no time ordering, leading to the second equality. The staple-shaped gauge-link ${\cal W}_n$ is defined in Eq.~(\ref{eq:staplen}),
while the staple-shaped gauge-link ${\cal W}_p$ is defined similarly as:
\begin{align}
	&{\cal W}_p(\xi)=W^{\dagger}_{p}(\xi)W_{\perp}W_{p}(0) \label{eq:staplep} \ , \\
	&W_{p}(\xi)= {\cal P}{\rm exp}\left[-ig\int_{0}^{-\infty} d\lambda p\cdot A(\xi+p \lambda)\right] \ .
\end{align}
The soft function is shown in Fig.~\ref{Fig:soft1} as a Wilson loop in Minkowski space.

If the rapidity divergences are multiplicative, one can use $S$ as the rapidity renormalization factor for the TMDPDF defined in \eq{beam}. In on-light-cone schemes such as the $\delta$ regularization, it has been argued in~\cite{Vladimirov:2017ksc} based on conformal transformation that the rapidity divergences are indeed multiplicative in position space. For each of the staple-shaped light-like gauge-links, the rapidity divergence is proportional to $\exp\left[-(1/2)K(b_\perp,\mu)\ln \left(\mu^2/2(\delta^{\pm})^2\right)\right]$ where $K(b_\perp,\mu)$ is the nonperturbative Collins-Soper evolution kernel~\cite{Collins:1981uk}. Thus at small $\delta^{\pm}$, we can write
\begin{equation}
	S(b_\perp,\mu,\delta^+,\delta^-)=e^{\ln \frac{\mu^2}{2\delta^+\delta^-}K(b_\perp,\mu)+{\cal D}_2(b_\perp,\mu)} \ \ ,
\end{equation}
where ${\cal D}_2(b_\perp,\mu)$ is a $b_\perp$-dependent but rapidity-independent function. Notice that our definitions of $\delta^{\pm}$ differ from those in Ref.~\cite{Echevarria:2015byo} by a factor of $\sqrt{2}$ due to our normalization of light-cone vectors.

The soft-function in $\delta$ regularization satisfies the renormalization group equation
\begin{align}
	&\mu^2\frac{d}{d\mu^2}\ln S(b_\perp,\mu,\delta^+,\delta^-)\nonumber \\
	&=-\Gamma_{\rm cusp}(\alpha_s)\ln \frac{\mu^2}{2\delta^+\delta^-} + \gamma_s(\alpha_s)\ ,
\end{align}
where $\Gamma_{\rm cusp}(\alpha_s) $ is the light-like cusp anomalous dimension~\cite{Polyakov:1980ca,Korchemsky:1987wg} and the $\gamma_s(\alpha_s)$ is the soft anomalous dimension~\cite{Korchemskaya:1992je}. The Collins-Soper kernel and the rapidity-independent part ${\cal D}_2$ satisfy the RGEs:
\begin{align}
	\mu^2\frac{d}{d\mu^2} K(b_\perp,\mu)=-\Gamma_{\rm cusp}(\alpha_s) \label{eq:cusp}\ ,\\
	\mu^2\frac{d}{d\mu^2} {\cal D}_2(b_\perp,\mu)=\gamma_s(\alpha_s)-K(b_\perp,\mu) \label{eq:ad_d2}\ .
\end{align}

At one-loop, the soft function $S(b_\perp,\mu,\delta^+,\delta^-)$ is given by~\cite{Echevarria:2012js} :
\begin{align}
	&S(b_\perp,\mu,\delta^+,\delta^-)\nonumber \\ &=1+\frac{\alpha_s C_F}{2\pi}\left(L_b^2-2L_b \ln \frac{\mu^2}{2\delta^+\delta^-}+\frac{\pi^2}{6}\right) \ ,
\end{align}
where $L_b=\ln\left(\mu^2 b_\perp^2e^{2\gamma_E}/4\right)$.
Therefore, we have at the leading order,
\begin{align}
	K(b_\perp,\mu)&=-\frac{\alpha_sC_F}{\pi} L_b  \,,\\
	{\cal D}_2(b_\perp,\mu)&=\frac{\alpha_s C_F}{2\pi}\left(L_b^2+\frac{\pi^2}{6}\right) \ ,
\end{align}
and $\Gamma_{\rm cusp}=\alpha_s C_F/\pi +{\cal O}(\alpha_s^2)$, $\gamma_s={\cal O}(\alpha_s^2)$. It is worth pointing out that $K$~\cite{Li:2016ctv,Vladimirov:2016dll} and ${\cal D}_2$~\cite{Li:2016ctv} are known to 3-loop order in the exponential regularization scheme.

With the above soft function, we can take its square root to perform rapidity renormalization for the bare TMD correlator. The square root can be explained as follows: $S$ contains two staples, while $f$ contains one, thus the rapidity divergences as well as scheme dependencies in $S$ are twice as those in $f$. This leads to the following definition of the renormalized physical TMDPDF~\cite{Echevarria:2012js,Collins:2012uy}:
\begin{align}\label{eq:TMD}
	f^{\rm TMD}(x,b_\perp,\mu,\zeta)=\lim_{\delta^- \rightarrow 0}\frac{f(x,b_\perp,\mu,\delta^-/P^+)}{\sqrt{S(b_\perp,\mu,\delta^- e^{2y_n},\delta^-)}} \ ,
\end{align}
where the rapidity scale reads
\begin{align} \label{zeta}
	\zeta=2(xP^+)^2 e^{2y_n} \ .
\end{align}
The rapidity dependence in the numerator of the right-hand side of Eq.~(\ref{eq:TMD}) has the form $\exp[-\frac{1}{2}K(b_\perp,\mu)\ln \frac{(\delta^-)^2}{(xP^+)^2}]$, while in the denominator it behaves as $\exp[\frac{1}{2}K(b_\perp,\mu)\ln \frac{\mu^2}{2(\delta^-)^2e^{2y_n}}]$. The $\delta^-$ dependence thus cancels out in the ratio, leaving a dependence on the rapidity scale as $\exp[-\frac{1}{2}K(b_\perp,\mu)\ln \frac{\mu^2}{2(xP^+)^2e^{2y_n}}]$, which is controlled by the so-called Collins-Soper evolution equation:
\begin{align}\label{eq:cs}
	2\zeta \frac{d}{d\zeta} \ln  f^{\rm TMD}(x,b_\perp,\mu,\zeta)=K(b_\perp,\mu) \ .
\end{align}
The $\zeta$-dependence comes from the initial-state quark radiation and is intrinsically nonperturbative for large $b_\perp$. $f^{\rm TMD}(x,b_\perp,\mu,\zeta)$ is the standard object to be matched to in LaMET.

We should emphasize that although $f^{\rm TMD}$ is free from rapidity divergences, it does contain soft radiation from the
charged particles in the initial state. This can be seen clearly by considering Feynman diagrams for the unsubtracted $f$
and applying soft approximation to gluons. ``One-half'' of the soft contribution in $f$
is subtracted to define the physical $f^{\rm TMD}$ due to the requirement of factorization of
physical processes. The remaining soft radiation also has a natural rapidity cut-off
associated with $\ln(xP^+)$, reflected in the $\zeta$-dependence. What is remarkable,
however, is that $f^{\rm TMD}$ is rapidity-regulator independent. Although a general proof to all
orders in perturbation theory is beyond the scope of this review, it is due to
factorization and exponentiation of the soft physics in $f$ and thus the scheme cancellation
can be done systematically in the exponent. It worth mentioning that in old-fashioned or SCET-like approaches, one can define the ``subtracted'' TMDPDF or ``beams functions'' that contains only collinear physics. However, they are generically scheme dependent and must be combined with an extra soft functions in factorization theorems. At one-loop level, the scheme-independent one-loop TMDPDF for an external quark state reads,
\begin{align}\label{eq:ftmd_1loop}
	&f^{\rm TMD}(x,b_\perp,\mu,\zeta) = \delta(1-x)\nonumber \\
	&+\frac{\alpha_sC_F}{2\pi}F(x,\epsilon_{\tiny\mbox{IR}},b_\perp,\mu)\theta(x)\theta(1-x)+\frac{\alpha_sC_F}{2\pi}\delta(1-x)\nonumber \\
	&\times \left[-\frac{1}{2}L_b^2+\left(\frac{3}{2}-\ln \frac{\zeta}{\mu^2}\right)L_b+\frac{1}{2}-\frac{\pi^2}{12}\right] \ ,
\end{align}
where
\begin{align}
	F(x,\epsilon_{\tiny\mbox{IR}},b_\perp,\mu)=\left[-\left(\frac{1}{\epsilon_{\tiny\mbox{IR}}}+L_b\right)\frac{1+x^2}{1-x}+1-x\right]_+\,.
\end{align}
Two-loop order results for the TMDPDFs can be found in~\cite{Catani:2011kr,Catani:2012qa,Gehrmann:2014yya,Luebbert:2016itl,Echevarria:2016scs,Luo:2019hmp} and three-loop order results can be found in~\cite{Luo:2019szz}.

The physical TMDPDF also satisfies the RG equation,
\begin{align}
	\gamma_\mu(\mu,\zeta)&=\mu^2\frac{d}{d\mu^2}\ln f^{\rm TMD}(x,b_\perp,\mu,\zeta) \nonumber \\
	&\equiv \frac{1}{2}\Gamma_{\rm cusp}(\alpha_s)\ln \frac{\mu^2}{\zeta}-\gamma_H(\alpha_s)  \ ,
\end{align}
where $\gamma_H$ is called the hard anomalous dimension. At one-loop, the cusp and hard anomalous dimensions read
\begin{align}\label{eq:ad_hard}
	\Gamma_{\rm cusp}(\alpha_s) =\frac{\alpha_s C_F}{\pi}; ~~~~~
	\gamma_H(\alpha_s) =-\frac{3\alpha_sC_F}{4\pi}  \ .
\end{align}
Recently the cusp anomalous dimension have been calculated to 4-loops~\cite{Henn:2019swt,vonManteuffel:2020vjv}.

Combining the RGE and the rapidity evolution equation for the TMDPDF, one obtains the consistency condition :
\begin{align}
	\mu^2 \frac{d}{d\mu^2} K(b_\perp,\mu)
	&= -2 \zeta \frac{d}{d\zeta} \gamma_\mu(\mu,\zeta) =- \Gamma_{\rm cusp}(\alpha_s(\mu)) \ ,
\end{align}
from which one finds a resummed form for the Collins-Soper kernel:
\begin{align}
	K(b_\perp,\mu) &= -2 \int_{1/b_\perp}^\mu \frac{d\mu'}{\mu'} \Gamma_{\rm cusp}(\alpha_s(\mu')) + K(\alpha_s(1/b_\perp)) \ .
\end{align}
Here $K(\alpha_s(1/b_\perp))$ contains both perturbative and non-perturbative contributions. The TMDPDFs at different scales are then related by
\begin{align} \label{eq:tmd_evolution}
	&f^{\rm TMD}(x, b_\perp, \mu, \zeta) = f^{\rm TMD}(x, b_\perp, \mu_0, \zeta_0) \\
	& \times\exp\biggl[ \int_{\mu_0}^\mu \frac{d\mu'}{\mu'} \gamma_\mu(\mu',\zeta_0) \biggr]
	\exp\biggl[ \frac12 K(b_\perp,\mu) \ln\frac{\zeta}{\zeta_0} \biggr]  \ . \nn
\end{align}
The double-scale evolution in the $\mu-\zeta$ plane for phenomenology has been recently studied in~\cite{Scimemi:2018xaf}.
With the scheme-independent physical TMDPDF defined above, the DY cross section at $s=(P_A+P_B)^2$ and small $Q_\perp$ can be factorized as
\begin{align}
	&\frac{d\sigma}{dQ_\perp^2}=\int dx_Adx_B d^2b_\perp e^{i\vec{b}_\perp \cdot \vec{Q}_\perp}\hat \sigma(x_Ax_Bs,\mu)\nonumber \\ &\times f_A^{\rm TMD}(x_A,b_\perp,\mu,\zeta_A)f_B^{\rm TMD}(x_B,b_\perp,\mu,\zeta_B)+ ... \ .
\end{align}
The rapidity scales satisfy $\zeta_A\zeta_B= Q^4\equiv(x_Ax_Bs)^2$. The remaining term at large but finite $Q^2$ are called power corrections or ``higher-twist'' contributions. A detailed study of the power corrections to TMD factorization is beyond the scope of this review. Without mention we will omit all the power-corrections in equations. The QCD part of the hard cross section $\hat{\sigma}$ at one-loop level reads
\begin{align}
	\hat \sigma(x_A,x_B)=\left|1+\frac{\alpha_sC_F}{4\pi}\left(-L^2_{Q}+3L_Q-8+\frac{\pi^2}{6}\right)\right|^2 \ ,
\end{align}
where $L_Q=\ln \frac{-Q^2-i0}{\mu^2}$, and the result is now known up to three loops (see~\cite{Moch:2005tm,Baikov:2009bg,Lee:2010cga,Gehrmann:2010ue} and the references therein). Similarly for the SIDIS process we have
\begin{align}
	\frac{d\sigma}{dQ_\perp^2}&=\int dxdz d^2b_\perp e^{i\vec{b}_\perp \cdot \vec{Q}_\perp}H(x,z,\mu,Q)\nonumber \\
	&\quad \times f^{\rm TMD}(x,b_\perp,\mu,\zeta_A)d^{\rm TMD}(z,b_\perp,\mu,\zeta_B) \ ,
\end{align}
where $d^{\rm TMD}(z,b_\perp,\mu,\zeta_B)$ is the TMD fragmentation function and $H$ is the hard kernel.


\subsection{Lattice Quasi-TMDPDFs and Matching}
\label{sec:quasitmd}
Before LaMET, there had been efforts to access TMD physics from lattice QCD by calculating the ratios of the $x$-moments of TMDPDFs~\cite{Hagler:2009mb,Musch:2010ka,Musch:2011er,Engelhardt:2015xja,Yoon:2017qzo}, which are free from complications associated with the soft function and can be compared to certain experimental observables. In LaMET, we are more interested in obtaining the full $x$ and $\vec{k}_\perp$ dependence of the TMDPDFs~\cite{Ji:2014hxa,Ji:2018hvs,Ebert:2018gzl,Ebert:2019okf,Ji:2019sxk,Ji:2019ewn}. Therefore, a proper treatment of the soft function subtraction and matching is essential.
The earliest suggestion of a bent soft function in~\cite{Ji:2014hxa} and the follow-up work~\cite{Ebert:2019okf} has the correct IR logarithms at one-loop order, but this is expected to break down at higher-loop orders~\cite{Ji:2019sxk}, thus not allowing for a perturbative matching. Another suggestion which uses a naive rectangle-shaped Wilson loop~\cite{Ji:2018hvs,Ebert:2019okf} does not possess the correct IR physics, either. Nevertheless, in~\cite{Ebert:2018gzl} important progress was made for calculating the nonperturbative
Collins-Soper kernel $K(b_\perp,\mu)$ from the ratio of quasi-TMDPDFs at two different large momenta.
Recently, some of the authors showed~\cite{Ji:2019sxk,Ji:2019ewn} that the quasi-TMDPDF combined with a reduced soft function capture the correct IR physics to all-orders and thus allow for a perturbative matching to the physical TMDPDF.

To construct such quasi-TMDPDFs, the collinear part can be treated in a way similar to the collinear PDFs, while the soft
piece is more challenging. Our starting point is that the physical $f^{\rm TMD}$ is independent of the rapidity regulator, so one can use a scheme in which the gauge-links in both $f$ and $S$
are off the light-cone, such as that used in~\cite{Collins:2011zzd}. In this case, one can use Lorentz symmetry to
boost the staple-shaped gauge-link ${\cal W}_n$ in $f$ to a purely space-like staple
with no time dependence. However, one can only use
this trick for one of the staples in $S$, say ${\cal W}_n$, whereas the other one
${\cal W}_p$ is still time-dependent. In other words, there is no way to get rid of the time dependence
in $S$ entirely with Lorentz boost alone. This is natural because $S$ in fact represents the square of
an $S$-matrix, which appears to be intrinsically Minkowskian. However, using the LaMET principle that time
dependence of an operator can be simulated through external physical states at large momentum, we find that
$S$ can indeed be calculated on the lattice in the off-light-cone scheme as a form factor. A detailed discussion will be given in
the next subsection. Here we assume that this is true, and discuss the matching between
quasi- and physical TMDPDFs.

\begin{figure}[htb]
	\centering
	\includegraphics[width=0.8\columnwidth]{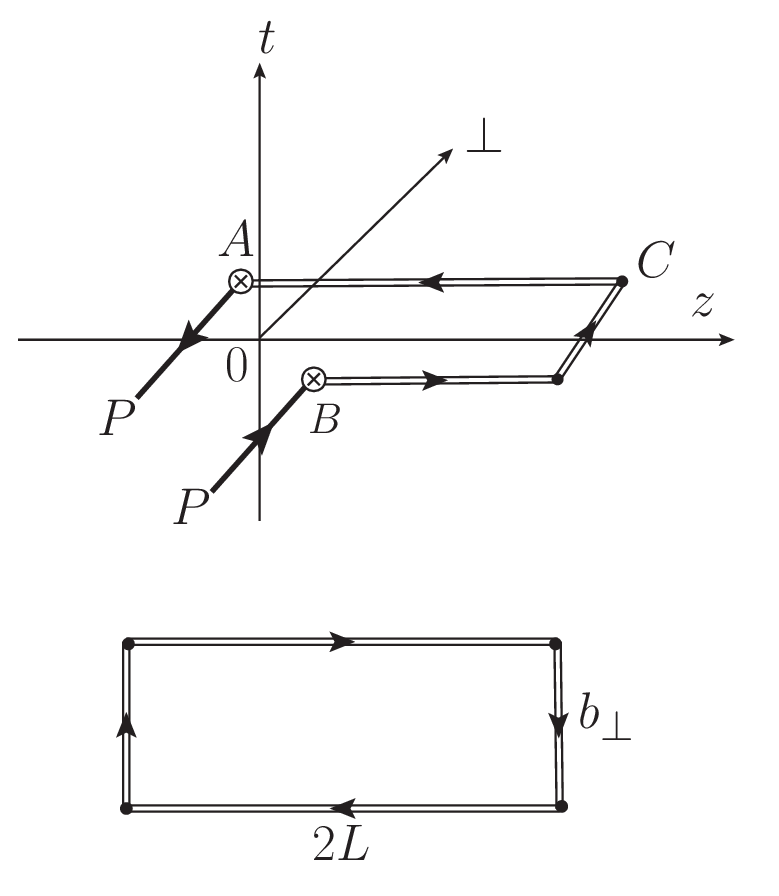}.
	\caption{The quasi-TMDPDF (upper) and the Euclidean Wilson-loop $Z_E(2L,b_\perp,\mu,0)$ (lower). In the figure, $A=\lambda n_z/2+\vec{b}_\perp/2$, $B=-\lambda n_z/2-\vec{b}_\perp/2$ and $C=Ln_z+{\vec b}_\perp$. The crosses denote the quark-link vertices.}
	\label{fig:quasiTMD}
\end{figure}

First, we define the quasi-TMDPDF with a staple–shaped gauge-link along the $z$ direction~\cite{Ji:2014hxa,Ji:2018hvs,Ebert:2019okf,Ji:2019ewn} as
\begin{align}\label{eq:quasi_TMD}
	& \tilde f(\lambda ,b_\perp,\mu,\zeta_z) \\
	&=\! \lim_{L \rightarrow \infty}  \frac{\langle P| \bar \psi\big(\frac{\lambda n_z }{2}\!+\!\vec{b}_\perp\big)\gamma^z{\cal W}_{z}(\frac{\lambda n_z}{2}\!+\!\vec{b}_\perp;L)\psi\big(\!-\!\frac{\lambda n_z}{2}\big) |P\rangle}{\sqrt{Z_E(2L,b_\perp,\mu)}} \ , \nonumber
\end{align}
where the $\MS$ renormalization is implied, and
\begin{align}\label{eq:staplez}
	&{\cal W}_z(\xi;L)=W^{\dagger}_{z}(\xi; L)W_{\perp}W_{z}(-\xi^zn_z;L)  \ ,\\
	&W_{z}(\xi;L)= {\cal P}{\rm exp}\Big[-ig\int_{\xi^z}^{L} d\lambda n_z\cdot A(\vec{\xi}_\perp\!+\!n_z \lambda)\Big] \ .
\end{align}
Here $\xi^z=-\xi\cdot n_z$ and $\zeta_z=(2xP^z)^2$ is the Collins-Soper scale of the quasi-TMDPDF. $W_{\perp}$ is inserted at $z=L$ to maintain explicit gauge invariance. $\sqrt{Z_E(2L,b_\perp,\mu,0)}$ is the square root of the vacuum expectation value of a flat rectangular Euclidean Wilson-loop along the $n_z$ direction with length $2L$ and width $b_\perp$:
\begin{align}\label{eq:Z_E}
	Z_E(2L,b_\perp,\mu)=\frac{1}{N_c}{\rm Tr}\langle 0|W_{\perp}{\cal W}_z(\vec{b}_\perp;2L)|0\rangle \ .
\end{align}
Again, $\gamma^z$ can be replaced by $\gamma^t$ as in the collinear quasi-PDF.
For a depiction of $\tilde f$ and $Z_E$ see Fig.~\ref{fig:quasiTMD}.

The purpose of the factor $Z_E$ is as follows. At large $L$, the naive quasi-TMD correlator in the numerator of Eq.~(\ref{eq:quasi_TMD}) contains divergences that go as $e^{-LE(b_\perp,\mu)}$ where $E(b_\perp)$ is the ground state energy of a pair of static heavy-quarks. $E(b_\perp,\mu)=2\delta m+ V(b_\perp,\mu)$ contains both the linear divergent mass corrections $2\delta m$ and the heavy-quark potential $V(b_\perp,\mu)$ due to mutual interactions. In literature the $LV(b_\perp,\mu)$ part was sometimes called the ``pinch pole singularity.'' Therefore, we introduce the square root of a rectangular Wilson-loop $Z_E(2L,b_\perp,\mu)$ with twice the length to cancel all these divergences and guarantee the existence of the $L\to\infty$ limit after the subtraction. The introduction of $\sqrt{Z_E}$ also removes additional contributions from the transverse gauge link. An alternative approach to avoid the pinch-pole singularity was proposed in~\cite{Li:2016amo}. We should mention that although the $\sqrt{Z_E}$ subtraction removes all the linear divergences, the logarithmic UV divergences are still present. Therefore, a non-perturbative renormalization of $\tilde f$ on the lattice is still required, which has been studied in the RI/MOM scheme~\cite{Shanahan:2019zcq}, and its matching to the $\MS$ scheme has been calculated at one-loop order~\cite{Constantinou:2019vyb,Ebert:2019tvc}.

The quasi-TMDPDFs defined above satisfy the following RGE~\cite{Collins:1981uk,Ji:2014hxa,Ji:2019ewn}
\begin{align}
	\mu^2 \frac{d}{d\mu^2}\ln \tilde f(x,b_\perp,\mu,\zeta_z)=\gamma_{F}(\alpha_s(\mu)) \ ,
\end{align}
where $\gamma_F$ is the anomalous dimension for the heavy-to-light current in \sec{renorm-nl}. This is due to the fact that the quasi-TMDPDF, after the self-energy subtraction, contains only logarithmic UV divergences associated with quark-Wilson-line vertices. In the $\MS$ scheme, the one-loop quasi-TMDPDF in an external quark state with momentum $(p^z,0,0,p^z)$ reads~\cite{Ji:2018hvs,Ebert:2019okf}
\begin{align}
	&\tilde f(x,b_\perp,\mu,\zeta_z)=\nonumber \\ &1+\frac{\alpha_sC_F}{2\pi}F(x,\epsilon_{\tiny \mbox{IR}},b_\perp,\mu)\theta(x)\theta(1-x)+\frac{\alpha_sC_F}{2\pi}\delta(1-x) \nonumber \\
	& \times \Bigg[-\frac{1}{2}L_b^2+L_b\Big(\frac{5}{2}-L_z\Big)-\frac{3}{2}-\frac{1}{2}L_z^2+L_z\Bigg]\ ,
\end{align}
where $L_z=\ln(\zeta_z/\mu^2)$. As expected, the $L$ dependence has been cancelled in the large $L$ limit.

As there is no light-like gauge-link in $\tilde f$, no additional rapidity regulator is needed. Instead, there is an explicit dependence on the hadron momentum (or energy), which is similar to the momentum RGE for collinear quasi-PDF. The momentum (rapidity) evolution equation for $\tilde f$ reads~\cite{Collins:1981uk,Ji:2014hxa,Ji:2019ewn},
\begin{align}
	P^z\frac{d}{dP^z}\ln \tilde f(x,b_\perp,\mu,\zeta_z) \!=\! K(b_\perp,\mu)\!+\!{\cal G}\Big(\frac{(P^z)^2}{\mu^2}\Big) \ ,
\end{align}
where ${\cal G}(\zeta_z/\mu^2)$ is perturbative and $K(b_\perp,\mu)$ is the Collins-Soper kernel. A similar equation was proven for off-light-cone TMD-fragmentation functions in~\cite{Collins:1981uk}. From this equation, it is clear that a correct matching to $f^{\rm TMD}(x,b_\perp,\mu,\zeta)$ with arbitrary $\zeta$ must include $K(b_\perp,\mu)$ to compensate the $P^z$ dependence.

There is actually one more requirement for the matching: there is a rapidity scheme dependence which must
be removed, since the quasi-TMDPDF can be viewed as defined with an off-light-cone regulator along the $z$ direction. To understand this dependence, let us consider $f$ again in the off-light-cone regularization, where there are rapidity divergences. The divergence is cancelled by the square root of an off-light-cone soft function $S_{\rm DY}(b_\perp,\mu,Y,Y')$, with $Y,Y'$ being the rapidities of the off-light-cone space-like vectors $p\rightarrow p_Y= p-e^{-2Y}(p^+)^2n $ and $n \rightarrow n_{Y'}=n-e^{-2Y'}p/(p^+)^2$. Schematically, we have:
\begin{align}\label{eq:S_DY}
	S_{\rm DY}(b_\perp,\mu,Y,Y')= \frac{{\rm tr}\langle 0|{\cal W}_{n_{Y'}}(\vec{b}_\perp) {\cal W}^{\dagger}_{p_Y}(\vec{b}_\perp)|0\rangle}{N_c\sqrt{Z_E}\sqrt{Z_E}}  \ ,
\end{align}
where ${\cal W}_{n_{Y'}}(\vec{b}_\perp)$ and ${\cal W}^{\dagger}_{p_Y}(\vec{b}_\perp)$ are staple-shaped gauge-links in $n_{Y'}$, $p_Y$ directions, respectively. $\sqrt{Z_E}$ is introduced to subtract the pinch pole singularities for the off-light-cone staple-shaped gauge-links. In terms of $\ln \rho^2=2(Y+Y')$ sometimes we also write this soft function as $S_{\rm DY}(b_\perp,\mu,\rho)$. At large $\rho $, we have
\begin{align}
	S_{\rm DY}(b_\perp,\mu,Y,Y')= e^{(Y+Y') K(b_\perp,\mu)+{\cal D}(b_\perp,\mu)}+... \,.
\end{align}
We can perform a Lorentz boost of ${\cal W}_{n_{Y'}}(\vec{b}_\perp) {\cal W}^{\dagger}_{p_Y}(\vec{b}_\perp)$ in \eq{S_DY} such that one of the gauge-links, say ${\cal W}_{n_{Y'}}$, is boosted to the equal-time version ${\cal W}_z$ in $\tilde f$, whereas the other gauge-link ${\cal W}_{n_{Y}}$ is boosted to ${\cal W}_{n_{Y+Y'}}$. The soft function becomes $S_{\rm DY}(b_\perp,\mu,Y+Y',0)$ which
contains light-cone divergence for the $p_{Y+Y'}$ direction, but is still the same $S_{\rm DY}(b_\perp,\mu,Y,Y')$ due to boost invariance. The square-root of the finite part $e^{{\cal D}(b_\perp,\mu)}$
is exactly what is needed to cancel the rapidity-scheme dependence. We define the rapidity-independent part as
the reduced soft function:
\begin{align}\label{eq:reduced}
	S_r(b_\perp,\mu)\equiv e^{-{\cal D}(b_\perp,\mu)} \,.
\end{align}
Based on the renormalization property of non-light-like Wilson-loops, the reduced soft function satisfies the RG equation
\begin{align}
	\mu^2 \frac{d}{d\mu^2} \ln S_r(b_\perp,\mu)=\Gamma_S(\alpha_s) \,,
\end{align}
where $\Gamma_S$ is the constant part of the cusp-anomalous dimension at large hyperbolic cusp angle $Y+Y'$ for the off-light-cone soft function:
\begin{align}
	&\mu^2 \frac{d}{d\mu^2} \ln S_{\rm DY}(b_\perp,\mu,Y,Y')\nonumber \\
	&\qquad\qquad=-(Y+Y')\Gamma_{\rm cusp}(\alpha_s)-\Gamma_S(\alpha_s)\,.
\end{align}
At one-loop level~\cite{Ebert:2019okf},
\begin{align}
	S^{(1)}_{\rm DY}(b_\perp,\mu,Y,Y')=\frac{\alpha_sC_F}{2\pi}\big[2-2(Y+Y')\big]L_b \ ,
\end{align}
and $\Gamma_S^{(1)} (\alpha_s)= \alpha_s C_F/\pi$.
Based on RGE, at two-loop level ${\cal D}(b_\perp,\mu)$ can be predicted to be
\begin{align}
	{\cal D}^{(2)}(b_\perp,\mu)=c_2+\Gamma_S^{(2)}L_b-\frac{\alpha_s^2\beta_0C_F}{2\pi}L_b^2 \ ,
\end{align}
where
\begin{align}
	\Gamma_S^{(2)}=-\frac{\alpha_s^2}{\pi^2}\Big[C_FC_A\big(-\frac{49}{36}+\frac{\pi^2}{12}-\frac{\zeta_3}{2}\big)+C_FN_F\frac{5}{18}\Big]\nn
\end{align}
is the two-loop anomalous dimension for $S_r$ which can be extracted from~\cite{Grozin:2015kna}, $\beta_0=-\left(\frac{11}{3}C_A-\frac{4}{3}N_f T_F\right)/(2\pi)$ is the coefficient of one-loop $\beta$-function, and $c_2$ is a constant to be determined by explicit calculation.

After taking into account the reduced soft function, we can now write down
the matching formula between the quasi-TMDPDF and the scheme-independent TMDPDF\cite{Ji:2019ewn}:
\begin{align}\label{eq:quasiTMDmatch}
	&f^{\rm TMD}(x,b_\perp,\mu,\zeta)\\
	&=H\left(\frac{\zeta_z}{\mu^2}\right) e^{-\ln (\frac{\zeta_z}{\zeta})K(b_\perp,\mu)}\tilde f(x,b_\perp,\mu,\zeta_z)S_r^{\frac{1}{2}}(b_\perp,\mu)+...\ ,\nn
\end{align}
where the power-corrections of order ${\cal O}\left(\Lambda_{\rm QCD}^2/\zeta_z,M^2/(P^z)^2,1/(b^2_\perp\zeta_z) \right)$.
The above relation except for the definition of $S_r(b_\perp,\mu)$ was argued to hold in~\cite{Ebert:2019okf}, where the unknown function $g_q^S$ in Eq.~(5.3) should be identified as the reduced soft function here; it has also been confirmed recently in~\cite{Vladimirov:2020ofp}.

We now explain the individual factors of the formula.

\begin{enumerate}
	\item The factor $ H(\zeta_z/\mu^2)$ is the perturbative matching kernel, which is a function of $\zeta_z/\mu^2=(2xP^z)^2/\mu^2$. The kernel is responsible for the large logarithms of $P^z$ generated by the ${\cal G}(\zeta_z/\mu^2)$ term of the momentum RG equation. Unlike the case of quasi-PDFs, the momentum fractions of the quasi-TMDPDF and the TMDPDF are the same. This is due to the fact that at leading power in $1/\zeta_z$ expansion, the $k_\perp$ integral is naturally cut off by the transverse separation around $k_\perp \sim 1/b_\perp \ll P^z$. Therefore, the momentum fraction can only be modified by collinear modes for which there are no distinction between $x=k^z/P^z$ and $x=k^+/P^+$. In comparison, for the $\vec{k}_\perp$ integrated quasi-PDF, the $k_\perp \ge P^z$ region leads to non-trivial $x$ dependence outside the physical region. This is also consistent with the fact that the momentum evolution equation for quasi-TMDPDF is local in $x$ instead of being a convolution.
	\item The factor $\exp\big[\ln (\frac{\zeta_z}{\zeta})K(b_\perp,\mu)\big]$ is the part involving the Collins-Soper evolution kernel. From the momentum evolution equation, it is clear that at large $P^z$ there are logarithms of the form $K(b_\perp,\mu)\ln \frac{\zeta_z}{\mu^2}  $ with $\zeta_z$ being the natural Collins-Soper scale. Therefore, to match to the TMDPDF at arbitrary $\zeta$, a factor $\exp\big[\ln (\frac{\zeta_z}{\zeta})K(b_\perp,\mu)\big]$ is required to compensate the difference. An important implication of this property is that one can obtain the Collins-Soper kernel $K(b_\perp,\mu)$ by constructing the ratio of quasi-TMDPDFs at two different momenta or $\zeta_z$'s~\cite{Ebert:2018gzl},
	\begin{align}
		\label{eq:TMD-CS_kernel}
		\ \ \quad\frac{\tilde f(x,b_\perp,\mu,\zeta_{z,1})}{\tilde f(x,b_\perp,\mu,\zeta_{z,2})}=\frac{H\left(\frac{\zeta_{z,2}}{\mu^2}\right)}{H\left(\frac{\zeta_{z,1}}{\mu^2}\right)}\left(\frac{\zeta_{z,1}}{\zeta_{z,2}}\right)^{K(b_\perp,\mu)} \ .
	\end{align}
	Thus given the $\tilde f$'s at the two rapidity scales, the Collins-Soper kernel $K(b_\perp)$ can be obtained.
	
\end{enumerate}

Combining the RGEs of the quasi-TMDPDF $\tilde f$, reduced soft function $S_r$ and physical TMDPDF $f^{\rm TMD}$, we obtain the RGE of the matching kernel $H\left(\frac{\zeta_z}{\mu^2}\right)$ \cite{Ji:2019ewn},
\begin{align}
	\mu^2 \frac{d}{d\mu^2} \ln H^{-1}\left(\frac{\zeta_z}{\mu^2}\right)=\frac{1}{2}\Gamma_{\rm cusp}(\alpha_s)  \ln\frac{\zeta_z}{\mu^2}\!+\!\frac{\gamma_C(\alpha_s)}{2}  \ ,
\end{align}
where $\gamma_C(\alpha_s)=2\gamma_F(\alpha_s)+\Gamma_S(\alpha_s)+2\gamma_H(\alpha_s)$. The matching kernel is closely related to the perturbative part of the rapidity evolution kernel ${\cal G}\left(\frac{\zeta_z}{\mu^2}\right)$ through
\begin{align}
	2\zeta_z\frac{d}{d\zeta_z}\ln H^{-1}\left(\frac{\zeta_z}{\mu^2}\right)={\cal G}\left(\frac{\zeta_z}{\mu^2}\right) \ .
\end{align}
Again, we can see that the anomalous dimension of ${\cal G}\left(\frac{\zeta_z}{\mu^2}\right)$ is $\Gamma_{\rm cusp}(\alpha_s)$.

It is convenient to write $H$ in the exponential form, $H=e^{-h}$. Collecting all the above results, one obtains at one-loop level~\cite{Ji:2018hvs,Ebert:2019okf}
\begin{align}
	h^{(1)}\left(\frac{\zeta_z}{\mu^2}
	\right)=\frac{\alpha_sC_F}{2\pi}\left(-2+\frac{\pi^2}{12}-\frac{L_z^2}{2}+L_z\right) \ .
\end{align}
Similar as before, the two loop contribution $h^{(2)}$ is predicted to be
\begin{align}
	&h^{(2)}\left(\frac{\zeta_z}{\mu^2}\right)=c'_2-\frac{1}{2}\left(\gamma^{(2)}_C-\alpha_s^2\beta_0 c_1\right)\ln\frac{\zeta_z}{\mu^2}\\
	&-\frac{1}{4}\left(\Gamma^{(2)}_{\rm cusp}-\frac{\alpha_s^2\beta_0 C_F}{2\pi}\right)\ln^2\frac{\zeta_z}{\mu^2}-\frac{\alpha_s^2\beta_0C_F}{24\pi}\ln^3\frac{\zeta_z}{\mu^2} \ ,\nn
\end{align}
where $c_1=\frac{C_F}{2\pi}\left(-2+\frac{\pi^2}{12}\right)$ and $c'_2$ is again a constant to be determined in perturbation theory at two-loop level.

Finally, we compare the current formulation with previous approaches to lattice TMDPDF. First, we comment on the developments in~\cite{Hagler:2009mb,Musch:2010ka,Musch:2011er,Engelhardt:2015xja,Yoon:2017qzo} in which the $x$-moments of TMDPDF are extracted from ratio of quasi-TMDPDF.  From Eq.~(\ref{eq:quasiTMDmatch}), it is clear that both the matching kernel $H$ and the exponential factor of Collins-Soper kernel depends on $x$ non-trivially. Therefore, simply taking the ratio of moments for quasi-TMDPDF will not be sufficient to reproduce the same ratio for TMDPDF, although the soft function does cancel. This observation is also made recently in Ref.~\cite{Ebert:2020gxr}. Second, the quasi-TMDPDF defined with the naive rectangle-shaped soft function, i.e. $Z_E$, is $\tilde f$ in \eq{quasi_TMD}, so it is obvious that it still needs the reduced soft function $S_r$ to be matched to $f^{\rm TMD}$. As for the other proposal in~\cite{Ji:2014hxa,Ebert:2019okf}, it replaces $Z_E$ in $\tilde f$ with $S_{\rm bent}$ which is the vacuum matrix element of a spacelike bent-shaped Wilson loop with angle $\pi/2$ at each junction, and does not include the function $S^{1\over2}_r$ in \eq{quasiTMDmatch}. Although $\sqrt{S_{\rm bent}/Z_E}$ agrees with $S_r^{-1\over2}$ at one-loop order~\cite{Ji:2018hvs,Ebert:2019okf}, it is expected to be different at higher orders. In fact, for the anomalous dimension $\Gamma_{\pi\over2}$ defined through
\begin{align}
	\Gamma_{\frac{\pi}{2}}(\alpha_s) \equiv \mu^2\frac{d}{d\mu^2} \ln \left(\frac{S_{\rm bent}(L,b_\perp,\mu)}{Z_E(2L,b_\perp,\mu)}\right) \,,
\end{align}
it starts to deviate from $\Gamma_S(\alpha_s)$ at two-loop order~\cite{Grozin:2015kna}, as
\begin{align}
	-\Gamma_S(\alpha_s)&=\frac{\alpha_sC_F}{\pi} \\
	&+\frac{\alpha_s^2}{\pi^2}\left[C_FC_A\left(-\frac{49}{36}+\frac{\pi^2}{12}-\frac{\zeta_3}{2}\right)+C_FN_F\frac{5}{18}\right]  \,, \nn \\
	\Gamma_{\frac{\pi}{2}}(\alpha_s)&=\frac{\alpha_sC_F}{\pi} \\
	&+\frac{\alpha_s^2}{\pi^2}\left[C_FC_A\left(-\frac{49}{36}+\frac{\pi^2}{24}\right)+C_FN_F\frac{5}{18}\right] \,. \nn
\end{align}
In the equation, $\zeta_3=\sum_{n=1}^{\infty}(1/n^3) \ne \pi^2/12$, therefore the two anomalous dimensions are different. The differences in the anomalous dimension will result in different logarithmic behaviors in $b_\perp$, as the soft functions are dimensionless and depend on $b_\perp$ and $\mu$ only. At large $b_\perp$, it will lead to different IR physics that cannot be controlled by perturbation theory.

Combining the reduced soft function and the quasi-TMDPDF, one can effectively factorize the DY cross section,
\begin{align}
	&\sigma = \int dx_A dx_B d^2b_\perp e^{i\vec{Q}_\perp \cdot \vec{b}_\perp}\hat \sigma(x_A,x_B,Q^2,\mu)\nonumber \\ &\times \tilde f(x_A,b_\perp,\mu,\zeta_A) \tilde f(x_B,b_\perp,\mu,\zeta_B) S_r(b_\perp,\mu) \ .
\end{align}
where all non-perturbative quantities do not involve the light cone,
and can be calculated on lattice.

Spin-dependent TMDPDFs are also physically important. They can be computed
using LaMET theory~\cite{Ebert:2020gxr}. Again one can define quasi distributions just like the spin-independent ones.
For a general proton target $|PS\rangle$ and the general spin structure $\Gamma$ of the parton, the parent TMDPDF can be defined as :
\begin{align}\label{eq:parent_TMD}
	&f_{[\Gamma]}^{\rm TMD}(x,\vec{k}_\perp,\mu,\zeta)=\frac{1}{2P^+}\int\frac{d\lambda}{2\pi}\int\frac{d^2 \vec{b}_\perp}{(2\pi)^2}e^{-i\lambda x + i\vec{k}_\perp\cdot \vec{b}_\perp}\nonumber\\
	&\times \lim_{\delta^- \rightarrow 0}\frac{\langle PS|\bar\psi(\lambda n+\vec{b}_\perp)\Gamma\, {\cal W}_n(\lambda n +\vec{b}_\perp)|_{\delta^-}\psi(0)|PS\rangle}{\sqrt{S(b_\perp,\mu,\delta^-e^{2y_n},\delta^-)}} \,,
\end{align}
where the $\zeta=2(P^+)^2e^{2y_n}$ is the rapidity scale, see \sec{tmd} for more detail of the soft function subtraction. The individual spin-dependent TMD distributions can then be obtained through Lorentz decompositions ~\cite{Tangerman:1994eh,Ralston:1979ys,Mulders:1995dh}:
\begin{align}\label{eq:TMDPDF}
	f_{[\gamma^+]}^{\rm TMD}&=f_1-\frac{\epsilon^{ij}k^i S^j_\perp}{M}f_{1T}^\perp   \,,\\
	f_{[\gamma^+\gamma_5]}^{\rm TMD}&=S^+ g_1+\frac{\vec{k}_\perp\cdot \vec{S}_\perp}{M}g_{1T} \,, \\
	f_{[i\sigma^{i+}\gamma_5]}^{\rm TMD}&=S_\perp^i h_1+\frac{(2k^i k^j-\vec{k}_\perp^2\delta^{ij})S^j_\perp}{2M^2}h_{1T}^\perp\nn\\
	&\qquad+\frac{S^+ k^i}{M}h_{1L}^\perp+\frac{\epsilon^{ij} k^j}{M}h_1^\perp \,,
\end{align}
where we suppress the arguments $(x,\vec{k}_\perp,\mu,\zeta)$ in all distributions;
$f_1$, $g_1$, and $h_1$ are unpolarized, helicity and transversity TMDPDFs, respectively;
the indices $i$ and $j$ are in transverse space of $\vec{k}_\perp$; $S^+$ and $S^i_\perp$ are longitudinal and transverse spin components.

Note that the Sivers function $f^\perp_{1T}$~\cite{Sivers:1989cc} and the Boer-Mulders function $h^\perp_1$~\cite{Boer:1997nt} are $T$-odd. The orientation of the gauge-link have important effects on these two functions~\cite{Collins:2002kn,Collins:2011zzd}, such that they change sign between the DY and SIDIS processes.  In the light-cone gauge, these contributions arise from the transversal gauge-link at infinities~\cite{Belitsky:2002sm}. They are related to the phenomenologically interesting single transverse-spin asymmetry~\cite{Boer:1997nt,Efremov:2004tp,Collins:2005ie,Collins:2005wb}.

\subsection{Off-light-cone Soft Function}
\label{sec:offlight}
In previous subsections, the soft function has been introduced to define rapidity-scheme-independent
TMDPDFs. The major motivation of introducing the soft function is to capture nonperturbative effects due to
soft-gluon radiations from fast moving color-charges.
For many inclusive processes the soft radiations cancel in the total cross section, but for certain processes where a small transverse momentum is measured, such cancellation can be incomplete and result in measurable consequences.
In such cases, the TMD soft function is introduced to account for the soft-gluon effects
and appears in factorization theorems for the Drell-Yan (DY) process~\cite{Collins:1984kg,Collins:1988ig}
and semi-inclusive DIS (SIDIS)~\cite{Ji:2004wu,Ji:2004xq}.

To calculate the TMD physics nonperturbatively, formulating
a Euclidean version of the soft function is critical. Since the soft function in fact is a cross section and hence real and positive,
it satisfies the necessary condition for a Monte Carlo simulation. In this subsection, we present an approach to calculate it
in heavy-quark effective theory (HQET)~\cite{Ji:2019sxk}. There is also another method proposed to extract the reduced soft function $S_r$ from a light-meson form factor~\cite{Ji:2019sxk}, where many subtleties of HQET can be avoided. The first lattice calculation of the reduced soft function based on the light-meson formalism has been performed in the recent work~\cite{Zhang:2020dbb}.

Due to the different space-time pictures of the DY and SIDIS processes, the soft functions for the two processes also differ from each other as shown in Fig.~\ref{Fig:soft1}. To define the soft function, one also needs to specify a time-ordering prescription. Since it is a cross section, it involves a time order and an anti-time order (or cut diagrams). However, in the light-cone limit, the time order does not matter.
What really matters is the rapidity regularization scheme. It has been proven for the $\delta$ regulator in ~\cite{Vladimirov:2017ksc} that the time ordering is not quite relevant up to overall phase factors, and the soft functions for the two processes are equal. The method therein can be modified to apply to the off-light-cone scheme too. Therefore, our first step is to convert the cut-diagrams into Feynman diagrams by imposing just
the single time order. In this way, the soft function can be viewed as a scattering amplitude.

In the off-light-cone scheme, there are further complications caused by the space-like or
time-like choices for off-light-cone vectors. Fortunately, one can show that in the light-cone limit, the space-like and time-like choices
are equivalent up to overall phase factors~\cite{future}. Thus we will use the notation $S(b_\perp,\mu,Y,Y')$ to denote a generic off-light-cone soft function that satisfies our demands.

With these in mind, we show that the off-light-cone soft function $S(b_\perp,\mu,Y,Y')$ is equivalent to an equal-time form factor of fast-moving color sources and can be formulated on the Euclidean lattice. From the matching formula Eq.~(\ref{eq:quasiTMDmatch}) in the last subsection, once the off-light-cone soft function is known, we can combine it with the lattice calculated quasi-TMDPDF to obtain the physical TMDPDF. Therefore, the cross section of DY processes in the low transverse-momentum region~\cite{Collins:1984kg} becomes predictable from first principles~\cite{Ji:2019sxk}.

\begin{figure}[htb]
	\centering
	\includegraphics[width=0.7\columnwidth]{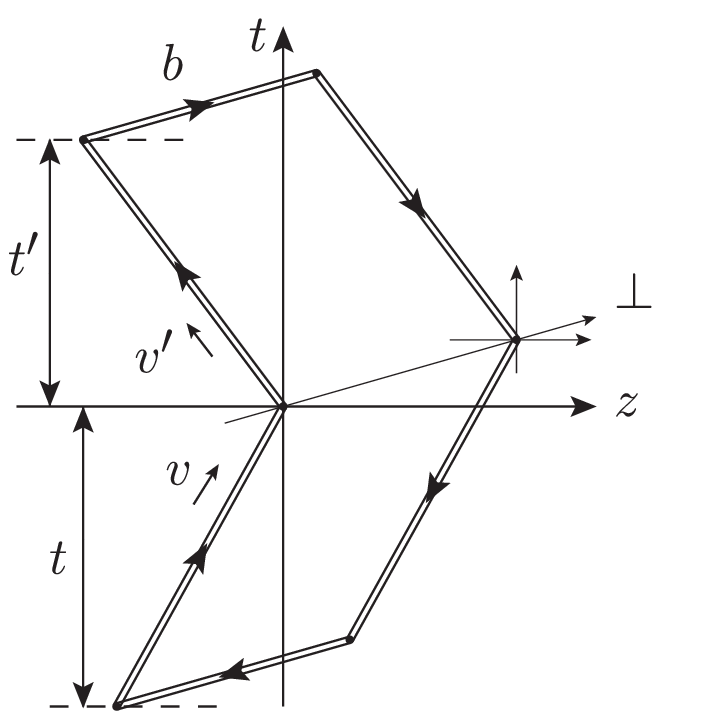}
	\caption{The Wilson-loop ${\cal W}$ showing a pair of quark and antiquark scattering at $t=0$.}
	\label{fig:S}
\end{figure}

To begin with, we define the scattering amplitude of a Wilson loop as shown in Fig.~\ref{fig:S}:
\begin{align}\label{eq:W_t}
	&W(t,t',b_\perp,Y,Y')\nonumber \\
	&=\frac{1}{N_c}{\rm Tr\,}\langle0|{\cal T} \left[{\cal W}^{\dagger}_{v'}(\vec{b}_\perp,t'){\cal W}_{v}(\vec{b}_\perp,t)\right]|0\rangle
\end{align}
where $|0\rangle$ is the QCD vacuum state and $N_c$ is number of colors and ${\rm Tr}$ is the color-trace.
Timelike four-vectors $v^\mu=\gamma(1,\beta,\vec 0_\perp)$ and $v'^\mu=\gamma'(1,-\beta',\vec 0_\perp)$ approach the lightcone as $\beta$ and $\beta' \to 1$. The rapidity $Y$ and the speed $\beta$ are related through $\beta = {\rm tanh} Y$, in terms of the light-cone vectors $p$ and $n$, the velocities read $v=\frac{e^Y}{\sqrt{2}}\left(\frac{p}{p^+}+e^{-2Y}p^+n \right)$ and $v'=\frac{e^Y}{\sqrt{2}}\left(e^{-2Y}\frac{p}{p^+}+ p^+ n \right)$. The ${\cal W}_{v}(\vec{b}_\perp,t)$ is a staple-shaped gauge-link along $v$ direction similar to those defined in Eqs.~(\ref{eq:staplep}) and (\ref{eq:staplez}). $t$ and $t'$ are the lengths of the $t$-components of the staples.
The single time-order prescription for $S$ allows physical interpretation as a chronological process.
Similar to the quasi-TMDPDF, the Wilson-loop in Eq.~(\ref{eq:W_t}) contains pinch-pole singularities associated to time evolution of initial and final states at large $t$ and $t'$.
Therefore, we need to subtract them out in Eq.~(\ref{eq:W_t}) with rectangular Wilson-loops~\cite{Collins:2008ht,Ji:2018hvs}.
This leads to an off-light-cone realization of the soft function:
\begin{align}\label{eq:S_t}
	& S(b_\perp,\mu,Y,Y')\nonumber \\
	&=\lim_{\substack{t\to\infty\\t'\to\infty}}\frac{W(t,t',b_\perp,\mu,Y,Y')}{\sqrt{Z(2t,b_\perp,\mu,Y)Z(2t',b_\perp,\mu,Y')}} \ ,
\end{align}
where $Z(2t,b_\perp,Y)$ is the vacuum expectation value of rectangular Wilson loop which is similar to $W$ by setting $v'=v$ and $t'=t$, i.e. $Z(2t,b_\perp,Y)=W(t,t,b_\perp,Y,-Y)$.
The factor $Z$ has a clear physical interpretation: It can be viewed as the wave function renormalization for incoming or outgoing color sources.
After the subtraction through $Z$, the only remaining UV divergences for $S(b_\perp,\mu,Y,Y')$ are the cusp divergences with hyperbolic angle $Y+Y'$.

We should mention that a more common definition of the soft function $S_{\rm DY}(b_\perp,\mu,Y,Y')$ for the DY process was proposed in~\cite{Collins:2011ca,Collins:2011zzd}. The space-like vectors $u^\mu=\gamma(\beta,1,0,0)$ and $u'^\mu=\gamma'(-\beta',1,0,0)$ were chosen instead of time-like $v $ and $v'$ to define the soft function for the DY process. This soft function has already been defined in the last subsection in Eq.~(\ref{eq:S_DY}). $u$ and $u'$ are equal to $p_Y$, $n_{Y'}$ up to overall normalization factors.

While $S$ and $S_{\rm DY}$ are defined differently, we can show that
\begin{align}\label{eq:S_t=S}
	S(b_\perp,\mu,Y,Y')=S_{\rm DY}(b_\perp,\mu,Y,Y')
\end{align}
using analyticity property~\cite{future}. Here we focus on $S$ in Eq.~(\ref{eq:S_t}), which has a simple Euclidean realization.

After defining the soft function $S$, we now show that it is equal to a form factor. In HQET, the propagator of a color source is equivalent to a gauge-link along its moving direction.
Thus $W(t,t',b_\perp,\mu,Y,Y')$ can be expressed by fields in HQET with the Lagrangian
\begin{align}
	{\cal L_{\rm HQET}}=Q_v^{\dagger}(x)(iv \cdot D)Q_v(x)+ \bar Q_v^{\dagger}(x)(iv \cdot D)\bar Q_v(x) \ ,
\end{align}
where $Q_v$ and $\bar Q_v$ are quark and anti-quark in the fundamental and anti-fundamental representations, respectively; $v^\mu=\gamma(1,\beta,\vec 0_\perp)$ is the four velocity; $D$ is the covariant derivative.
Note that quarks in HQET can be viewed as color sources.
If the gluon soft function is considered, the heavy quarks should be in the adjoint representation.

In HQET, a color-singlet heavy-quark pair separated by $\vec b$ generates a heavy quark potential $V(|\vec b|)$ in the ground state, and the spectrum includes a gapped continuum above it.
The state can also have a residual momentum $\delta\vec P$, which is arbitrary due to reparameterization invariance~\cite{Luke:1992cs,Manohar:2000dt}, and for simplicity we always consider $\delta\vec P=0$.
When the sources move with a velocity $v$, the ground state can be labeled by
$|\overline Q Q,\vec{b},\delta \vec P\rangle_v$, where the residue momentum $\delta\vec P=\vec{P}_{\rm total}-2m_Q\gamma\vec\beta$ is the difference between the total momentum $\vec{P}_{\rm total}$ and the kinetic momentum of the heavy-quarks.
The residual energy of the state is $E=\gamma^{-1}V(|\vec b_{\perp}|)+\vec{\beta}\cdot \delta\vec P$.

Consider a process with incoming and outgoing states being heavy-quark pairs separated by ${\vec b}_\perp$ and at velocity $v$ and $v'$, respectively.
Such a state is created by the interpolating fields
\begin{align}
	{\cal O}_v(t,{\vec b}_\perp)=\int d^3\vec r \, Q_v^{\dagger}(t,\vec r\,) {\cal U}(\vec r,\vec r\,',t) \bar Q_v^{\dagger}(t,\vec r\,')  \ ,
\end{align}
where $\vec r\,'=\vec r+\vec b_\perp$ and ${\cal U}(\vec r,\vec r\,',t)$ is a gauge-link connecting $\vec r\,'$ to $\vec r$ at time $t$.
The heavy-quark pair created by ${\cal O}_v$ is forced to be at relative separation $\vec{b}_\perp$ and to have vanishing residual momentum $\delta \vec P=0$.
Between the incoming and outgoing states, a product of two local equal-time operators
\begin{align}
	J(v,v',\vec{b}_\perp)= \bar Q^\dagger_{v'}(\vec{b}_\perp)\bar Q_v(\vec{b}_\perp)Q_{v'}^\dagger(0)Q_v(0)
\end{align}
is inserted at $t=0$.
Then $W$ can be expressed in terms of HQET propagators which are gauge-links in the $v,v'$ directions.
After integrating out the heavy-quark fields, we obtain up to an overall volume factor~\cite{Ji:2019sxk}
\begin{align}\label{eq:W_HQ}
	&W(t,t',b_\perp,\mu,Y,Y')  \\
	&=\frac{1}{N_c} \langle0|{\cal O}^{\dagger}_{v'}(t',{\vec b}_\perp) J(v,v',\vec{b}_\perp) {\cal O}_v(-t,\vec{b}_\perp)|0\rangle \nonumber\\
	&\xrightarrow[\substack{t\to\infty\\t'\to\infty}]{} \frac{1}{N_c}\Phi^\dagger(b_\perp,\mu) S(b_\perp,\mu,Y,Y') \Phi(b_\perp,\mu)e^{-iE't'-iEt}  \ , \nn
\end{align}
where
\begin{align}\label{eq:S_HQ}
	\Phi(b_\perp,\mu)&=\lim_{T\to\infty}{}_v\langle \overline QQ,\vec{b}_\perp|{\cal O}_v(T,\vec{b}_\perp)|0\rangle\,,\\
	S(b_\perp,\mu,Y,Y')&={{}_{v'}}\langle\overline QQ,\vec{b}_\perp|J(v,v',\vec{b}_\perp)|\overline QQ,\vec{b}_\perp\rangle_v \ . \nn
\end{align}
In the last line of Eq.~(\ref{eq:W_HQ}), we have inserted a complete set of heavy-quark pair states before and after $J$.
At large $t$ and $t'$, the contribution from the continuum spectrum is damped out due to the Riemann-Lebesgue lemma~\cite{zuazo2001fourier}, while the contribution from $|\overline QQ,\vec{b}_\perp,\delta \vec{P}=0\rangle_v$ with residual energy $E=\gamma^{-1}V(|\vec b_\perp|)$ survives.
As a result we obtain Eqs.~(\ref{eq:W_HQ})---(\ref{eq:S_HQ}), where we have omitted the state label $\delta \vec{P}=0$ for simplicity.
Alternatively, we can also give $t$ and $t'$ a small negative imaginary part, which is consistent with the time order, to damp out all states except $|\overline QQ,\vec{b}_\perp\rangle_v$ at large $t$ and $t'$.
Note that $\Phi(\vec{b}_\perp,\mu)$ is independent of $Y$ because it is boost invariant.

Similarly, $Z$ can also be formulated in HQET as
\begin{align}\label{eq:Z_HQ}
	Z(2t,b_\perp,Y)&=\frac{1}{N_c}\langle0|{\cal O}^\dagger_{v'}(t,{\vec b}_\perp){\cal O}_v(-t,\vec{b}_\perp)|0\rangle \nonumber\\
	&\xrightarrow[\substack{t\to\infty}]{}\frac{1}{N_c}\Phi^\dagger({\vec b}_\perp,\mu)\Phi({\vec b}_\perp,\mu)e^{-2iEt}\,,
\end{align}
whose $t$-component has length $2t$.
The $Y$ dependence of $Z$ is implicit in the energy $E$.
Combining Eqs.~(\ref{eq:W_HQ}) and (\ref{eq:Z_HQ}), we obtain $S$ defined in Eq.~(\ref{eq:S_t}). We emphasize that Eq.~(\ref{eq:S_t}) can be seen as a LSZ reduction formula, in which we amputate the external heavy-quark pair states $|\overline QQ,{\vec b}_\perp\rangle_v$.

Being an equal-time observable, $S(b_\perp,\mu,Y,Y')$ can be straightforwardly realized in Euclidean time as:
\begin{align}\label{eq:S_lattice}
	&S(b_\perp,\mu,Y,Y') \\
	&=\lim_{\substack{T\to\infty\\T'\to\infty}}\frac{W_E(T,T',b_\perp,\mu,Y,Y')}{\sqrt{Z_E(2T,b_\perp,\mu,Y)Z_E(2T',b_\perp,\mu,Y')}}  \ , \nn
\end{align}
where the subscript $E$ indicates the quantity is defined in Euclidean time, with corresponding
variables $T$ and $T'$. Due to boost invariance, the factor $Z_E(T,b_\perp,\mu,Y)$ relates to the rectangular Wilson-loop defined in Eq.~(\ref{eq:Z_E}) along the $n_z$ direction through the relation $Z_E(2T,b_\perp,\mu,Y)=Z_E(2\gamma^{-1}T,b_\perp,0)$.
The relevant matrix elements are now calculated by a lattice version of HQET with the Lagrangian~\cite{Aglietti:1993hf,Hashimoto:1995in,Horgan:2009ti}
\begin{align}\label{eq:L_HQ}
	&{\cal L}_{{\rm HQET}}^E  \\
	&=Q_v^\dagger(x) (i\tilde v\cdot D_E)Q_v(x)+\bar Q_v^\dagger(x) (i\tilde v\cdot D_E)\bar Q_v(x) \ , \nn
\end{align}
where the subscript $E$ denotes the Euclidean space, $i\tilde v\cdot D_E=\gamma(D^\tau-i\beta )D^z$ with $\tilde v^\mu=\gamma(-i,-\beta,\vec 0_\perp)$. We have explicitly verified Eq.~(\ref{eq:S_lattice}) in Euclidean perturbation theory to the one-loop order.

The soft function cannot be calculated on the lattice by simply replacing the Minkowskian gauge-links in Eq.~(\ref{eq:W_t}) with a finite number of Euclidean gauge-links. Through HQET, we find a time-independent formulation of the soft function, which opens up the possibility of direct lattice calculations.

\subsection{Light-Front Wave-Function Amplitudes And Soft Function from Meson Form Factor}
\label{sec:others}

Light-front quantization (LFQ) or formalism is a natural language for parton physics
in which partons are made manifest at all stages of calculations. It
favors a Hamiltonian approach to QCD like for a non-relativistic quantum
mechanical system, i.e., to diagonalize the Hamiltonian
\begin{equation}
	\hat P^-|\Psi_n\rangle = \frac{M^2_n+{\vec P_\perp}^2}{2P^+}|\Psi_n\rangle \ ,
\end{equation}
to obtain wave functions for the QCD bound states~\cite{Brodsky:1997de}.
The light-front wave functions (LFWFs) thus obtained can, in principle, be used to calculate
all the partonic densities and correlations functions. Moreover, like in
condensed matter systems, knowing quantum many-body wave-functions
allows one to understand interesting aspects of quantum coherence and
entanglement, as well as the fundamental nature of quantum systems.
Therefore, a practical realization of light-front quantization program clearly would be a big step
forward in understanding the fundamental structure of the proton.

However, from a field theory point of view, wave functions are not the most natural
objects to consider due to the non-trivial vacuum, UV divergences as well as the requirement of Lorentz symmetry, according to which the space and time should be treated on equal footing. The proton or other hadrons are
excitations of the QCD vacuum which by itself is very complicated
because of the well-known phenomena of chiral symmetry breaking and color confinement.
To build a proton on top of this vacuum, one naturally has a question of what part of the
wave-function reflects the property of the proton and what reflects the vacuum:
It is the difference that yields the properties
of the proton that are experimentally measurable. There is no clean way to make
this separation unless one builds the proton out of elementary excitations or quasi-particles that
do not exist in the vacuum, as often done in condensed matter systems.

The partons in the IMF avoids the above problems to a certain extent.
In fact, due to the kinematic effects, in the IMF all partons in
the vacuum have longitudinal momentum $k^+=0$, and to some degree of accuracy,
the proton is made of partons with $k^+\ne 0$. This natural separation of
degrees of freedom (DOF) is particularly welcome, making
a wave-function description of the proton more natural and interesting in IMF
than in any other frame.

To implement the above DOF separation, one possibility is to
assume triviality of the light-front vacuum. The question that to what extent this holds
has been continuously debated over the years.
One knows {\it a priori} that in relativistic QFT, the vacuum state is
boost invariant and frame-independent. In fact, it was proven in~\cite{Nakanishi:1976yx,Nakanishi:1976vf}
that not only the vacuum can not be trivial, even the Green's functions of the full theory cannot
pose generic meaningful restrictions to the null-planes $\xi^+=c$.  In fact, the vacuum
zero modes do contain non-trivial dynamics and contribute to the properties of the proton~\cite{Ji:2020baz}.
Nevertheless, one can adopt an effective theory point of view to simply cut off the zero-modes
and relegate their physics to renormalization constants. In some simple cases, these zero-modes can be
treated explicitly~\cite{Heinzl:1991myy,Yamawaki:1998cy}.

By imposing an IR cut-off on the $k^+\ge \epsilon$ in the effective Hilbert space,
all physics below $k^+=\epsilon$ are taken into account through
renormalization constants. We then obtain an effective LF theory with trivial
vacuum,
\begin{equation}
	a_{k\lambda}|0\rangle = b_{p\sigma}|0\rangle = d_{p\sigma}|0\rangle=0 \ .
\end{equation}
where $|0\rangle $ is the vacuum of LFQ.
Therefore, the proton can be expanded in terms of
the superposition of Fock states in the LF gauge $A^+=0$~\cite{Brodsky:1997de},
\begin{align}
	|P\rangle=\sum_{n=1}^{\infty} \int d\Gamma_n \psi^0_n(x_i,\vec{k}_{i\perp})\prod a^{\dagger}_i(x_i, \vec{k}_{i\perp})|0\rangle \ .
\end{align}
where $a^\dagger$ are generic quarks and gluon quanta on the light-front, the phase-space integral reads $d\Gamma_n=\prod \frac{dk^+d^2k_\perp}{2k^+(2\pi)^3}$. The $\psi_n(x_i,\vec{k}_{i\perp})$
are LFWF amplitudes or simply WF amplitudes, where $x_i$ to denote the set of momentum fractions from $x_1$ to $x_n$.
They are a complete set of non-perturbative quantities which describe the partonic
landscape of the proton. The above amplitudes can in principle be calculated through Hamiltonian
diagonalization. However, as explained in Sec.~\ref{sec:lamet-frame}, a direct systematic
solution in LFQ is impractical.

LaMET offers an alternate route to calculate these WF amplitudes. Thanks to the triviality
of the vacuum after the truncation $k^+\ge \epsilon$, they can then be written in terms of the
invariant matrix elements by inverting the above expansion,
\begin{align}
	\psi^0_n(x_i,\vec{k}_{i\perp})=\langle 0|\prod a_i(x_i,\vec{k}_{i\perp})|P\rangle \ .
\end{align}
After properly restoring gauge-invariance and imposing regularizations, they
become the matrix elements of light-cone correlators, the same type as those in the
TMDPDFs. Therefore, the LaMET method applies to them, which allows one to effectively
obtain the results of light-front quantization through instant quantization in a large momentum frame.

To realize the goal, the LFWF amplitudes also need a rapidity renormalization, as in the case of TMDPDFs.
In this section, we explain how the reduced soft function $S_r$ can be obtained by combining the LFWF amplitudes
and a special light-meson form factor, instead of as the form factor in HQET discussed in the previous section. A lattice calculation based on the light-meson framework has been performed in~\cite{Zhang:2020dbb}.

Let us consider the following form factor of a pseudoscalar light-meson state with constituents $\overline\psi\eta$,
\begin{align}
	F(b_\perp,P,P',\mu)=\langle P'|\overline\eta(\vec b_\perp)\Gamma'\eta(\vec b_\perp) \overline\psi(0)\Gamma\psi(0)|P\rangle
\end{align}
where $\psi$ and $\eta$ are light quark fields of different flavors; $P^\mu=(P^t,0,0,P^z)$ and $P'^\mu=(P^t,0,0,-P^z)$ are two large momenta which approach two opposite light-like directions in the limit $P^z\to\infty$;
$\Gamma$ and $\Gamma'$ are Dirac gamma matrices, which can be chosen as $\Gamma=\Gamma'=1$, $\gamma_5$ or
$\gamma_\perp$ and $\gamma_\perp\gamma_5$, so that the quark fields have leading components
on the respective light-cones.

\begin{figure}
	\includegraphics[width=0.7\columnwidth,angle=90]{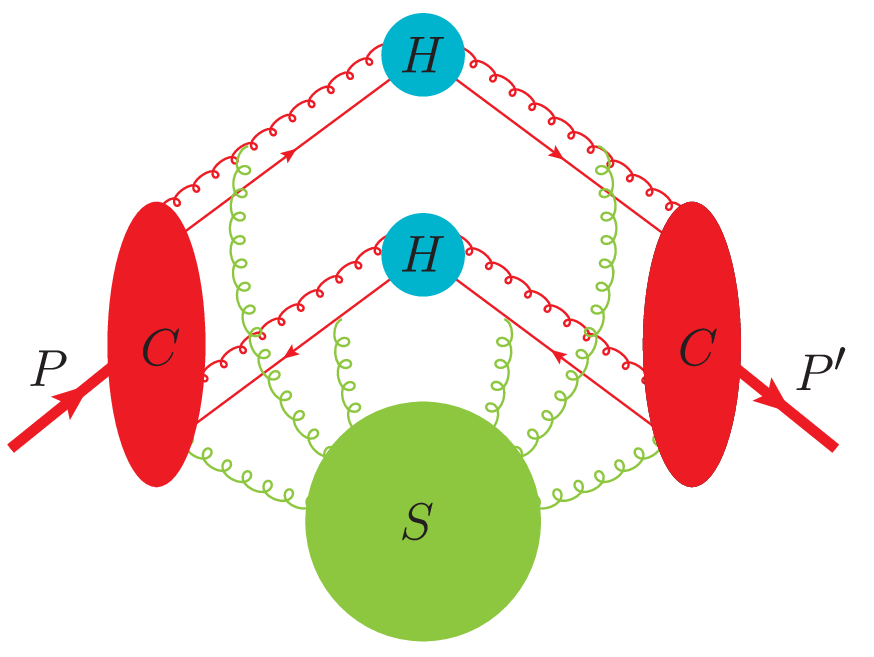}
	\caption{\label{fig:reduced_form} The reduced diagram for the large-momentum form factor $F$ of
		a meson. Two $H$ denote the two hard cores separated in space by $\vec{b}_\perp$, $C$ are collinear sub-diagrams
		and $S$ denotes the soft sub-diagram.}
\end{figure}

At large momentum, the form factor factorizes through TMD factorization into LFWF amplitudes. To motivate the factorization,
we need to consider the leading region of IR divergences in a similar way for SIDIS and Drell-Yan~\cite{Ji:2004wu,Collins:2011ca},
and the result is shown in Fig.~\ref{fig:reduced_form}.
There are two collinear sub-diagrams responsible for collinear modes in $+$ and $-$ directions, and a soft sub-diagram responsible for soft
contributions. Besides, there are two IR-free hard cores localized around $(0,0,0,0)$ and $(0,\vec{b}_\perp,0)$. In the covariant gauge, there are arbitrary numbers of longitudinally-polarized collinear and soft gluons that can connect to the hard and collinear sub-diagrams.
Based on the region decomposition, we now follow the standard procedure to make factorization into
LF quantities~\cite{Collins:2011ca}.

We first factorize the soft divergences. This can be done with the soft function $S(b_\perp,\mu,\delta^+,\delta^-)$. It re-sums the soft gluon radiations from fast-moving color-charges. Intuitively, soft gluons have no impact on the velocity of the fast-moving color charged partons, and the propagators of partons eikonalize to straight gauge links along their moving trajectory.

We then factorize the collinear divergences. For the incoming direction, the collinear divergences is captured by the LFWF amplitude for the incoming parton $\psi_{\bar q q}(x,b_\perp,\mu,\delta^{'-})$ defined with future-pointing gauge-links.
\begin{align}
	&\psi_{\bar q q}(x,b_\perp,\mu,\delta^{'-})=\int\frac{d\lambda}{4\pi}e^{-ix\lambda}\\
	&\langle 0|\bar \psi(\lambda n/2 +\vec{b}_\perp)\gamma^+{\cal W}_n(\lambda n/2+\vec{b}_\perp)|_{\delta^{'-}}\psi(-\lambda n/2)|P\rangle \ ,\nn
\end{align}
where the staple-shaped gauge-link $W_n$ is defined similar to that in Eq.~(\ref{eq:staplen}), the only exception being the gauge-links $W_n$ should point to $+\infty$ instead of  $-\infty$.

However, the naive LFWF amplitude contains soft divergences as well, to avoid double-counting, we must subtract out the soft contribution from the bared collinear WF amplitude with the soft function $S(b_\perp,\mu,\delta^+,\delta^{'-})$. This leads to the collinear function for the incoming direction: $\psi_{\bar q q}(x,b_\perp,\mu,\delta^{'-})/S(b_\perp,\mu,\delta^+,\delta^{'-})$. Similarly, for the out-going direction one obtains the collinear function $\psi^{\dagger}(x',b_\perp,\mu,\delta^{'+})/S(b_\perp,\mu,\delta^{'+},\delta^{-})$.

Here we briefly comment on the choices for the gauge-link directions in the soft functions and the WF amplitudes. Naively, the gauge-links along the $p$ direction have to be past-pointing. However, similar to the arguments in~\cite{Collins:2004nx} for the SIDIS process, based on the space-time picture of collinear divergences, one can chose future pointing gauge-links along $p$ direction as well. With all the gauge-links being future pointing, the soft function equals to $S^-$ which is manifestly real, and the WF amplitudes for the incoming and outgoing hadrons are in complex conjugation to each other.

Besides the collinear and soft functions, we still need the hard core $H_F(Q^2,\bar Q^2,\mu^2)$ where $Q^2=xx'P\cdot P'$, $\bar Q^2=\bar x\bar x' P\cdot P'$ and an integral over the momentum fractions $x$,$x'$ is assumed. Taking together, we have the TMD factorization of the form factor into hard, collinear and soft functions:
\begin{align}\label{eq:form_fac_bare}
	&F(b_\perp,P,P',\mu)=\int dx dx' H_F(Q^2,\bar Q^2,\mu^2) \\ &\times \left[\frac{\psi_{\bar q q}^{\dagger}(x',b_\perp,\mu,\delta^{'+})}{S(b_\perp,\mu,\delta^{'+},\delta^-)}\right]\left[\frac{\psi_{\bar q q}(x,b_\perp,\mu,\delta^{'-})}{S(b_\perp,\mu,\delta^+,\delta^{'-})}\right]\nonumber \\
	&\times S(b_\perp,\mu,\delta^+,\delta^-) \ . \nn
\end{align}
All the rapidity regulators in all the WF amplitudes and the soft functions are cancelled.

Let us consider a one-loop example. The incoming hadron state consists of a free quark with momentum $x_0P^+$ and a free anti-quark with momentum $\bar x_0 P^+$. Similarly the outgoing state consists of a pair of free quark and anti-quark with momentum $x'_0P^{'-}$, $\bar x'_0 P^{'-}$, respectively. The spin projection operator for the incoming state is proportional to $\gamma^5 \gamma^-$ and for the out-going state is proportional to $\gamma^5 \gamma^+$. The tree level form factor is normalized to 1. At one-loop level, the pseudo-scalar form factor with vector currents $\Gamma=\gamma^{\mu}$, $\Gamma'=\gamma_{\mu}$ where a summation over $\mu$ is assumed reads:
\begin{align}
	F(b_\perp,P,P',\mu)=1+\frac{\alpha_s C_F}{2\pi}F^{(1)}(b_\perp,Q^2,\bar Q^2,\mu^2)\ ,
\end{align}
where $Q^2=2x_0x_0'P^+P^{'-}$, $\bar Q^2=2\bar x_0\bar x_0'P^+P^{'-}$ and
\begin{align}
	&F^{(1)}(b_\perp,Q^2,\bar Q^2,\mu^2) \\
	&=-7+\left(-\frac{1}{2}\ln^2 b^2_\perp Q^2+\frac{3}{2}\ln b^2_\perp Q^2+\left(Q \rightarrow \bar Q\right)\right) \ . \nn
\end{align}
This result can be obtained from the one-loop DY structure function~\cite{DAlesio:2014mrz} using the substitution $\ln^2(-Q^2b^2_\perp)\rightarrow \frac{1}{2}\ln^2 Q^2b^2_\perp+\ln^2 \bar Q^2b^2_\perp$ and $\ln(-Q^2b^2_\perp)\rightarrow \frac{1}{2}\ln Q^2b^2_\perp+\ln \bar Q^2b^2_\perp$. Similar to the TMD factorization for SIDIS and DY process, one should also notice that the hard kernel $H_F(Q^2,\bar Q^2,\mu^2)$ can be obtained from that of the space-like Sudakov form factor:
\begin{align}
	H_F(Q^2,\bar Q^2,\mu^2)=H^{\rm sud}(-Q^2)H^{\rm sud}(-\bar Q^2) \ ,
\end{align}
where $H^{\rm sud}(-Q^2)$ is given in~\cite{Collins:2017oxh}. At one-loop level, we then obtain:
\begin{align}
	& H_F(Q^2,\bar Q^2,\mu^2)=1\nonumber \\
	&+\frac{\alpha_s}{4\pi}\left(-16+\frac{\pi^2}{3}+3L_Q+3L_{\bar Q}-L_Q^2-L_{\bar Q}^2\right)  \ ,
\end{align}
where $L_Q=\ln \frac{Q^2}{\mu^2}$ and $L_{\bar Q}=\frac{\bar Q^2}{\mu^2}$.

Now we construct the Euclidean version of the factorization in terms of the quasi-WF amplitudes,
the reduced soft function, and hard contribution. The quasi-WF amplitudes are defined in a way similar to Eq.~(\ref{eq:quasi_TMD}):
\begin{align}
	&\tilde \psi_{\bar q q}(x,b_\perp,\mu,\zeta_z)=\int \frac{d\lambda}{4\pi}e^{ix\lambda}\nonumber\\
	&\frac{\langle 0| \bar \psi\big(\frac{\lambda n_z }{2}\!+\!\vec{b}_\perp\big)\gamma^z{\cal W}_{z}(\frac{\lambda n_z}{2}\!+\!\vec{b}_\perp;L)\psi\big(\!-\!\frac{\lambda n_z}{2}\big) |P\rangle}{Z_E(2L,b_\perp,\mu)} \ ,
\end{align}
in which the staple-shaped gauge-link ${\cal W}_z$ is defined in Eq.~(\ref{eq:staplez}). The gauge-links should point to $+z$ direction in accordance to the $+\infty$ choice on the light-cone side.

The factorization to the LFWF amplitude follows a similar reasoning to that of the quasi-TMDPDFs presented in previous sections.  Alternatively, we can factorize it using quantities defined in on-light-cone rapidity scheme,
\begin{align}\label{eq:quasifac_bare}
	\tilde \psi_{\bar q q}(x,b_\perp,\mu,\zeta_z)&=H^+_1\left(\zeta_z/\mu^2,\bar \zeta_z/\mu^2\right)\\ &\times \left[\frac{\psi_{\bar q q}(x,b_\perp,\mu,\delta^-)}{S(b_\perp,\mu,\delta^+,\delta^-)}\right]S(b_\perp,\mu,\delta^+) \ . \nn
\end{align}
This factorization is the result of applying a similar leading-region analysis to the quasi-WF amplitude. The $\psi_{\bar q q}(x,b_\perp,\mu,\delta^-)/S(b_\perp,\mu,\delta^+,\delta^-)$ re-sums all the collinear divergences, while the soft function $S(b_\perp,\mu,\delta^+)$ contains an off-light-cone direction along ${n_z}$. It re-sums the soft divergences of the quasi-WF amplitude. The soft functions $S(b_\perp,\mu,\delta^+,\delta^-)$ and $S(b_\perp,\mu,\delta^+)$ subtract away the regulator dependencies introduced in the bare LFWF amplitude. The overall combination in the right-hand side of Eq.~(\ref{eq:quasifac_bare}) is rapidity-scheme independent. Similar to the case of the form factor, we can chose all the gauge-links along the incoming collinear direction to be future-pointing.

Combining together Eqs.~(\ref{eq:form_fac_bare}) and (\ref{eq:quasifac_bare}) and using the relation $\zeta \zeta'= \zeta_z \zeta'_z$,
one obtains the form factor factorization,
\begin{align}
	& F(b_\perp,P,P',\mu)\\
	&= \int dx dx' H(x,x')\tilde \psi_{\bar q q}^{\dagger}(x',b_\perp)\tilde \psi_{\bar q q}(x,b_\perp)S_{r}(b_\perp,\mu)\ ,\nn
\end{align}
where we have only kept the $x,b_\perp$ dependencies of the WF amplitudes with other variables being omitted
, and the hard kernel $H$ is given by:
\begin{align}
	&H(x,x')=H(\zeta_z,\zeta'_z,\bar \zeta_z,\bar \zeta'_z,\mu^2)\nonumber \\
	&\qquad=\frac{H_F(Q^2,\bar Q^2,\mu^2)}{H^+_1\left(\zeta_z/\mu^2,\bar \zeta_z/\mu^2\right)H^+_1\left(\zeta'_z/\mu^2,\bar \zeta'_z/\mu^2\right)} \ ,
\end{align}
where $Q^2=\sqrt{\zeta_z\zeta'_z}$ and $\bar Q^2=\sqrt{\bar \zeta_z\bar\zeta'_z}$. And the reduced soft function is
\begin{align}
	S_{r}(b_\perp,\mu)=\lim_{\delta^+,\delta^-\rightarrow 0}\frac{S(b_\perp,\mu,\delta^+,\delta^-)}{S(b_\perp,\mu,\delta^+)S(b_\perp,\mu,\delta^-)} \ .
\end{align}
It can be shown based on property of off-light-cone soft functions that $S_r$ defined here agrees with the one defined in Eq.~(\ref{eq:reduced}).

Therefore, with non-perturbative quantities $F$ and $\psi^+$, we obtain the reduced soft function,
\begin{equation}
	S_{r}(b_\perp,\mu)=\frac{F(b_\perp,P,P',\mu)}{\int dx dx' H(x,x')\tilde \psi_{\bar q q}^{\dagger}(x',b_\perp)\tilde \psi_{\bar q q}(x,b_\perp)} \ ,
	\label{eq:form_qfac}
\end{equation}
where $H$ can be obtained perturbatively.

Based on the one-loop results for the form factor, the quasi-WF amplitudes and the reduced soft function, the one-loop matching kernel for the vector current can be extracted as:
\begin{align}
	&H(\zeta_z,\zeta'_z,\bar \zeta_z,\bar \zeta'_z,\mu^2)=1+ \frac{\alpha_sC_F}{2}i\ln \frac{\sqrt{\zeta_z\bar\zeta_z}}{\sqrt{\zeta'_z\bar\zeta'_z}}  \\
	&+\frac{\alpha_sC_F}{4\pi}\left(-8+\ln^2\frac{\sqrt{\zeta_z}}{\sqrt{\zeta'_z}}+\ln \frac{\sqrt{\zeta_z\zeta'_z}}{\mu^2}+(\zeta \rightarrow \bar \zeta)\right)\nonumber \ ,
\end{align}
and the renormalization group equation for $H$ reads:
\begin{align}
	\mu^2\frac{d}{d\mu^2}\ln H(\zeta_z,\zeta'_z,\bar \zeta_z,\bar \zeta'_z,\mu^2)=-2\gamma_F(\alpha_s)-\Gamma_{S}(\alpha_s) \ ,
\end{align}
where $\gamma_F$ and $\Gamma_S$ have been defined before.

Here we briefly comment on the end-point behavior. As $x\sim 0$, the hard kernel diverges logarithmically near the end point as $ 1+\alpha_s \ln^2 x$, but the quasi-WF amplitudes approach zero at large or small $x$ linearly, thus the end point regions behave as $x\ln^2 x$, which is free from those problems for the $k_T$ factorization for electromagnetic form factor~\cite{Li:1992nu}. Moreover, we can fix the $z$-component momentum transfer at each of the vertices to be $P^z$, which indicating that $x+x'=1$. In this case the end-point behavior is improved to $x^2\ln^2 x$.

\section{Lattice Parton Physics with LaMET}
\label{sec:lattice}

Lattice gauge theory simulates continuum QCD in imaginary time on a discretized 4D Euclidean lattice. The method is characterized by the finite lattice spacing $a$ and volume $L_1\times L_2\times L_3\times T$, and input parameters such as the strong coupling and quark masses. To calculate physical quantities, one usually expects to take the continuum ($a\to0$) and infinite volume $L_i, T\to\infty$ limits, as well as tuning the quark masses so that observables such as the pion mass $m_\pi$ agrees with the physical value of $\sim 140$ MeV. There are different methods to implement the fermions on the lattice~\cite{Rothe:1992nt}, which leads to different properties of the lattice action such as chiral symmetry breaking for Wilson fermions. In the lattice calculation of hadron matrix elements, the initial and final states are generated by acting the source and sink interpolation operators on the vacuum, and the ground-state contributions are filtered out by propagating over a sufficiently large Euclidean time. A boosted hadron state can be obtained by inserting momentum into the source and sink operators through Fourier transform in the 3D spatial coordinates.

The lattice QCD calculations of parton physics using LaMET started with the exploratory studies on the simplest PDFs and the gluon helicity~\cite{Lin:2014zya,Alexandrou:2015rja,Yang:2016plb}, which yielded fairly encouraging results, demonstrating that LaMET is a viable approach.
In subsequent studies, more attention has been paid to the systematics,
including establishing a proper renormalization and matching procedure, simulating at the physical pion mass, removing the excited-state contamination, etc. Such studies have greatly improved the precision of the calculations, with the latest results exhibiting a reasonable agreement with phenomenological PDFs~\cite{Alexandrou:2018pbm,Lin:2018qky,Alexandrou:2018eet,Izubuchi:2019lyk,Gao:2020ito}. In the meantime, explorations have also been made on similar large momentum data using the coordinate-space factorization methods including the pseudo-PDF~\cite{Orginos:2017kos,Joo:2019bzr,Joo:2019jct,Joo:2020spy} and current-current correlation~\cite{Bali:2019dqc,Sufian:2019bol,Sufian:2020vzb}.
Nevertheless, dedicated large-scale efforts with the state-of-art resources are yet to be seen. Lattice parton physics with LaMET is just at its dawn. With EIC
in the US going forward, a new era of lattice calculations is to come.

In this section, we summarize the current status of lattice calculations using LaMET and discuss future prospects.
We will begin with a general discussion on what kind of lattice setups are best suited for
LaMET calculations, and then briefly summarize relevant lattice techniques that facilitate such
calculations. After that, we review the lattice calculations that have been carried out so far and
point out future improvements. A nice complementary discussion about
lattice calculations has been made in~\cite{Cichy:2018mum}. Other reviews that summarize the recent developments in the lattice calculation of PDFs can be found in~\cite{Monahan:2018euv,Zhao:2020vll,Constantinou:2020pek}.

\subsection{Special Considerations for Lattice Calculations}
\label{sec:lattice-lattice}

In this subsection, we discuss the challenges for lattice calculations in LaMET, and estimate the required lattice
requirements by taking the collinear PDFs as an example.

\subsubsection{Challenges due to large momentum}
\label{sec:lattice-lattice-challenges}

\begin{figure}
	\includegraphics[width=0.9\linewidth]{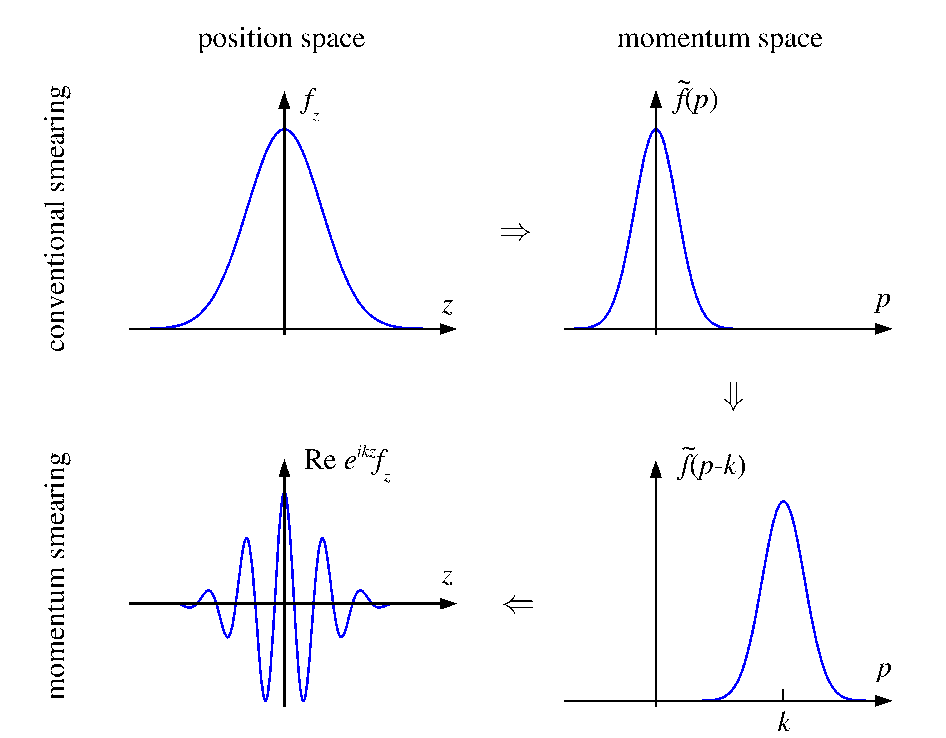}
	\caption{
		Conventional smearing (left) versus momentum smearing (right)~\cite{Bali:2016lva}:
		Conventional smearing has small overlap with high momentum state.
		Momentum smearing shifts momentum to peak at nonzero value in momentum space.
	}
	\label{fig:lattice-momentum-smearing}
\end{figure}

In addition to common challenges with other lattice calculations, such as taking the continuum and infinite volume limits, simulating at or extrapolating to the physical pion mass, etc., LaMET applications
require generating large-momentum hadron on lattice. For LaMET expansion,
$1/(xP^z)$ is the expansion parameter, and for the coordinate-space factorizations, large
quasi-light-cone distance $\lambda$ requires even bigger hadron momentum.
However, realizing this faces a number of practical challenges.
First, it has been difficult to generate large-momentum hadron states on the lattice, until the technique of momentum smearing~\cite{Bali:2016lva} was proposed.
The conventional smearing method in coordinate space is designed to increase the overlap with ground state hadron at rest. Thus, it is not surprising that such a smearing is not efficient when the hadron has a large momentum.
The momentum smearing technique introduces an extra phase factor $e^{i\vec{k}\cdot \vec{z}}$ to the quark field,
such that it is peaked at nonzero momentum $\vec{k}$ in Fourier space, as shown in~\fig{lattice-momentum-smearing}.
In this way, the overlap with high momentum state is vastly increased after Euclidean time evolution. Recently, the momentum smearing technique has been incorporated into the framework of distillation~\cite{Egerer:2020hnc} to improve the extraction of ground-state energy and matrix elements at momentum $\lesssim$ 3 GeV.
Although there are other proposed methods to generate large momentum~\cite{Wu:2018tvt}, the momentum smearing has become a standard technique in LaMET applications.

Second, the proton size is frame-dependent and changes with its momentum.
In the proton's rest frame, simulating its structure requires that the lattice spacing be much smaller than the QCD confinement scale, i.e. $a\ll\Lambda_{\rm QCD}^{-1}$.
When the proton is moving fast, it undergoes Lorentz contraction by a boost factor $\gamma$ in the momentum direction, thus a finer lattice spacing $a\ll(\gamma \Lambda_{\rm QCD})^{-1}$ is needed.
If $a\leq 0.2$ fm is the minimum requirement to investigate a static proton, one will need at least $a\leq 0.04$ fm to have the same resolution for a proton at 5 GeV.
A smaller lattice spacing is difficult to achieve with current computing resources,
for it suffers from the well-known critical slowing down problem, i.e., the auto-correlation times of observables such as the topological charge increase when approaching the continuum limit~\cite{Schaefer:2010hu}, which can be much longer than the Monte-Carlo simulation times.
A lattice with open (Neumann) boundary condition on gauge fields in the Euclidean time direction~\cite{Luscher:2011kk}, which allows topological charge to flow in and out at boundaries of time, may overcome this problem.

Third, the gaps between the ground state and the excited state energies become smaller because of the time dilation effect.
In the proton's rest frame, the excited state contamination exponentially decays with the mass gap $\Delta M$ and evolution time $\tau$ in the form of $e^{-\Delta M \tau}$.
In the boosted frame, the mass gap $\Delta M$ in the decay factor is replaced by the energy gap $\Delta E \sim \Delta M/\gamma$, and the decay changes like $e^{-\Delta M \tau} \to e^{-\Delta E \tau}=e^{-\Delta M \tau / \gamma}$ under Euclidean time evolution.
Therefore, with a boosted state, a longer time evolution (source-sink separation) is needed.
For example, if a source-sink separation of 1 fm is needed to separate the excited state of proton with 2 GeV momentum, a proton with 5 GeV momentum will require a source-sink separation of 2.5 fm.
Even if the two-state fit technique is used, a longer time evolution is still required so that only the ground and first excited states dominate.

Last but not the least, lattice calculation requires $P^z\ll 1/a$, so that discretization effects of ${\cal O}((aP^z)^n)$ are under control. Therefore, one has to go to smaller lattice spacing in order to reach larger momentum. The quantification of ${\cal O}((aP^z)^n)$ effects alone in LaMET calculations has not been done, as all discretization errors are treated on equal footing in continuum extrapolation.

To summarize, to achieve a precision calculation of boosted hadron structure on lattice, a fine lattice spacing (at least in the longitudinal direction) and a large box size in the time direction are essential, which of course will also require control over the signal-to-noise ratio at large Euclidean times.

\subsubsection{Considerations for lattice setup}
\label{sec:lattice-lattice-special}

In practical calculations, a correlation function is first obtained on lattice in coordinate space, and then Fourier transformed to momentum space with the phase factor $e^{i\lambda x}$ where $\lambda=zP^z$.
Therefore, the smallest $x$ one can reach can be roughly estimated from the largest $\lambda$ as $x\sim 1/\lambda$.
However, a more stringent constraint comes from requiring that the higher-twist contribution ${\cal O}(\Lambda_{\rm QCD}^2/(xP^z)^2)$ be small so that the factorization is still valid, which implies $x\gg \Lambda_{\rm QCD}/P^z$.
This also provides a rough estimate for the largest attainable $x$ ($x\ll 1-\Lambda_{\rm QCD}/P^z$) since the momentum fraction carried by other partons is $\sim (1-x)$ which should also be bounded from below by the above estimate.

For the current state-of-the-art simulations, the lattice spacing can reach 0.04 fm~\cite{Fan:2020nzz,Gao:2020ito}, which implies $P^z_{\rm max}\sim 5$ GeV and the effective resolution in longitudinal direction is about $\gamma a\sim 0.2$ fm.
Thus the valid $x$ region that can be extracted from lattice is roughly 0.1 to 0.9.
On the other hand, to avoid finite volume effects, it is believed that $m_\pi L\gtrsim 4$.
For physical pion mass, the box size in spatial direction $L$ should be at least 6 fm, which means the box size is 150 lattice spacing. So far, the largest box size in LaMET calculations is 5.8 fm~\cite{Lin:2018qky}.
As discussed in Sec.~\ref{sec:lattice-lattice-challenges}, the source-sink separation of 2.5 fm is needed for $P^z=5$ GeV.
So the box size in time direction $T$ does not need to be particularly longer than $L$, and $T=L$ is sufficient in this lattice setup.
In summary, with $a=0.04$ fm at physical pion mass, one need a
$L^3\times T=150^3\times 150$ lattice to reliably extract $0.1<x<0.9$ region, which
could be possible in an exa-scale computer.

There are potential tricks to reduce the computational cost.
First, the required source-sink separation can be shorter if one uses a multi-state instead of two-state fit with enough statistics.
However, since the number of fitting parameters in $n$-state fit grows as $n^2$, such a fitting will become infeasible for too large $n$.
Second, note that the resolution required for transverse proton structures is not affected by the Lorentz boost, one may use a coarse lattice in the transverse directions, $a_\perp=0.1$ fm.
The required box size is then $L_\parallel\times L_\perp^2\times T=150\times 60^2\times 150$.
This asymmetric lattice can greatly reduce the resources needed for large momentum since the transverse box size is fixed.
However, generating configurations and renormalization on such a lattice might bring new problems and should be further studied.

In the near future, exascale supercomputers may help to reach higher momentum, as large as 5 GeV for the proton,
and improve the precision of LaMET calculations. Further theoretical developments and new ideas on the technique and algorithms are also needed to overcome the simulation difficulties.

\subsection{Non-Singlet PDFs}
\label{sec:lattice-pdf}

In this subsection, we review current status of lattice calculations of flavor non-singlet (isovector) PDFs in the proton and
pion.
The non-singlet case has the advantage that the mixing with gluons as well as the lattice calculation of disconnected diagrams can be avoided, thus greatly reducing the computational challenge. It is the most extensively studied parton observable with LaMET so far.

\subsubsection{Proton}
\label{sec:lattice-pdf-proton}

The pioneering lattice studies for the isovector quark PDF in the proton were carried out in~\cite{Lin:2014zya,Alexandrou:2015rja}.
These are proof-of-principle studies as the renormalization of quasi-PDFs was not well understood at that time. Nevertheless, their results encouraged the follow-up theoretical works on LaMET, including a proper renormalization and matching suitable for lattice implementations.

Certain lattice artifacts have also been studied. For example, although there is no power-divergent mixing for the quasi-PDF operators on the lattice, additional operator mixings that are not seen in the continuum can still occur if a non-chiral lattice fermion such as the Wilson-type fermion is used.
In~\cite{Constantinou:2017sej,Chen:2017mie} it was shown that at ${\cal O}(a^0)$ the operator for the unpolarized quark quasi-PDF, $O_{\gamma^z}(z)$, can mix with the scalar operator $O_{\bf 1}(z)$, whereas $O_{\gamma^t}(z)$ does not.
To reduce the systematic uncertainty from such mixing, $\Gamma=\gamma^t$ has been used since then for lattice calculations of the unpolarized quark PDF, e.g. in~\cite{Alexandrou:2017huk,Green:2017xeu,Chen:2017mzz}. Similarly, for helicity and transversity cases, one should choose $\Gamma=\gamma^5\gamma^z$ and $\Gamma=i\sigma^{z\perp}=\gamma^\perp\gamma^z$, respectively, in order to avoid the mixing.
It should be noted that at ${\cal O}(a)$ all $\tilde{O}_\Gamma(z)$'s can mix with others~\cite{Chen:2017mie}. Nevertheless, a fine lattice spacing can reduce these effects.

\begin{figure}
	\includegraphics[width=0.45\textwidth]{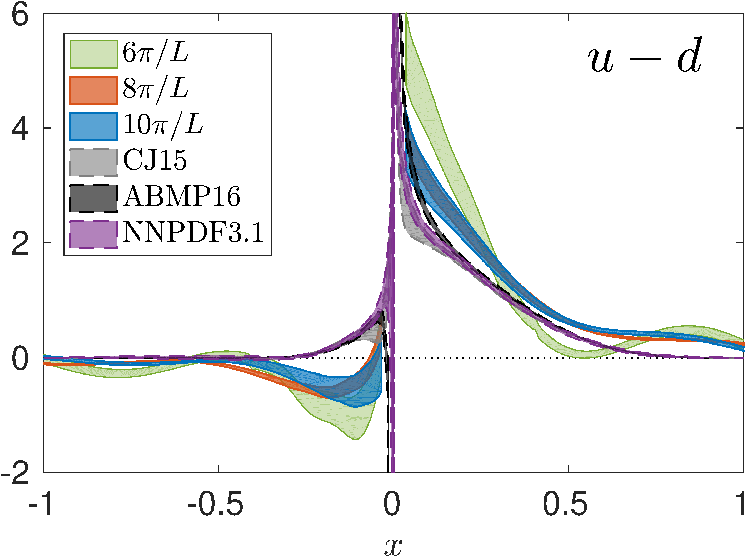}
	\includegraphics[width=0.45\textwidth]{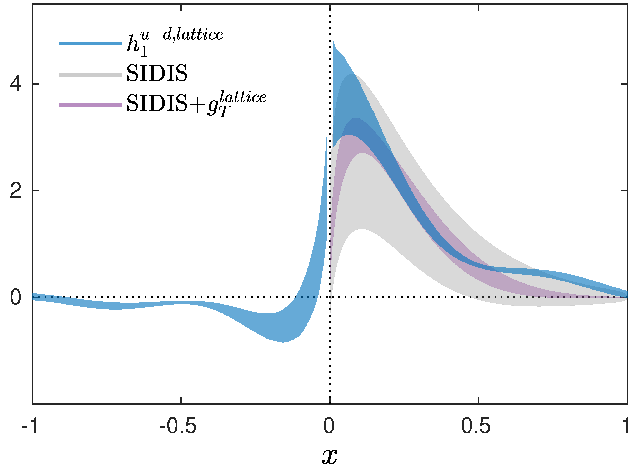}
	\caption{
		Proton isovector quark PDF~\cite{Alexandrou:2018pbm,Alexandrou:2018eet}:
		The unpolarized PDF with $P^z$ from 0.82 to 1.4 GeV and the transversity PDF with $P^z=1.4$ GeV are in upper and lower figures.
		CJ15~\cite{Accardi:2016qay}, ABMP16~\cite{Alekhin:2017kpj}, and NNPDF3.1~\cite{Ball:2017nwa} are global fits.
		SIDIS is global fit and SIDIS$+g_T^{lattice}$ is global fit with lattice constraint on tensor charge $g_T^{lattice}$~\cite{Lin:2017stx}.
	}
	\label{fig:lattice-proton_ETMC_2018}
\end{figure}

\begin{figure}
	\includegraphics[width=0.45\textwidth]{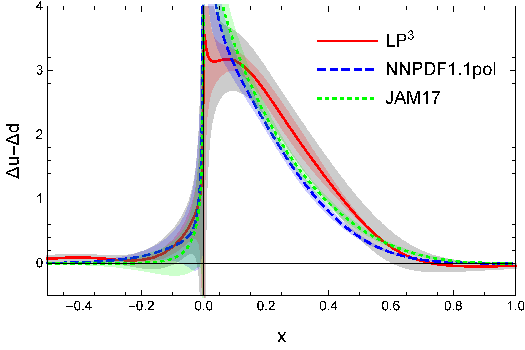}
	\caption{
		The proton isovector quark helicity PDF ($P^z=3.0$ GeV)~\cite{Lin:2018qky} with red band for statistic error and grey band for statistic and systematic errors.
		NNPDF1.1pol~\cite{Nocera:2014gqa} and JAM17~\cite{Ethier:2017zbq} are global fits.
	}
	\label{fig:lattice-proton_LP3_2018}
\end{figure}

In~\cite{Alexandrou:2017huk,Chen:2017mzz,Green:2017xeu}, the nonperturbative renormalization (NPR) of the quasi-PDFs was studied in the RI/MOM scheme~\cite{Martinelli:1994ty}.
This scheme has several advantages: The lattice regularization scheme can be converted to $\overline{\rm MS}$ scheme through RI/MOM renormalization condition,
the computation cost is affordable,
the systematic errors can be reduced or quantified more easily, etc.
The works before 2018 did not include NPR and the systematics were not accurately quantified.
The later works have implemented the RI/MOM scheme and the corresponding perturbative matching~\cite{Constantinou:2017sej,Stewart:2017tvs,Liu:2018uuj}.
The coordinate-space method is also developed in parallel in~\cite{Orginos:2017kos,Cichy:2019ebf,Joo:2019jct,Joo:2020spy,Bhat:2020ktg}.
In Figs.~\ref{fig:lattice-proton_ETMC_2018} and~\ref{fig:lattice-proton_LP3_2018}, we select some most recent lattice results.
ETMC published the proton unpolarized, helicity and transversity PDFs with $P^z=1.4$ GeV at physical pion mass~\cite{Alexandrou:2018pbm,Alexandrou:2018eet}, and ${\rm LP}^3$ published the proton helicity PDF with unprecedented momentum $P^z=3.0$ GeV at physical pion mass~\cite{Lin:2018qky}. Recently, calculations on fine lattices~\cite{Fan:2020nzz,Alexandrou:2020qtt} and an extrapolation to the continuum limit~\cite{Alexandrou:2020qtt} have become available. The finite volume effects, which was first studied in a model~\cite{Briceno:2018lfj}, have also been investigated on the lattice lately in~\cite{Lin:2019ocg}, where no sizeable volume-dependence was observed at $P^z=1.3$ and $2.6$ GeV.

The PDFs extracted from LaMET can be useful for phenomenology by providing input in kinematic regions that are difficult to measure in experiments.
It has attracted attention from global fit community~\cite{Lin:2017snn,Hobbs:2019gob,Lin:2020rut,Bringewatt:2020ixn}.
For example, it has been found that in the large-$x$ region of unpolarized PDF the lattice result will lead to significant improvement on global fit result if it reaches an accuracy of about $10\%$~\cite{Lin:2017snn}.
The sea quark asymmetry~\cite{Geesaman:2018ixo} is also possible to be investigated now directly on lattice.
For the transversity PDF, due to the difficulty of measurement in experiment, lattice results can already have impact on improving global fit and even making predictions. In addition to the isovector cases, calculations of the strange and charm unpolarized distributions~\cite{Zhang:2020dkn}, as well as the flavor separation of light quarks in the helicity PDF~\cite{Alexandrou:2020uyt}, have also been carried out recently.
From early exploratory results showing qualitative behavior of PDFs to the latest results which are comparable with global fits, it has come a long way in developing new techniques (momentum smearing, renormalization, matching, etc.) and the computation resources have been steadily increased over time.
The systematic uncertainties in the lattice calculation of PDFs have been thorougly investigated by the ETMC~\cite{Alexandrou:2019lfo}.
Further studies on systematics such as the discretization effects and finite volume effects on various lattice ensembles are still necessary.
In the future, lattice QCD is expected to make a significant impact on nuclon structure.

To conclude this subsection, we would like to mention that there are also lattice studies of the isovector PDF of other baryons, $\Delta^+$ to be more concrete, using LaMET~\cite{Chai:2020nxw}.

\subsubsection{Pion}
\label{sec:lattice-pdf-pion}

The pion valence quark distribution has been extracted from various Drell-Yan data for pion-nucleon/pion-nucleus scattering, while theoretical predictions do not yield consistent results with the experimental extraction, especially in large-$x$ region~\cite{Holt:2010vj}.
LaMET calculations will be able to shed valuable light on how to resolve this disagreement, provided that all systematics are well under control.

\begin{figure}
	\includegraphics[width=0.45\textwidth]{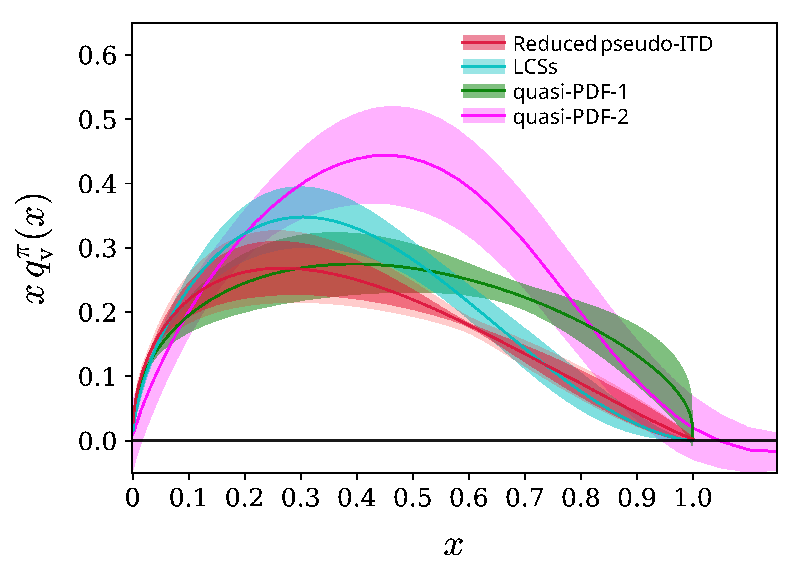}
	\caption{Pion valence quark PDFs in various approach:
		Compare the results of pesudo-PDF [Reduced pseudo-ITD~\cite{Joo:2019bzr}], quasi-PDF [quasi-PDF-1~\cite{Izubuchi:2019lyk} and quasi-PDF-2~\cite{Chen:2018fwa}], and the current-current correlator approach [LCSs~\cite{Sufian:2019bol}].
	}
	\label{fig:lattice-pion_valence_pseudo_2019}
\end{figure}

In principle, calculating the pion valence PDF is easier than the proton PDF. First, the pion state is easier to produce and the quark contractions are fewer. Second, the energy gap between the first excited and ground state of the pion is much bigger than the energy gap of the proton.
Therefore, the excited-state contamination is easier to control.
The simulation was first performed in~\cite{Chen:2018fwa} with the same lattice setup and procedure used in exploratory studies of the proton PDF.
A more thorough study on the pion valence quark PDF was done by the lattice QCD group of BNL~\cite{Izubuchi:2019lyk}.
It is worth pointing out that the excited state contamination was thoroughly studied using multi-state fits, with the ground and first excited states both agreeing with the expected dispersion relations,
indicating that the excited contamination is well under control.
The comparison of the lattice results from quasi-PDF, pseudo-PDF and current-current correlator approach are shown in Fig.~\ref{fig:lattice-pion_valence_pseudo_2019}.
Note that the $\rm LP^3$~\cite{Chen:2018fwa} result was obtained using Fourier transformation and inversion of factorization formula, while other three groups used parameterization models to fit the lattice data. More dedicated effort is needed to reduce the errors, and a meaningful comparison between different operators and analysis methods should be made.

For other mesons, we would like to mention that there is a study of kaon valence quark PDF using MILC configurations~\cite{Lin:2020ssv}.

\subsection{Gluon Helicity and Other Collinear Parton Properties}
\label{sec:lattice-other_distributions}

In this subsection, we summarize the applications of LaMET to other collinear parton observables, including the gluon helicity, the gluon PDFs, meson DAs and GPDs.

\subsubsection{Total gluon helicity}
\label{sec:lattice-gluon-helicity}

The total gluon helicity $\Delta G$ is a key component in understanding the proton spin structure. It has been intensively explored at RHIC and will be dedicatedly pursued at EIC in the future. However, a theoretical lattice calculation of $\Delta G$ had not been possible until the proposal of LaMET.

\begin{figure}
	\includegraphics[width=0.45\textwidth]{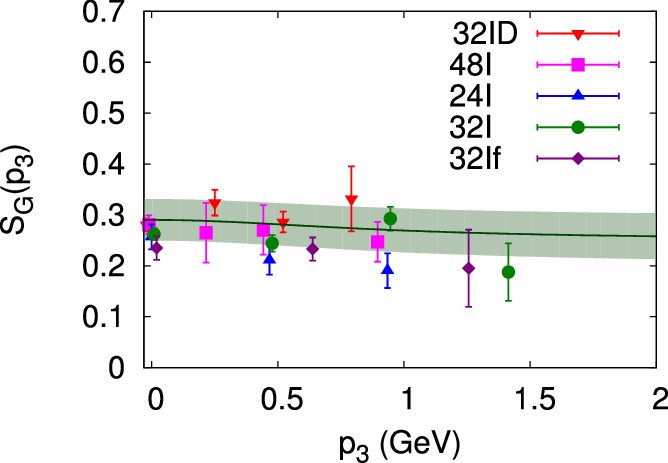}
	\caption{Total gluon helicity~\cite{Yang:2016plb}:
		The results are extrapolated to the physical pion mass and continuum as a function of the proton momentum $p_3$ on all the five ensembles indicated by different colors of the data points.
	}
	\label{fig:lattice-gluon_helicity_2016}
\end{figure}

The first such effort was made by $\chi$QCD collaboration in~\cite{Yang:2016plb}.  The calculation was carried out with valence overlap fermions on $2+1$ flavor domain-wall fermion gauge configurations, using ensembles with multiple lattice spacings and volumes including one with physical pion mass.
The authors simulated proton matrix elements of the free-field operator $(\vec E\times \vec A)^3$ in the Coulomb gauge at various momenta, and then converted them to the $\overline{\rm MS}$ scheme with one-loop lattice perturbation theory. The $\overline{\rm MS}$ matrix elements at each lattice momentum are shown in~\fig{lattice-gluon_helicity_2016}. Though a LaMET matching is necessary to match the results to the physical gluon helicity, the authors did not apply it due to the concern of perturbative convergence of the matching coefficient~\cite{Ji:2014lra}. Instead, as the $\overline{\rm MS}$ matrix elements show rather mild momentum dependence up to the maximum momentum $\sim$1.5 GeV, they extrapolated the results to infinite momentum, as well as physical pion mass and continuum limits, with a model motivated by chiral EFT. Their final result is $\Delta G(\mu^2 = 10\ \text{GeV}^2) = 0.251(47)(16)$, or 50(9)(3)\% of the total proton spin, which agrees with the truncated moment of $\Delta g(x)$~\cite{deFlorian:2014yva,Nocera:2014gqa} within uncertainties.

Despite such progress, one should be cautious that this calculation still needs further improvements in the future. Among others, the most important ones are simulations at larger proton momentum, performing an NPR and investigating perturbative convergence of LaMET matching and its implementation.

\subsubsection{Gluon PDF}
\label{sec:lattice-Gluon}

The gluon PDF is of great interest not only for precision physics at LHC, but also for understanding the gluonic structure of the proton and nuclei---as well as the small-$x$ dynamics---at the future EIC.
With the recent progress on the renormalization and matching for gluon quasi-PDFs~\cite{Wang:2017eel,Wang:2017qyg,Zhang:2018diq,Li:2018tpe,Wang:2019tgg} or the coordinate-space ``pseudo distributions''~\cite{Balitsky:2019krf}, a systematic lattice calculation of the gluon PDFs can be carried out in principle.

Before the above theoretical developments, an exploratory lattice study of the proton and pion unpolarized gluon PDFs were carried out in~\cite{Fan:2018dxu}. The authors calculated quasi gluon LF correlations and compared them to the LF correlations for the gluon PDFs. Later on, based on the multiplicative renormalizability of certain choice of the quasi gluon LF correlator~\cite{Zhang:2018diq}, the same authors used the ratio scheme~\cite{Balitsky:2019krf} in coordinate space to renormalize the lattice matrix elements, and fitted the proton unpolarized gluon PDF with a simple two-parameter model ~\cite{Fan:2020cpa}. Although the results show agreement with the global analyses in the large-$x$ region, the systematics from the model-dependence of the fit remains to be quantified for a controlled calculation of the gluon PDF.

\subsubsection{DA}
\label{sec:lattice-DA}

\begin{figure}
	\includegraphics[width=0.45\textwidth]{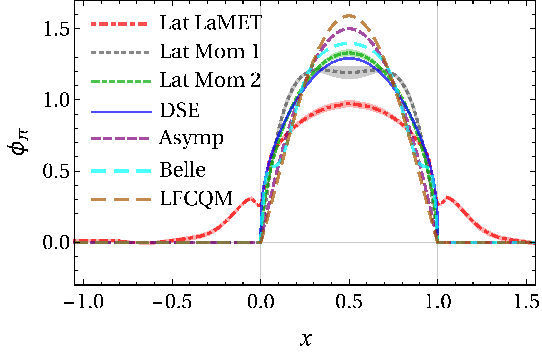}
	\caption{Pion DA~\cite{Chen:2017gck}:
		Comparison of $\phi_\pi$ (Lat LaMET) to previous determinations in the literature.
		Lat Mom 1 and 2 are parameterized fits to the lattice moments~\cite{Braun:2015axa};
		DSE is Dyson-Schwinger equation calculations~\cite{Chang:2013pq};
		Asymp is the asymptotic form $6x(1-x)$;
		Belle is a fit to the Belle data~\cite{Agaev:2012tm};
		LFCQM is light-front constituent quark model~\cite{deMelo:2015yxk}.
	}
	\label{fig:lattice-pion_DA_LP3_2017}
\end{figure}

According to \sec{others-da}, LaMET can be readily applied to calculating DAs, and the lattice resource needed is expected to be cheaper than that for PDFs since there is one less external state, which reduces the number of contractions for the quark propagators.
So far there are a few exploratory investigations on meson DAs, in particular, on pion~\cite{Zhang:2017bzy} and kaon DAs~\cite{Chen:2017gck}.
The lattice calculations of pion~\cite{Zhang:2017bzy} and kaon DAs~\cite{Chen:2017gck} were first explored without the NPR and the corresponding matching. Recently, the pion and kaon DAs from the RI/MOM scheme analysis are extrapolated to the continuum limit~\cite{Zhang:2020gaj}, where the authors eventually adopted a two-parameter model to fit the final result.
The above results are shown in~\fig{lattice-pion_DA_LP3_2017}.
Apart from LaMET, the current-current correlation methods~\cite{Braun:2007wv,Braun:2015axa,Detmold:2005gg} have also made much progress on the pion DA~\cite{Bali:2019dqc,Detmold:2018kwu,Detmold:2020lev}.
%

\subsubsection{GPD}
\label{sec:lattice-GPD}

As discussed in \sec{gpo}, the global fitting of GPDs still faces challenges from their complicated kinematic dependence and limited information from the experimental observables despite the progress made~\cite{Kumericki:2016ehc,Favart:2015umi}.
On the other hand, previous lattice QCD method is only able to calculate the lowest few moments of the GPDs~\cite{Hagler:2009ni}, which is far from sufficient to reconstruct their full kinematic dependence.
Applying LaMET to GPD calculations will provide important information on the GPDs, especially in kinematic regions that are not accessible in currently available experiments.
In addition, on the lattice one can study the GPD dependence on one kinematic variable by fixing the others. All these will help differentiate commonly used
models in GPD parameterization.

\begin{figure}
	\includegraphics[width=0.45\textwidth]{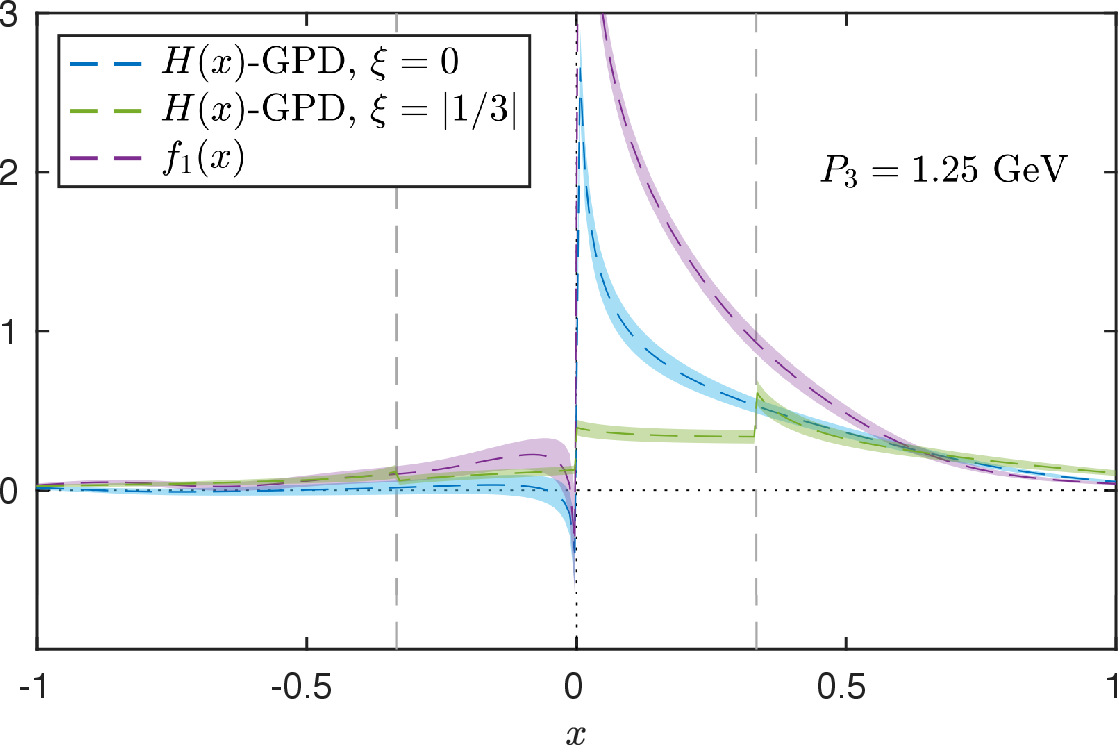}
	\caption{Proton unpolarized isovector quark GPD~\cite{Alexandrou:2020zbe} $H(x,\xi,t)$ for $t=-0.69\ {\rm GeV^2}$ extracted from quasi-GPDs at $P_3=1.25$ GeV, which is compared to the unpolarized PDF $f_1(x)$.
	}
	\label{fig:gpd}
\end{figure}

Calculating the quasi-GPDs requires more resources than quasi-PDF, but does not need further techniques in principle.
Besides, the lattice renormalization factors for the quasi-PDFs can be used here, as has been argued in \sec{gpo}.
The first lattice calculation of the pion unpolarized isovector quark GPD was carried out in~\cite{Chen:2019lcm}, though the results are not yet able to differentiate models or compare to experiments.
Recently, ETMC completed the first proof-of-principle calculation of the proton unpolarized and helicity GPDs~\cite{Alexandrou:2020zbe}, as shown in \fig{gpd}, which demonstrates that it is feasible to extract the GPDs with controlled systematics on available computational resources.

%

\subsubsection{Higher-twist PDF}

The higher-twist PDFs probe multi-parton correlations, and their contribution at $x=0$ can shed light on the LF zero modes~\cite{Ji:2020baz}. As we discussed in \sec{gpo}, such distributions can also be calculated on the lattice with the LaMET approach.

The first attempt to calculate the isovector twist-three PDF $g_T(x)$ has been carried out by ETMC~\cite{Bhattacharya:2020cen} using the one-loop matching coefficient they computed in~\cite{Bhattacharya:2020xlt,Bhattacharya:2020jfj}. Their results show agreement with the Wandzura-Wilczeck approximation~\cite{Wandzura:1977qf}, which ignores the contribution from dynamical twist-three contributions, and the Burkhardt-Cottingham sum rule~\cite{Burkhardt:1970ti}. Nevertheless, the mixing between $g_T(x)$ and other twist-three distributions was not considered, and further study is still required for an accurate matching to the light-cone PDF.

\subsection{TMDs}
\label{sec:lattice-TMD}

With tremendous experimental focus on the TMDPDFs for studying 3D proton structures and gluon saturation at EIC, their first-principle calculation from lattice QCD will significantly boost this direction by providing useful nonperturbative inputs for all the phenomenological analyses.

In this subsection, we discuss the status and prospects of calculating the quasi-TMDPDF and soft function with LaMET. Besides, we note that before LaMET there had already been efforts to extract information of TMDs by studying ratios of the lattice correlators~\cite{Hagler:2009mb,Musch:2010ka,Musch:2011er,Engelhardt:2015xja,Yoon:2017qzo}, which has made a series of progress in the past decade. We begin with a brief review of them.

\subsubsection{Pre-LaMET study --- ratio of lattice correlators}

By employing Lorentz covariance, the $x$-moments of TMDPDFs are related to the form factors of spacelike staple-shaped gauge link operators, which can be directly simulated on the lattice. Although the lattice calculation of the soft function was not available during that time, ratios of the spin-dependent and the unpolarized matrix elements were formed to cancel it, thus providing useful information of different TMDPDFs.
For example, the time-reversal odd TMDPDFs can be studied with the staple-shaped gauge link operator in a transversely polarized proton state, thus helping understand properties related to single-spin asymmetry (SSA), which was measured experimentally at STAR~\cite{Adamczyk:2015gyk}
and COMPASS~\cite{Aghasyan:2017jop}.
In~\cite{Musch:2011er,Engelhardt:2015xja}, the Sivers and Boer-Mulders functions of proton and pion were studied; Other time-reversal even functions, such as the worm-gear function $g_{1T}$~\cite{Tangerman:1994eh}, were also studied~\cite{Yoon:2017qzo}.

\subsubsection{Quasi-TMDPDF and Collins-Soper kernel}

The lattice calculation of the quasi-TMDPDF defined in \eq{quasi_TMD} is straightforward.
The matrix element of the staple-shaped quark Wilson line operator can be simulated the same way as the quasi-PDF case, except that the geometry of the gauge-link is different, while the calculation of Wilson loop $Z_E$ is standard practice in lattice QCD. The more challenging part, however, is the renormalization of the quasi-TMDPDF and its matching to the $\overline{\rm MS}$ scheme.

Using the auxiliary field theory formalism, one can argue that staple-shaped quark Wilson line operator is also multiplicatively renormalizable~\cite{Ebert:2019tvc,Green:2020xco}. On a non-chiral lattice, it suffers from finite mixing with other quark bilinear operators, as was predicted by one-loop lattice perturbation theory~\cite{Constantinou:2019vyb}. The full mixing pattern for such operators with different Dirac matrices have been studied in the RI/MOM scheme on three quenched lattice ensembles with different spacings~\cite{Shanahan:2019zcq}, and a diagonalization of the mixing matrix is adopted to renormalize these operators. Meanwhile, the one-loop conversion factors that convert the RI/MOM matrix elements to the $\overline{\rm MS}$ scheme have been calculated in continuum perturbation theory for both the $z=0$~\cite{Constantinou:2019vyb} and $z\neq0$~\cite{Ebert:2019tvc} cases.

Although the soft function is still needed to fully determine the physical TMDPDF, the $\overline{\rm MS}$ quasi-TMDPDF can already be used to extract the Collins-Soper kernel according to \eq{TMD-CS_kernel}~\cite{Ji:2014hxa,Ebert:2018gzl,Ebert:2019okf}. Since the Collins-Soper kernel can be defined from both the bare TMDPDF and the soft function, it is independent of the external state and can be calculated in a pion which is the least expensive on the lattice. Up to mass corrections suppressed by the momentum in \eq{quasiTMDmatch}, this calculation also allows for using an unphysical valence pion mass, as long as the sea quark masses are physical.

\begin{figure}[htb]
	\includegraphics[width=0.48\textwidth]{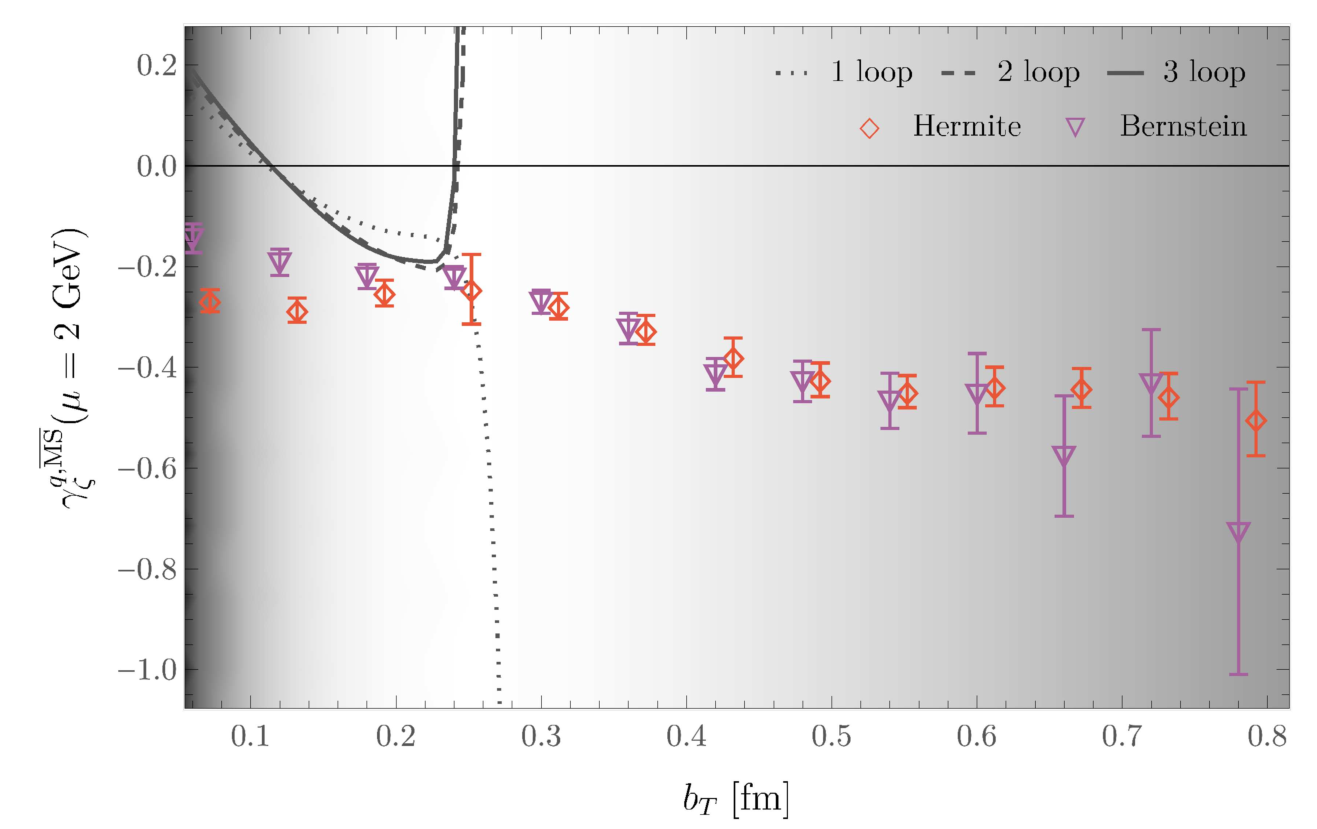}
	\caption{The Collins-Soper kernel from the first exploratory calculation on a quenched lattice~\cite{Shanahan:2020zxr}. The results are obtained by using fits to the $\MS$ unsubtracted quasi-TMDPDFs with Hermite and Bernstein polynomial bases. The solid and dashed lines are the perturbative predictions~\cite{Li:2016ctv,Vladimirov:2017ksc}, which is hit the Landau pole near $b_\perp\sim 0.25$ fm. The background shading density is proportional to a naive estimate of the power corrections $1/(b_\perp P^z) + b_\perp /L$.
	}
	\label{fig:lattice-cs}
\end{figure}

With the method developed in~\cite{Ebert:2018gzl}, the first exploratory lattice calculation of the Collins-Soper kernel was performed in~\cite{Shanahan:2020zxr} on a quenched lattice with heavy valence pion mass $m_\pi\sim 1.2$ GeV, and the result is shown in~\fig{lattice-cs}. As one can see, the lattice prediction is robust for $0.1\ {\rm fm} < b_\perp < 0.8\ {\rm fm}$, which covers the nonperturbative region that is important for TMD evolution in global analyses. Besides, at small $b_\perp$, the perturbative calculation can serve as a calibration for estimating the systematic uncertainties, as there are power corrections of ${\cal O}(1/(P^z b_\perp))$ which can only be reduced with larger $P^z$. In~\cite{Zhang:2020dbb}, the Collins-Soper kernel has also been extracted from a pion quasi-TMD DA, where the lattice renormalization was left out, and the result is in agreement with~\cite{Shanahan:2020zxr} within errors for a wide range of $b_\perp$.
With improved lattice ensembles and systematic corrections in the future, it is promising to have a precise determination of the Collins-Soper kernel for TMD phenomenology.

\subsubsection{Soft function}
\label{sec:lattice-SF}

As the remaining piece towards physical TMDPDFs, the soft function must be calculated in lattice QCD. In particular, the reduced soft function in \eq{reduced} eliminates the regulator-scheme-dependence of the off-the-light-cone quasi-TMDPDF, so its calculation alone has great physical significance. According to \secs{offlight}{others}, two methods have been proposed to calculate the off-the-light-cone soft function or reduced soft function on the lattice~\cite{Ji:2019sxk}, as we discuss in the following. One relies on simulating HQET on the lattice, while the other requires calculating a light-meson form factor of transversely-separated current products.

\begin{figure}[htb]
	\includegraphics[width=0.48\textwidth]{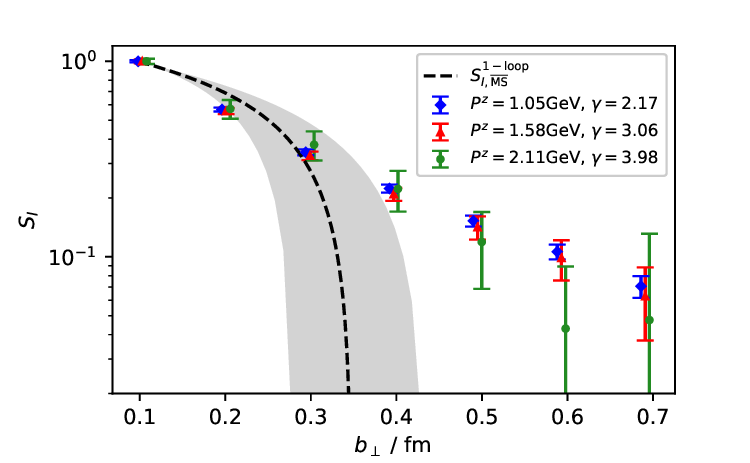}
	\caption{The reduced soft factor as a function of $b_\perp$ extracted from the light-meson form factor in \sec{tmd}~\cite{Zhang:2020dbb}. The results are obtained with quasi-TMD DAs at different pion momentum $P^z$, with perturbative matching and power corrections ignored. The dashed line is one-loop prediction in perturbation theory, which hits the Landau pole at $b_\perp\sim 0.3$ fm, and the grey band is the error by varying $\mu$ by a factor of $1/\sqrt{2}$ and $\sqrt{2}$.
	}
	\label{fig:lattice_soft}
\end{figure}

The latter method has been implemented in the first exploratory lattice calculation of the reduced soft function~\cite{Zhang:2020dbb}, which includes simulations of the pion form factor in two external states with opposite large momenta, as well as the pion quasi-TMD DA.
The results for the reduced soft function, which are obtained with tree-level matching and omission of lattice renormalization, are shown in \fig{lattice_soft}. As one can see, they agree with the perturbative prediction for small $b_\perp$ within errors, as expected, and start to deviate when $b_\perp$ becomes large. Since the quasi-TMD DA depends on the momentum $P^z$, the stability of results at different $P^z$ suggests the validity of \eq{form_qfac}. In the future, larger statistics and improved systematics in both lattice and perturbative matching will be necessary to achieve a precision calculation of this quanity.

\section{Conclusion and Outlook}
\label{sec:conclusion}

Since Feynman proposed the parton model more than fifty years ago, our understanding of the partonic structure of the proton has been greatly advanced. On one hand, a number of high-energy experiments carried out at facilities worldwide including SLAC, DESY, CERN, Fermi Lab, JLab, BNL, etc. allowed us to probe various aspects of hadronic structures at different energies and polarizations. On the other hand, many parton observables have been proposed in parallel
that provide a multi-dimensional description of the proton structure, including the collinear
PDFs, TMDPDFs, GPDs, parton DAs, LFWFs and so on.

Although QCD factorization theorems with RG improvement allow us to extract these parton observables through their connection to experimental observables, it is highly desirable to predict them from {\it ab initio}
calculations such as lattice QCD. Developments along this line have been rather slow due to
difficulties in simulating real-time dynamics. The situation, however, has changed since the proposal of LaMET a few years ago, which provides a systematically improvable method to calculate parton physics from first principles.

In this paper, we give an overview of LaMET formalism and its applications to observables which
can be accessed in lattice QCD and other non-perturbative methods.  By investigating the frame dependence of the structure of bound state hadrons, we explain how the IMF physics or parton physics naturally arises as an EFT description of the proton structure. Such an EFT description is most naturally formulated in SCET and LFQ,
but practical non-perturbative calculations of the proton matrix elements
have been difficult. LaMET in effect provides what is needed to realize LFQ. This is achieved by forming appropriate quasi parton observables in a large momentum state and match them to the true parton observables on the LF through factorization. In the case of PDFs, the former corresponds to finite-momentum distributions whose running is controlled by the momentum RGE, whereas the latter corresponds to IMF PDFs whose running is controlled by the usual RGE. It should be pointed out that
LaMET is a very general framework which can be applied to
large-momentum physical quantities calculated with any non-perturbative methods,
either Euclidean (with imaginary time) or Minkowskian (with real time).
Moreover, given a large momentum state, the same parton physics can be determined
from different quasi observables that form a universality class.

We then present how to calculate the parton observables in practice, with a particular focus on the collinear PDFs, GPDs, DAs, TMDPDFs and LFWFs. We also discuss the proton spin structure and show how the partonic contributions to proton spin can be obtained following the same approach. We finally summarize the lattice studies carried out so far with LaMET which, on one hand, demonstrate that LaMET is a promising approach to compute partonic structures of the proton, and on the other hand, clearly indicate that a lot of improvements are still required to reach such an accuracy that the lattice results can have considerable impact on phenomenology.

We complete this review with a few comments on improvements of lattice calculations for the future.
We recommend~\cite{Alexandrou:2019lfo} for more systematic discussion on some of the issues, for example, the continuum, infinite volume, and physical pion mass limits.

\begin{itemize}
	
	\item Large hadron momentum. Since the future of LaMET lies in larger momenta which naturally require smaller lattice spacings, it will be critical to address the challenges from using large momenta and small spacings for exa-scale computations, such as the excited state contamination or topological charge freezing problem.
	
	\item Renormalization. As discussed in \sec{renorm-npr}, the mass renormalization of Wilson line operators is favored for it is gauge invariant and does not introduce extra higher-twist effects or large statistical errors at long distance. However, its matching to the $\MS$ scheme, especially the renormalon ambiguities, still needs to be resolved for a full systematic application. Moreover, alternative schemes that include the above features are also highly desirable.
	
	\item Higher-order perturbative matching. In current LaMET calculations, one-loop perturbative matching
	has brought considerable corrections. Higher-order matching kernels will be necessary to control the systematics from this procedure.
	
	\item Power corrections. They are important if the hadron momentum is not very large or when $x$ is close to 0 or 1.
	Little progress has been made toward a model-independent determination of the power corrections so far. One contingent strategy is to extrapolate to $P^z\to\infty$ limit after implementing matching and target-mass corrections, but the ultimate solution relies on the lattice calculation of higher-twist distributions that has been discussed in \sec{others-ht}.
	
\end{itemize}

The above discussion of systematics is generic and applies to all quasi-observables. The rich theoretical developments in the past years have paved the way for calculating a wide range of parton observables using LaMET. With the rapid increase in computing resources and progress in developing new techniques and algorithms, we expect to see the above systematics to be kept under control step by step in the future. That would be important in establishing LaMET as a systematic approach to computing parton physics, and making lattice calculations play a crucial role in the EIC era.

\section*{Acknowledgments}

The authors are thankful for collaborations with G. Bali, V. M. Braun, J.-W. Chen, W. Detmold, M. Ebert, X. Gao, B. Gl\"assle, M. G\"ockeler, M. Gruber, Y. Hatta, F. Hutzler, T. Ishikawa, T. Izubuchi, L. Jin, N. Karthik, P. Korcyl, R. Li, C.-J. D. Lin, H.-W. Lin, K.-F. Liu, C. Monahan, S. Mukherjee, P. Petreczky, A. Sch\"{a}fer, C. Schugert, P. Shanahan, I. Stewart, P. Sun, S. Syritsyn, A. Vladimirov, M. Wagman, W. Wang, P. Wein, X. Xiong, J. Xu, Y. Xu, Y.-B. Yang, F. Yuan, I. Zahed, Q.-A. Zhang, R. Zhang, S. Zhao and R. Zhu. The authors are also indebted to the enlightening discussions with J. Chang, K. T. Chao, K. Cichy, T. Cohen, J. Collins, M. Constantinou, M. Engelhardt, J. Green, Y. Jia, K. Jansen, K. Lee, J. Karpie, H.-N. Li,
J. P. Ma, Y.-Q. Ma, A. Manohar, R. McKeown, S. Meinel, B. Mistlberger, J. Negele, K. Orginos, J. Qiu, A. Radyushkin, E. Shuryak, G. Sterman, R. Sufian, F. Steffens, A. Walker-Loud, and C.-P. Yuan. XJ has been partially supported by the U.S. Department of Energy, Office of Science, Office of Nuclear Physics, under contract No.~DE-FG02-93ER-40762. YZ is supported by the U.S. Department of Energy, Office of Science, Office of Nuclear Physics, under contract No.~DE-SC0012704, No.~DE-AC02-06CH11357, and within the framework of the TMD Topical Collaboration. YZ is also supported by the U.S. Department of Energy, Office of Science, Office of Nuclear Physics and Office of Advanced Scientific Computing Research within the framework of Scientific Discovery through Advance Computing (ScIDAC) award Computing the Properties of Matter with Leadership Computing Resources.
JHZ is supported in part by National Natural Science Foundation of China under grant No. 11975051, No. 12061131006, and by the Fundamental Research Funds for the Central Universities. YSL is supported by National Natural Science Foundation of China under Grant No. 11905126.

\section*{Appendix A: Acronyms, abbreviations and terminologies}
\label{sec:abbreviation}
Here we list some acronyms, abbreviations and terminologies used throughout this review:
\begin{equation}
\begin{array}{L@{\qquad}L}
AM & angular momentum \\
BFKL & Balitsky–Fadin–Kuraev–Lipatov \\
BPHZ & Bogoliubov-Parasiuk-Hepp-Zimmermann\\
BRST & Becchi-Rouet-Stora-Tyutin \\
DA & distribution amplitude \\
DGLAP & Dokshizer-Gribov-Lipatov-Altarelli-Parisi \\
DIS & deep-inelastic scattering \\
DR & dimensional regularization \\
DVCS & deeply-virtual Compton scattering\\
DVMP & deeply-virtual meson production \\
DY & Drell-Yan \\
EFT & effective field theory \\
EIC & Electron-Ion Collider \\
EOM & equation of motion \\
ERBL & Efremov-Radyushkin-Brodsky-Lepage \\
GCPO & generalized collinear parton observable \\
GPD & generalized parton distribution \\
GTMD & generalized transverse-momentum-\\
\ &dependent distribution\\
HQET & heavy-quark effective theory \\
IMF & infinite-momentum frame \\
IR & infrared \\
LaMET & large momentum effective theory \\
LC & light-cone \\
LF & light-front \\
LFWF & light-front wave function \\
$\overline {\rm MS}$ & modified minimal subtraction \\
NPR & non-perturbative renormalization \\
OAM & orbital angular momentum \\
OPE & operator product expansion \\
PDF & parton distribution function \\
QCD & quantum chromodynamics \\
QED & quantum electrodynamics \\
QFT & quantum field theory \\
RGE & renormalization group equation \\
RI/MOM & regularization-independent\\
\ &  momentum subtraction \\
SCET & soft-collinear effective theory \\
SIDIS & semi-inclusive deep-inelastic scattering \\
TMD & transverse-momentum-dependent \\
UV & ultraviolet
\end{array}\nonumber
\end{equation}
\begin{itemize}
	\setlength\itemsep{0em}
\item[] {\it \noindent Parton model}: a model proposed by R. Feynman in which hadrons are viewed as a collection of point-like quasi-free partons.
\item[] {\it Parton distribution function}: a probability function describing how the longitudinal momentum is distributed among the partons (quarks and gluons) in a hadron.
\item[] {\it Factorization theorem}: a theorem that separates hadronic observables into process-dependent short-distance partonic observables and universal long-distance functions characterizing the hadron structure.
\item[] {\it Light-front quantization}: a quantization program that is carried out at equal light-front time and yields a relativistic description of QCD bound states in terms of light-front wave functions.
\item[] {\it Bjorken $x_B$}: The variable proposed by J. D. Bjorken to characterize the kinematics in DIS. Its definition is given above \eq{diskin}.
\item[] {\it Scaling}: The behavior that an observable is independent of the scale at which it is probed.
\item[] {\it Effective field theory}: a theory framework that describes physical phenomena at a given length scale using only active degrees of freedom at that scale, while integrating out degrees of freedom at other length scales.
\item[] {\it Renormalization group equation}: an equation that describes how a physical system can be viewed and interpreted at different scales.
\item[] {\it HQET}: an effective field theory obtained from QCD by taking the infinite heavy quark mass limit.
\item[] {\it Gauge link or Wilson line}: a nonlocal quantity constructed as exponentials of integrals of gauge fields along a given path, used to connect fields at different spacetime points to maintain gauge invariance.
\item[] {\it Compton amplitude}: the quantum amplitude for scattering of a (virtual) photon by the proton.
\item[] {\it Auxiliary field approach}: an approach in which the nonlocal gauge link can be replaced by the two-point function of the auxiliary field.
\item[] {\it Matching}: a procedure used to relate full theory operators to effective field theory operators.
\item[] {\it Nonsinglet}: a combination accounting for the difference between quark distributions, e.g., the isovector combination $u-d$ discussed extensively in the context of this review.
\item[] {\it Universality class}: a collection of operators that flow into the same fixed point under momentum renormalization group running.
\item[] {\it Quasi-light-front correlations}: spatial correlations defining the finite momentum distributions.
\item[] {\it Collinear divergence}: divergence in a Feynman diagram when loop momentum of the internal line is collinear to that of the external massless particle.
\item[] {\it Two-particle-irreducible diagram}: a Feynman diagram that cannot be divided into disconnected parts by cutting two internal lines.
\item[] {\it Wilson fermion}: a way to discretize the QCD fermion action on the lattice, which breaks down the chiral symmetry.
\item[] {\it Generalized parton distribution}: generalization of PDFs to non-forward kinematics, i.e., the initial and final states have different momenta.
\item[] {\it Skewness}: defined to characterize the longitudinal momentum transfer in GPDs.
\item[] {\it Distribution amplitude}: transition matrix element between vacuum and hadron state, representing the probability amplitude of finding a given Fock state in the hadron.
\item[] {\it Twist}: defined as dimension - spin of the operator. Leading-twist (higher-twist) denotes the leading (nonleading) power behavior in the quantity under investigation.
\item[] {\it Zero-mode}: the degrees of freedom with zero longitudinal momentum in LFQ.
\item[] {\it Transverse momentum dependent(TMD) PDF}: defined in Eq.~(\ref{eq:TMDPDF}), the distribution function of both longitudinal and transverse momentum for partons.
\item[] {\it Staple-shaped gauge-link}: the pair of gauge-links separated along transverse directions that appear in the definition of TMD-PDFs, they are defined in Eq.~(\ref{eq:staplen}), Eq.~(\ref{eq:staplep}) and Eq.~(\ref{eq:staplez}).
\item[] {\it Rapidity divergence}: the divergence of TMDPDF and soft functions due to the presence of infinite rapidity scale introduced by the infinite-long gauge-links.
\item[] {\it Light-cone regulator}: regulators that regulates the rapidity divergence. First appeared in the paragraph of Eq.~(\ref{eq:RD}).
\item[] {\it On-light-cone}: rapidity regulator that maintain the presence of light-like separations in the gauge-link.
\item[] {\it Off-light-cone}: rapidity regulator that makes the separations of the gauge-link non-light-like.
\item[] {\it Soft function}:  functions that capture the factorable soft radiations of TMDPDF. Defined in Eq.~(\ref{eq:soft}) for on-light-cone and Eq.~(\ref{eq:S_DY}) for off-light-cone.
\item[] {\it Collins Soper kernel}: the kernel for rapidity evolution of TMDPDF, see Eq.~(\ref{eq:tmd_evolution}).
\item[] {\it Quasi-TMDPDF}: defined in Eq.~(\ref{eq:quasi_TMD}), similar to TMDPDF but with light-like separations replaced by space-like ones.
\item[] {\it Pinch-pole singularity}: the divergence due to infinite long gauge-link pair in the quasi-TMDPDF. Can be subtracted out by the factor $Z_E$, see discussion below Eq.~(\ref{eq:Z_E}).
\item[] {\it Off-light-cone soft function}: soft function using off-light-cone regulator. Defined in Eq.~(\ref{eq:S_DY}) and Eq.~(\ref{eq:S_t}). Required for matching quasi-TMDPDF to TMDPDF.
\item[] {\it Reduced soft function}: the rapidity independent part of off-light-cone soft function, see Eq.~(\ref{eq:reduced}).
\item[] {\it Light-Front wave function}: the wave function for hadron state in light-front quantization, expanded in the Free -Fock state.
\item[] {\it Reduced diagram}: the diagram showing the power-leading region of IR divergences. All the IR safe propagators are shrunk to blobs.
\item[] {\it Momentum smearing}: a lattice technique to increase the overlap of the field and nonzero-momentum state.
\item[] {\it Non-singlet}: transforms under the fundmental representation of $SU(N_f)$ with $N_f$ the quark flavor number.
\end{itemize}

\section*{Appendix B: Conventions}
\label{sec:notation}

We use the following convention for the metric tensor
\begin{align}\label{eq:metric}
g^{\mu\nu} = \text{diag}\,(1,-1,-1,-1)\,.
\end{align}

In ordinary coordinates, a generic four-vector is denoted as $v^\mu=(v^0, v^x, v^y, v^z)$ or $v^\mu=(v^0, {\vec v}_\perp, v^z)$. For example, the spacelike and timelike direction vector are written as $n_z=(0,0,0,1)$ and $n_t=(1,0,0,0)$, respectively. In light-cone coordinates $\xi^\pm ={1\over \sqrt{2}}(\xi^0\pm \xi^3)$, a vector is denoted as $v^\mu=(v^+, v^-, {\vec v}_\perp)$.

The hadron state $|P\rangle$ is normalized as
\begin{align}\label{eq:norm}
\langle P'|P\rangle = (2\pi)^3 2P^0 \delta^{(3)}(\vec{P}-\vec{P}')\,.
\end{align}

The covariant derivative and the Wilson line gauge link in the fundamental representation are defined as
\begin{align}
D^\mu \psi=(\partial^\mu+ig A^\mu)\psi=(\partial^\mu+ig t^a A_a^\mu)\psi,
\end{align}
and
\begin{align} \label{eq:wilsonline}
&W(x_2,x_1)=\nonumber \\
&\exp\left[-ig\int_{0}^{1} dt (x_2-x_1)_\mu A^\mu(x_1+(x_2-x_1)t)\right]\,.
\end{align}
The ones in the adjoint representation are completely analogous.

We use $O_\Gamma(s)$ to generically denote an operator defining the corresponding (quasi) parton observable, where $s$ can be a lightlike (for parton observables) or spacelike (for quasi parton observables) separation, and $\Gamma$ is a Dirac structure. The momentum fraction in a quasi-observable is denoted as $y$, while that in the usual parton observable is denoted as $x$.

The lightcone operator that defines the quark parton observable is
\begin{align}
O_\Gamma(\lambda n)=\bar\psi(0)\Gamma W(0,\lambda n)\psi(\lambda n)
\end{align}
with  $\Gamma$ denoting a Dirac matrix. If we take $\Gamma=\slashed{n}\equiv \gamma^+$, the unpolarized quark PDF is then given by
\begin{align}
q(x)=\frac{1}{2P^+}\int\frac{d\lambda }{2\pi}e^{ix\lambda }\langle P|O_{\gamma^+}(\lambda n)|P\rangle
\end{align}
with $n^\mu=1/\sqrt{2}(1/P^+,0,0,-1/P^+)$.

Accordingly, the quark quasi-observable is defined by
\begin{align}
{O}_\Gamma(z )=\bar\psi(zn_z/2)\Gamma W(zn_z/2,-zn_z/2)\psi(-zn_z/2) \ .
\end{align}
If we choose $\Gamma=\gamma^t$, the unpolarized quark quasi-PDF is then defined as
\begin{align}
{\tilde q}(y)=\frac{1}{2P^0}\int\frac{d\lambda }{2\pi}e^{iy \lambda}\langle P|{O}_{\gamma^t}(z )|P\rangle
\end{align}
with the quasi light-cone distance $\lambda =z P^z$.

The staple-shaped gauge link required for the TMDPDFs is defined as:
\begin{align}
{\cal W}_n(\lambda n/2+\vec{b}_\perp)=W^{\dagger}_n(\lambda n/2+\vec{b}_\perp)W_\perp W_n(-\lambda n/2) \ ,
\end{align}
where
\begin{align}
W_n(\xi)=W(\xi+\infty n,\xi) \ .
\end{align}

The un-subtracted unpolarized quark TMDPDF is then defined as:
\begin{align}
& f(x,\vec{k}_\perp,\mu,\delta^-/P^+) = \frac{1}{2P^+}\int\frac{d\lambda}{2\pi}\frac{d^2\vec{b}_\perp}{(2\pi)^2}
 e^{-i\lambda x+i\vec{k}_\perp \cdot \vec{b}_\perp}\nonumber \\
& \times \langle P| \bar \psi(\lambda n/2 +\vec{b}_\perp)\slashed{n}{\cal W}_n(\lambda n/2+\vec{b}_\perp)|_{\delta^-}\psi(-\lambda n/2)|P\rangle  \ ,
\end{align}
and the TMD soft function for DY process is defined as:
\begin{align}
& S(b_\perp,\mu,\delta^+,\delta^-)\nonumber \\
&=\frac{{\rm Tr}\langle 0|{\cal \bar T}W_p(\vec{b}_\perp)|_{\delta^+}W_n^{\dagger}(\vec{b}_\perp)|_{\delta^-}{\cal T} W_n(0)|_{\delta^-}W^{\dagger}_p(0)|_{\delta^+}|0\rangle }{N_c} \nonumber \\
&=\frac{{\rm Tr}\langle 0|{\cal W}_{n}(\vec{b}_\perp)|_{\delta^+} {\cal W}^{\dagger}_{p}(\vec{b}_\perp)|_{\delta^-}|0\rangle}{N_c} \,,
\end{align}
where $|_{\delta^{\pm}}$ denotes the rapidity regulator for the gauge links involved. In terms of these, the physical scheme independent TMDPDF is defiend as:
\begin{align}
&f^{\rm TMD}(x,b_\perp,\mu,\zeta)=\lim_{\delta^- \rightarrow 0}\frac{f(x,b_\perp,\mu,\delta^-/P^+)}{\sqrt{S(b_\perp,\mu,\delta^- e^{2y_n},\delta^-)}} \ ,
\end{align}
where $\zeta \equiv 2(xP^+)^2e^{2y_n}$ is the rapidity scale.

The staple-shaped gauge link for the quasi-TMDPDF is defined as:
\begin{align}
{\cal W}_{z}(\frac{\lambda n_z}{2}+\vec{b}_\perp;L)=W^{\dagger}_{z}(\xi; L)W_{\perp}W_{z}(-\xi^z n_z;L) \ ,
\end{align}
where
\begin{align}
W_z(\xi)=W(\xi+(L-\xi^z) n_z,\xi) \ .
\end{align}
The quasi-TMDPDF is then defined using ${\cal W}_{z}(\frac{\lambda n_z}{2}+\vec{b}_\perp;L)$ in exactly the same way as that for the un-subtracted TMDPDF:
\begin{align}
& \tilde f(\lambda ,b_\perp,\mu,\zeta_z) =\\
& \lim_{L \rightarrow \infty}  \frac{\langle P| \bar \psi\left(\frac{\lambda n_z }{2}+\vec{b}_\perp\right)\gamma^z{\cal W}_{z}(\frac{\lambda n_z}{2}+\vec{b}_\perp;L)\psi\left(-\frac{\lambda n_z}{2}\right) |P\rangle}{\sqrt{Z_E(2L,b_\perp,\mu)}} \ ,\nonumber
\end{align}
where $Z_E(2L,b_\perp,\mu)$ is a flat rectangular Euclidean Wilson-loop along the $n_z$ direction with length $2L$ and width $b_\perp$:
\begin{align}
Z_E(2L,b_\perp,\mu)=\frac{1}{N_c}{\rm Tr}\langle 0|W_{\perp}{\cal W}_z(\vec{b}_\perp;2L)|0\rangle \ .
\end{align}
The staple-shaped operators for LFWFs and quasi-LFWFs are the same as those for TMD-PDFs and quasi-TMDPDFs, and can be found in \sec{others}.

\newpage

\bibliography{lamet}

\end{document}